\PassOptionsToPackage{x11names,rgb,html,table}{xcolor}
\documentclass[acmsmall,nonacm,authorversion,screen,british,submission,appendix,final]{acmart}
\makeatletter  %
\PackageWarning{A}{\ACM@format@nr}
\ifnum\ACM@format@nr=4\relax %
	\@twocolumntrue
\fi
\makeatother

\input{preamble}
\usepackage[final]{listings}

\expandafter\patchcmd\csname \string\lstinline\endcsname{%
  \leavevmode
  \bgroup
}{%
  \leavevmode
  \ifmmode\mbox\fi
  \bgroup
}{}{%
  \typeout{Patching of \string\lstinline\space failed!}%
}

\lstdefinelanguage{Choral}{
  morekeywords={class,else,extends,if,implements,interface,new,null,private,%
      protected,public,abstract,final,static,return,super,this,try,catch,void, %
      enum,switch,case,default,throw,throws,with,forall,in},%
  otherkeywords={>>,::,@},
  sensitive,%
  escapechar=¦,
	morecomment=[l]//,%
	morecomment=[s]{/*}{*/},%
	morestring=[b]",%
  morestring=[b]',%
  moredelim=[is][]{\#}{\#}%
}[keywords,comments,strings]%

\lstdefinelanguage{comperr}{
  morekeywords={class,else,extends,if,implements,interface,new,null,private,%
      protected,public,abstract,final,static,return,super,this,try,catch,void, %
      enum,switch,case,default,throw,throws},%
  otherkeywords={>>,@},
  sensitive,%
	morecomment=[l]//,%
	morecomment=[s]{/*}{*/},%
	morestring=[b]",%
  morestring=[b]',%
  moredelim=[is][]{\#}{\#},
  moredelim=[is][\bfseries\color{sred}]{~}{~},
  moredelim=[is][basicstyle]{/~}{~/}
}[keywords,comments,strings]%

\usepackage[scaled=.85]{FiraMono}

\definecolor{sbase03}{HTML}{002B36}
\definecolor{sbase02}{HTML}{073642}
\definecolor{sbase01}{HTML}{586E75}
\definecolor{sbase00}{HTML}{657B83}
\definecolor{sbase0}{HTML}{839496}
\definecolor{sbase1}{HTML}{93A1A1}
\definecolor{sbase2}{HTML}{EEE8D5}
\definecolor{sbase3}{HTML}{FDF6E3}
\definecolor{syellow}{HTML}{B58900}
\definecolor{sorange}{HTML}{CB4B16}
\definecolor{sred}{HTML}{DC322F}
\definecolor{smagenta}{HTML}{D33682}
\definecolor{sviolet}{HTML}{6C71C4}
\definecolor{sblue}{HTML}{268BD2}
\definecolor{scyan}{HTML}{2AA198}
\definecolor{sgreen}{HTML}{859900}

\lstdefinestyle{solarized-light}{
  frame=none,
  breaklines=true,
  showstringspaces=false,
  tabsize=1,
  columns=fixed,
  mathescape=true,
  extendedchars=true,
  backgroundcolor=\color{sbase3},
  keywordstyle=\bfseries\color{sbase01},
  keywordstyle=[2]\bfseries\color{sbase01},
  stringstyle=\color{sviolet},
  numberstyle=\color{sviolet},
  identifierstyle=\color{sbase03},
  commentstyle=\color{sgreen},
  basicstyle=\color{sbase03}\ttfamily\lst@ifdisplaystyle\footnotesize\fi,
  moredelim=[is][\color{sblue}]{\#}{\#},
}

\usepackage[most]{tcolorbox}
\tcbuselibrary{listings}

\NewDocumentCommand{\newlang}{o m m m}{%
  \IfNoValueTF{#1}{\def\n{#2}}{\def\n{#1}}
  \expandafter\NewDocumentCommand\csname \n listing\endcsname{s O{} O{}}{%
	  \def\WithoutTitle{\tcblisting{
	          enhanced, %
	          before skip=\abovedisplayskip,
	          after  skip=\belowdisplayskip,
	          sharp corners=all,
	          boxrule=2pt,
	          boxsep=-.7em,
	          colframe=#4,
	          colback=sbase3,
	          title={},
	          listing only,
	          listing options={language=#2,numbers=left,style=solarized-light,##3},
	          ##2
 	  	}}%
  	\def\withTitle{\tcblisting{
         enhanced, %
         before skip=\abovedisplayskip,
         after  skip=\belowdisplayskip,
         sharp corners=all,
         boxrule=2pt,
         boxsep=-.7em,
         colframe=#4,
         colback=sbase3,
         detach title,
         finish={\node[anchor=south east, font=\footnotesize\itshape,
         text=sbase3,fill=#4] at (frame.south east) {#3 Code};},
         title={},
         listing only,
         listing options={language=#2,numbers=left,style=solarized-light,##3},
         ##2
   }}%
   \IfBooleanTF{##1}{\WithoutTitle}{\withTitle}}
  \expandafter\def\csname end\n listing\endcsname{\endtcblisting\noindent}
  \expandafter\def\csname\n\endcsname{\lstinline[language=#2,style=solarized-light]}
}

\newlang[chor]{choral}{Choral}{sblue}
\newlang[comm]{java}{Generated}{syellow}
\newlang[local]{java}{Local}{sbase01}
\newlang[blah]{java}{Java}{sbase1}
\newcommand{\annotation}[1]{{\color{sgreen!60!black}\texttt{#1}}}
\newcommand{\codeComment}[1]{{\color{sgreen}\texttt{#1}}}

\newcommand\code{\lstinline[language=Choral,style=solarized-light]}
\newcommand{\mcode}[1]{\hbox{\code{#1}}}

\newcommand{\Div}{\hspace{1.2ex}{\color{gray!80}\vline height 1em depth .3em width 1px}\hspace{1.2ex}}
\newcommand{\s}{{\color{lightgray}\cdot}}
\newcommand{\many}[1]{\overline{#1}}

\defann{hlNew}{\cb{green!30}\uline}
\defann{hlOpt}{\cb{gray!10}\dashuline}
\newcommand{\opt}[1]{\hlOpt*{\vphantom{\overline{\langle}}#1}} %
\newcommand{\diff}[1]{\hlNew*{\vphantom{\overline{\langle}}#1}}
\newcommand{\genW}{\mcode{\#A\#}}
\newcommand{\genWs}{\many{\genW}}
\newcommand{\at}{\mcode{@}}
\newcommand{\params}[1]{\langle#1\rangle}
\newcommand{\grammarTerm}[1]{\mathit{#1}}

\newcommand{\id}{\grammarTerm{id}}
\newcommand{\cst}{\grammarTerm{lit}}
\newcommand{\program}{\grammarTerm{P}}

\newcommand{\interface}{\grammarTerm{Interface}}
\newcommand{\class}{\grammarTerm{Class}}
\newcommand{\enum}{\grammarTerm{Enum}}
\newcommand{\Term}{\grammarTerm{Term}}

\newcommand{\ANN}{\grammarTerm{AN}}
\newcommand{\MOD}{\grammarTerm{MD}}

\newcommand{\FTP}{\grammarTerm{FTP}}
\newcommand{\TE}{\grammarTerm{TE}}
\newcommand{\FA}{\grammarTerm{FAcc}}

\newcommand{\MethodDef}{\grammarTerm{MDef}}
\newcommand{\VarDecl}{\TE\s \id}
\newcommand{\ClassConst}{\grammarTerm{CConst}}
\newcommand{\ClassField}{\grammarTerm{CField}}
\newcommand{\Stm}{\grammarTerm{Stm}}
\newcommand{\Exp}{\grammarTerm{Exp}}

\newcommand{\ExpChain}{\grammarTerm{EChain}}
\newcommand{\AsgOp}{\grammarTerm{AsgOp}}
\newcommand{\BinOp}{\grammarTerm{BinOp}}

\newcommand{\new}{\mcode{new}}
\newcommand{\case}{\mcode{case}}
\newcommand{\switch}{\mcode{switch}}

\definecolor{auxFnColor}{rgb}{0.56, 0.0, 1.0}
\newcommand{\auxFn}[1]{{ \color{auxFnColor}\textsc{#1} }}

\newcommand{\blank}{[blank]}

\def\solOpen{{\color{sblue}\llparenthesis}}
\def\solClose{{\color{sblue}\rrparenthesis}}
\newcommand{\sol}[1]{\solOpen #1 \solClose}

\newcommand{\unit}{\mcode{Unit}}
\newcommand{\uid}{{{\color{auxFnColor}\texttt{.id}}}}

\newcommand{\mergeSign}{{\color{red}{\sqcup}}}
\newcommand{\bigMergeSign}{{\color{red}{\dot\bigsqcup}}}
\newcommand{\solMerge}{\mathbin{\mergeSign}}

\newcommand{\unitSol}[1]{{\color{red}\text{[\![}} #1 {\color{red}\text{]\!]}}}
\newcommand{\unitSolChange}[1]{\unitSol{#1}^{\color{red}\star}}
\renewcommand{\paragraph}[1]{\noindent\textit{#1.}\ \ignorespaces}
\newcommand{\megapar}[1]{\noindent\textbf{#1.}\ \ignorespaces}

\acmJournal{TOPLAS}

\title{Choral: Object-Oriented Choreographic Programming}
\author{Saverio Giallorenzo}
\orcid{0000-0002-3658-6395}
\affiliation{
  \department{Department of Computer Science and Engineering}
  \institution{Universit\`a di Bologna}
  \streetaddress{Mura Anteo Zamboni 7}
  \city{Bologna}
  \postcode{40126}
  \country{Italy}}%
\affiliation{\institution{INRIA}
  \city{Sophia-Antipolis}
  \country{France}
}
\authornote{Work mainly performed while the author was employed at the University of Southern Denmark.}
\email{saverio.giallorenzo@gmail.com}

\author{Fabrizio Montesi}
\orcid{0000-0003-4666-901X}
\affiliation{
  \department{Department of Mathematics and Computer Science}
  \institution{University of Southern Denmark}
  \streetaddress{Campusvej 55}
  \city{Odense}
  \postcode{5230}
  \country{Denmark}
}
\authornote{Corresponding author.}
\email{fmontesi@imada.sdu.dk}

\author{Marco Peressotti}
\orcid{0000-0002-0243-0480}
\affiliation{
  \department{Department of Mathematics and Computer Science}
  \institution{University of Southern Denmark}
  \streetaddress{Campusvej 55}
  \city{Odense}
  \postcode{5230}
  \country{Denmark}
}
\email{peressotti@imada.sdu.dk}

\ccsdesc[500]{Computing methodologies~Concurrent programming languages}
\ccsdesc[500]{Software and its engineering~Multiparadigm languages}
\ccsdesc[500]{Software and its engineering~Classes and objects}
\ccsdesc[300]{Software and its engineering~Concurrent programming structures}

\keywords{Choreographies, Communication, Higher-Kinded Types}

\begin{document}

\begin{abstract}
Choreographies are coordination plans for concurrent and distributed systems, which define the roles of the involved participants and how they are supposed to work together.
In the paradigm of choreographic programming, choreographies are programs that can be compiled to executable implementations.

In this article, we present Choral, the first choreographic programming language based on mainstream abstractions.
The key idea in Choral is a new notion of data type, which allows for expressing that data is distributed over different roles.
We use this idea to reconstruct the paradigm of choreographic programming through object-oriented abstractions.
Choreographies are classes, and instances of choreographies are objects with states and behaviours implemented collaboratively by roles.

Choral comes with a compiler that, given a choreography, generates an implementation for each of its roles.
These implementations are libraries in pure Java, whose types are under the control of the Choral programmer.
Developers can then modularly compose these libraries in their own programs, in order to participate correctly in choreographies.
Choral is the first incarnation of choreographic programming offering such modularity, which finally connects more than a decade of research on the paradigm to practical software development.

The integration of choreographic and object-oriented programming yields other powerful advantages, where the features of one paradigm benefit the other in ways that go beyond the sum of the parts.
On the one hand, the high-level abstractions and static checks from the world of choreographies can be used to write concurrent and distributed object-oriented software more concisely and correctly.
On the other hand, we obtain a much more expressive choreographic language from object-oriented abstractions than in previous work.
This expressivity allows for writing more reusable and flexible choreographies.
For example, object passing makes Choral the first higher-order choreographic programming language, whereby choreographies can be parameterised over other choreographies without any need for central coordination.
We also extend method overloading to a new dimension: specialisation based on data location. Together with subtyping and generics, this allows Choral to elegantly support user-defined communication mechanisms and middleware.
\end{abstract}

\maketitle

\section{Introduction}
\label{sec:introduction}

\paragraph{Background}
Choreographies, broadly construed, are coordination plans for concurrent and distributed systems \citep{wscdl,bpmn}.
Examples of choreographies include distributed authentication protocols \citep{openid,webid}, cryptographic protocols \citep{DH76}, and multiparty business processes \citep{bpmn,wscdl}.
In software development, programmers use choreographies to agree on the interactions that communicating endpoints should enact to achieve a common goal; then, each endpoint can be programmed independently.
The success of this development process hinges on achieving \emph{choreography compliance}: when all endpoints are run together, they interact as defined by the choreographies agreed upon~\citep{M22}. %

\begin{figure}[h]%
	\begin{minipage}{.49\textwidth}%
		\includegraphics[width=.9\textwidth]{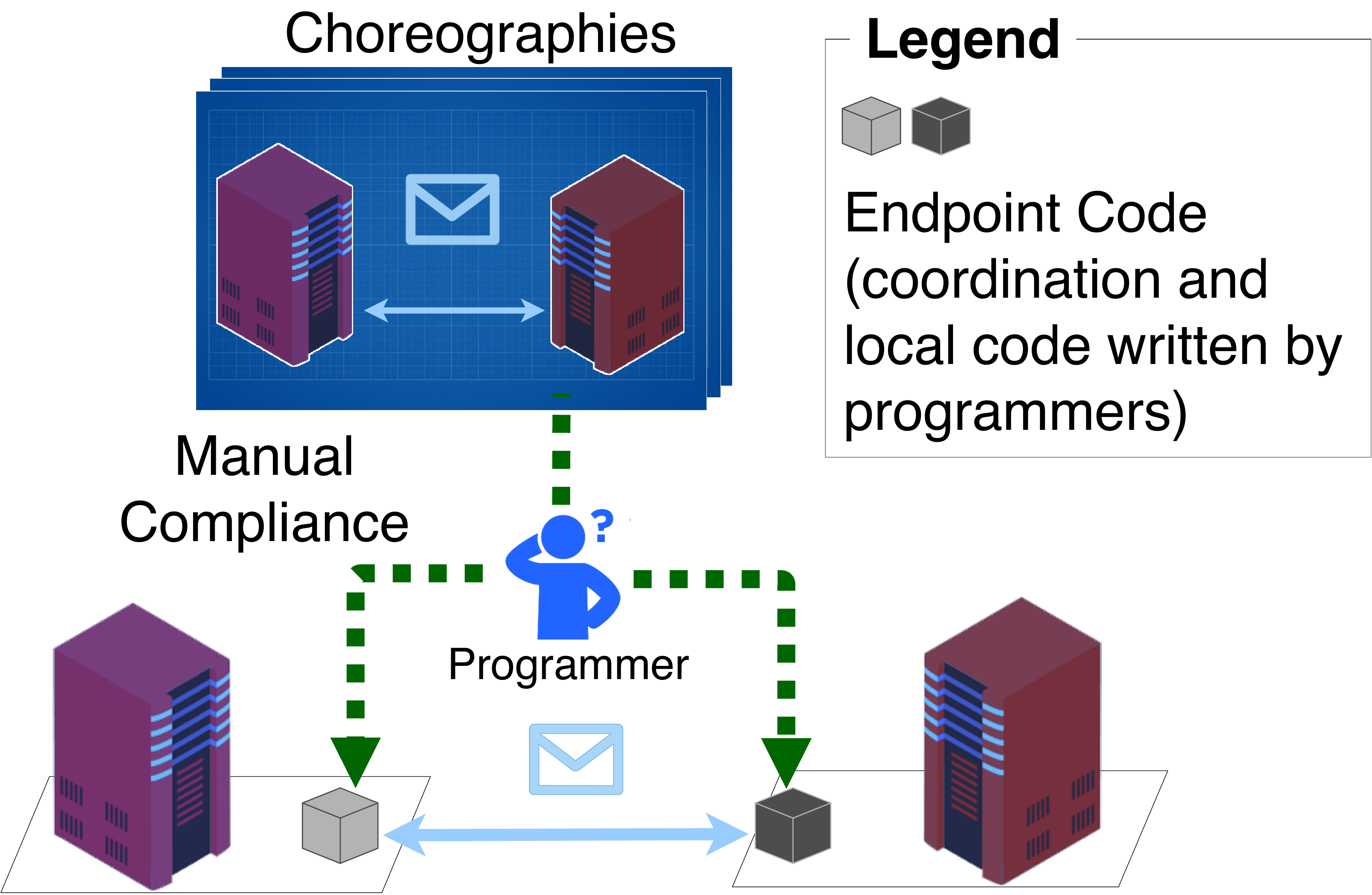}
		\caption{Choreography compliance: endpoints should communicate as intended by the choreographies that they engage in.}
		\label{fig:chor-compliance}
	\end{minipage}\hfill\textcolor{gray}{\vrule height 2.5cm}\hfill\begin{minipage}{.49\textwidth}%
		\includegraphics[width=.9\textwidth]{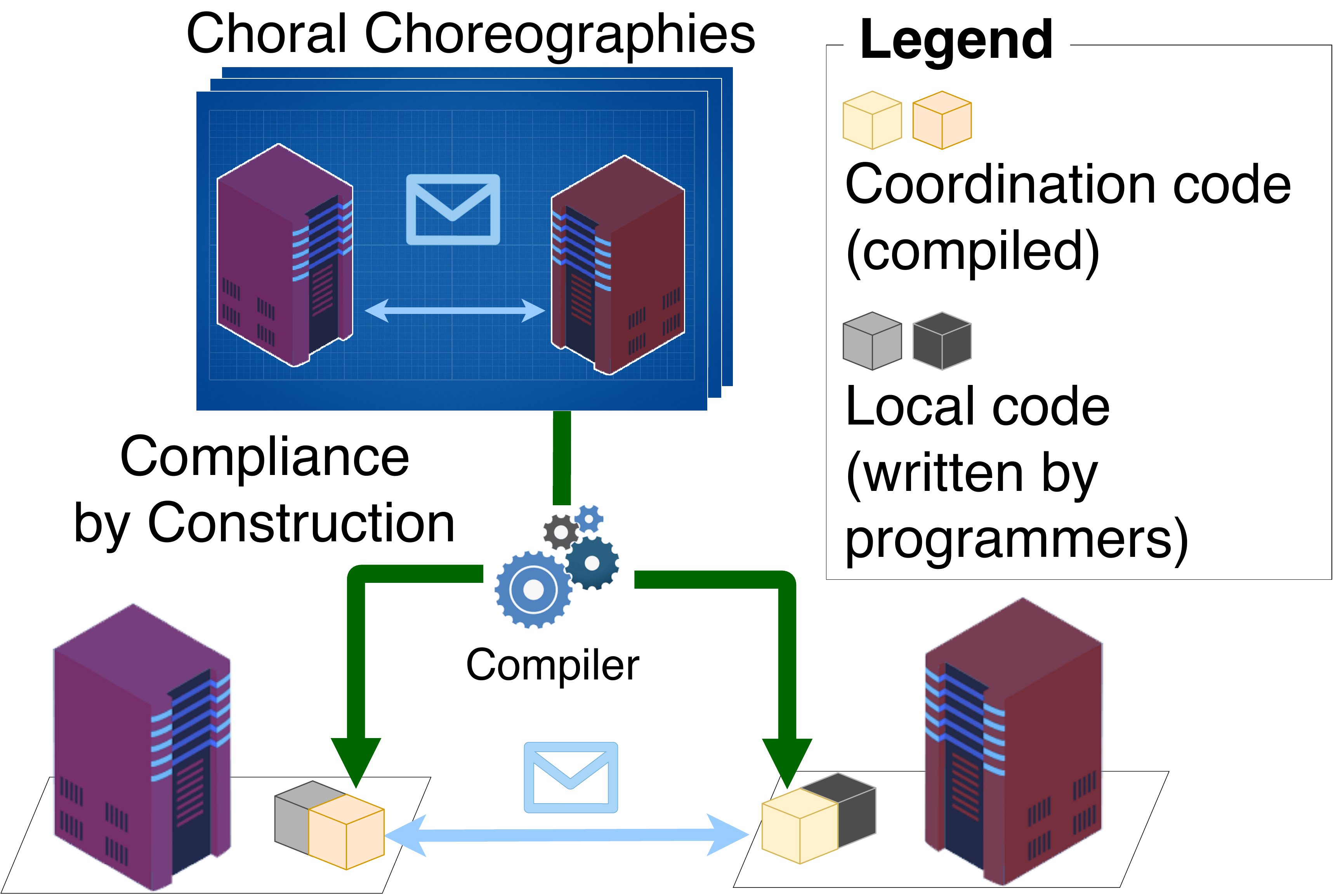}
		\caption{Our proposal: compliant-by-construction libraries are automatically compiled from choreographies in Choral.}
		\label{fig:choral}
	\end{minipage}%
\end{figure}

Achieving choreography compliance is hard, because of some usual suspects of concurrent and distributed programming: predicting how multiple programs will interact at runtime is challenging \citep{O18}, and mainstream programming languages do not adequately support programmers in reasoning about coordination in their code \citep{LPSZ08,LLLG16}.
Additionally, choreographies are complex. At a minimum, choreographies define the expected communication flows among their roles (abstractions of endpoints, like `Alice', `Bob', `Buyer', etc.) \citep{msc}. However, they often include also computational details of arbitrary complexity, for example: pre- or post-processing of data (encryption, validation, anonymisation, etc.), state information, and decision procedures to choose among alternative behaviours. These computational details are essential parts of many protocols, ranging from security protocols to business processes.
\cref{fig:chor-compliance} depicts the common situation where a programmer tries to ensure choreography compliance through their subjective interpretation of the choreography and the manual coding of endpoints.

In response to the challenge of choreography compliance, researchers investigated methods to relate choreographies to endpoint programs -- many are reviewed in \citep{Aetal16,Hetal16}. %
Initially, these methods focused on simple choreographic languages without computation that were inspired by, e.g., communicating automata, process calculi, and session types \citep{QZCY07,BZ07,BBO12,HYC16}.
Some of the ideas developed for choreographies were later adopted in the paradigm of \emph{choreographic programming} \citep{M13:phd,CM13}, where choreographies are written in a Turing-complete programming language that allows for defining arbitrary computation at endpoints. Thanks to the capability of combining computation with coordination, choreographic programming languages can capture realistic protocols that include data manipulation and decision procedures, including encryption strategies (as in security protocols), retry strategies (as in transport protocols), and marshalling (as in application protocols).
In choreographic programming, compliance is obtained by construction:
given a choreography, a compiler automatically translates it to a set of compliant endpoint implementations.
Choreographic programming has been shown to have promising potential in multiple contexts, 
including information flow \citep{LN15}, distributed algorithms \citep{CM16}, cyber-physical systems \citep{LNN16,LH17}, and self-adaptive systems \citep{DGGLM17}.

\paragraph{The problem}
Current approaches to choreographic programming come at a significant cost to modularity, and it remains unclear how the benefits of this paradigm can be applied to mainstream software development.

Typically, the endpoint code that implements a choreography comes in libraries that developers can use in their applications through an API \citep{M08,W12,AIM10}.
For example, a library that implements a choreography for user authentication might provide a method \code{authenticate} that a web service can invoke to run (its part of) the protocol in collaboration with a connected client.
The implementation of \code{authenticate} might involve a series of communications between the client and the web service. Potentially, a third-party identity provider might be involved as well, and messages might include passwords that should be hashed.
Usually, and ideally, all these details are hidden from the API (the signature of the method) exposed to the developer of the web service.
This allows the developer to minimise coupling between their implementation of the functionalities offered by the web service and the choreography used for authentication, bringing the expected benefit: the library implementing the choreography and the rest of the code of the web service can be updated independently and recombined, as long as the API provided by the former remains compatible with that expected by the latter.
This benefit is key to large-scale software development, and it is an initial assumption for modern development practices like microservices and DevOps \citep{DGLMMMS17}.

Unfortunately, previous frameworks for choreographic programming do not support modular software development \citep{M13:phd,DGGLM17}.
The code that these frameworks generate for each endpoint is a `black box' program without an API: it can be executed, but it is not designed to be composed by programmers within larger codebases. Thus, these frameworks fall short of providing the aforementioned benefit of modularity. Furthermore, the common scenario of programming an endpoint that participates in multiple choreographies is not supported, and neither is programming an endpoint where participating in a choreography is only part of what the endpoint does.

A major factor that makes modularity challenging is that current choreographic languages are based on behavioural models (process calculi, communicating automata, etc.), which makes translating choreographies into libraries based on mainstream abstractions (data, functions, objects, etc.) nontrivial.
For the simpler setting of choreographic languages without computation, we know that choreographies can be translated to libraries that offer a `fluid' API. For instance, \citet{SDHY17} produce object-oriented libraries whose APIs enforce the invocation of \code{send} and \code{receive} methods in the right order: if an endpoint should send, receive, and then send, the developer will be forced to write something like \code{o.send(..).receive(..).send(..)}.
However, this approach leaks the structure of the communication flow implemented by the library; thus, future versions of the library that adopt a different structure (e.g., an unnecessary communication might be removed, or some communications might be bunched together for efficiency) would require rewriting the application that uses the library.
Furthermore, this approach does not let the choreography programmer decide on how the generated API will look like; thus, we cannot use this method to generate drop-in replacements for existing libraries.

In summary, we still have to discover how the principles of choreographic programming can be applied to mainstream programming.
The aim of this article is to fill this gap.

\paragraph{This article}
We present Choral, a new choreographic programming language that supports modularity and is based on mainstream programming concepts (from object-oriented programming).
To demonstrate applicability, Choral is compatible with Java, but our ideas apply to most statically-typed object-oriented languages.

The fulcrum of Choral is a new interpretation of choreographies that builds naturally on top of existing language abstractions:
Choral is an \emph{object-oriented} choreographic programming language, where choreographies are classes and their instances are objects.The starting point for this interpretation is a generalisation of the key idea found in Lambda 5 \citep{MCHP04}, the model that inspired the research line on multitier programming \citep{NT05,CLWY06,SGL06,MCH07,WWS20}.
In Lambda 5, each data type is located at a (single) place, which enables reasoning on spatially-distributed computation.
Choral generalises these types from single to \emph{multiple} locations, which allows us to express that an object is implemented choreographically:
Choral objects have types of the form \code{T@(#A1#, ..., #An#)}, where \code{T} is the usual interface of the object, and \code{#A1#}, \ldots, \code{#An#} are the roles that collaboratively implement the object.

As an example, consider the case of a multiparty choreography for distributed authentication, where a service authenticates a client via a third-party identity provider.
We can define such a choreography as a Choral class with type \code{DistAuth@(#Client#, #Service#, #IP#)} (\code{#IP#} is short for identity provider).
The class can implement methods that involve multiple roles. For example, it can offer a method \code{authenticate} with the following signature.
\begin{chorlisting}[][numbers=none]
Optional@#Service#<AuthToken> authenticate(Credentials@#Client# credentials)
\end{chorlisting}
Invoking method \code{authenticate} with some \code{Credentials} located at \code{#Client#} returns an authorisation token at \code{#Service#} (\code{Optional}, since authentication might fail), denoting movement of data.

We leverage our object-oriented interpretation of choreographies to develop a methodology for choreography compliance that supports modularity and is compatible with mainstream programming. We depict this methodology in \cref{fig:choral}.
Given the code of a Choral class with some roles, a compiler produces a compliant-by-construction software library in pure Java for each role (`coordination code' in the figure): each library contains the local implementation of what its role should do to execute the choreography correctly.
These libraries offer Java APIs derived from the source choreographies, which reveal only the details pertaining the implemented role.
When a software developer programs an endpoint that should engage in a choreography, they can just take the library compiled for the role that they want to play and use it through its Java API. Through such APIs, developers can modularly compose multiple libraries with their own code (`local code' in the figure), thus gaining the ability to participate in multiple choreographies.

The Java code compiled from method \code{authenticate} for the roles
\code{#Client#} and \code{#Service#} has the following signatures, respectively.
\trivlist\item\relax \def\w{\dimexpr.5\textwidth-.7\parindent\relax}%
\begin{minipage}{\w}
\begin{commlisting}[][numbers=none]
// Compiled code for Client
Unit authenticate(Credentials credentials)
$\phantom{c}$
\end{commlisting}
\end{minipage}\hfill\begin{minipage}{\dimexpr \textwidth-\w-1.4\parindent\relax}
\begin{commlisting}[][numbers=none]
// Compiled code for Service
Optional<AuthToken> authenticate()
$\phantom{c}$
\end{commlisting}
\end{minipage}
\endtrivlist
The compiled signatures follow the principle that we have just mentioned -- that is, for each role, only what pertains that role is reported.
This is why the signature compiled for \code{#Client#} has a parameter of type \code{Credentials} and return type \code{Unit}: in the source choreography, the parameter is located at \code{#Client#} whereas the return type is located at another role.\footnote{We use a \code{Unit} return type that we provide with Choral instead of \code{void} for compositionality reasons. This is discussed in \cref{sec:interaction}.}
Following the same principle, the signature for \code{#Service#} has return type \code{Optional<AuthToken>} and no parameters.

\megapar{Contributions} We outline our main contributions.

\paragraph{Language}
We present Choral, the first choreographic programming language based on mainstream abstractions and interoperable with a mainstream language (Java).
The key novelty of the Choral language is that data types are higher-kinded types parameterised on roles.
We leverage this feature to bring the key aspects of choreographies to object-oriented programming (spatial distribution, interaction, and knowledge of choice).
Choral is also the first truly higher-order choreographic language, where choreographies passed as arguments can carry state and invocations of higher-order choreographies require no centralised coordination \citep{DH12,DGGLM17}.

Integrating object-oriented principles with choreographies brings key benefits
also in the other direction, in the sense that we gain a much richer
choreographic language than the state of the art. By using subtyping, we can
define abstract APIs for choreographies that can be implemented in different
ways and communication behaviours, bringing the usual substitution principle for
objects to choreographies. Method overloading allows
us to specialise computation based on the locations of arguments -- which is a
new dimension for overloading. Semantic parametricity enables the writing of
reusable choreographies that treat uniformly parameters that implement a shared
API. These features allow us to generalise choreographies from assuming a fixed
communication primitive to user-definable communication methods, thus freeing
Choral from commitments to any communication technology or middleware.
Furthermore, we can define in Choral the first hierarchy of `channel types'
for choreographies, which can be used to represent at the level of types the
topological assumptions of a choreography. In general, users are free to define
other `channel types' to support more topologies.

We implement a type checker that, in addition to the expected checks for an object-oriented language, detects also coding errors related to roles. For example, our checker can rule out computation at a role that erroneously accesses data at another role without proper communication. This makes distribution errors manifest to the programmer.
Our typing also supports the reuse of all existing Java classes and interfaces in Choral, because every such type can be lifted to a Choral type located at a single role.

In Choral, choreographies are concrete software artefacts. This poses questions such as what code should go in these artefacts and what code should remain local, and how these two parts of software should interact through APIs. We elicit these questions and report on our experience in addressing them throughout the article, where they become relevant.

\paragraph{Compiler}
We implement a compiler that translates Choral source into Java libraries that
comply with the source choreography: the code generated for each role performs
the actions prescribed by the choreography.
With our compiler, the programmer of the choreography is in control of the generated APIs: the APIs for each role follow the same structure found in the choreography, but where all data types and parameters in method signatures that do not pertain to the role are omitted, as we exemplified previously for the \code{authenticate} method.

\paragraph{Testing}
We present the first testing tool for choreographic programming.
Since choreography implementations are spread over multiple components (one for each role), testing choreographies can be difficult because it calls for integration testing.
Our testing tool leverages Choral to write `choreographic tests' that look like simple unit tests at the choreographic level, but are then compiled to integration tests that integrate the respective implementations of all roles.

\paragraph{Evaluation}
We explore the expressivity of Choral with use cases of different kinds, covering security protocols, cyber-physical systems connected to the cloud, and parallel algorithms. For these use cases, we discuss relevant code excerpts and development techniques induced by Choral.

We then move from our examples to real-world comparisons.

First, we show how Choral can be used to transition existing Java programs to choreographies. We reimplement in Choral the Java implementation of Karatsuba's algorithm for fast multiplication, which yields a parallel implementation.
Furthermore, we reimplement a clone of Twitter developed by the Spring team. Implementing this system in Choral requires identifying roles, which makes the original monolithic application much more modular.

Second, we compare Choral to a popular framework for concurrent programming in Java based on actors (Akka). In particular, we identify the key differences in the development processes and resulting architectures induced by the two technologies. We find that Choral provides concrete benefits in avoiding subtle concurrency bugs.

Third, we carry out a quantitative evaluation of how Choral impacts software development and runtime performance. Thanks to its choreographic approach, Choral consistently leads to smaller codebases. Moreover, the impact of our compiler on the speed of the edit-compile cycle is negligible (milliseconds). Finally, we show that the runtime performance of the code generated by Choral is not worse (and often better) than that of alternative implementations in Java and Akka.

\megapar{Outline}
We overview Choral with simple examples in \cref{sec:choral}, and give more realistic use cases in \cref{sec:usecases}. The syntax and implementation of Choral are discussed in \cref{sec:implementation}, and testing in \Cref{sec:testing}. We evaluate Choral in depth in \cref{sec:evaluation}. Related and future work are discussed in \cref{sec:discussion}. We draw conclusions in \cref{sec:conclusion}.

\section{Choral in Practice}
\label{sec:choral}

We start with an overview of the key features of Choral.
First, spatial distribution: the expression of computation that takes place at different roles (\cref{sec:roles-datatypes}). Second, interaction: the coding of data exchanges between roles (\cref{sec:interaction}). Third, knowledge of choice: the coordination of roles to choose between alternative behaviours (\cref{sec:knowledge-of-choice}).

The Choral language is quite big. Its usefulness depends on the capability to produce software libraries whose APIs look like `idiomatic' Java APIs, so we chose to incorporate a substantial set of features, which would commonly be considered necessary to use and produce Java APIs:
Choral supports classes, interfaces, generics, inheritance, and method
overloading. APIs generated by Choral support lambda expressions, in the sense that Java programmers can pass lambda expressions as arguments to our APIs. (Just as in Java, Choral sees these arguments as objects.) Supporting the Java syntax for lambda expressions inside Choral programs is not necessary for our objective, since they can be encoded as objects, so we leave it to future work on ergonomics.

In the rest of this section, we explain the key aspects of Choral by assuming that the reader is familiar with Java. The reader can assume that language constructs that have the same syntax as Java behave as expected (modulo our additions, which we explain in the text).

\subsection{Roles and data types}
\label{sec:roles-datatypes}
 
\paragraph{Hello roles}
All values in Choral are distributed over one or more roles, using the
\code{@}-notation seen in \cref{sec:introduction}. The degenerate case of values
involving one role allows Choral to reuse existing Java classes and interfaces,
lifted mechanically to Choral types and made available to Choral code. For
example, the literal \code{"Hello from A"@#A#} is a string value 
\code{"Hello from A"} \emph{located at role} \code{#A#}. Code involving 
different roles can be freely mixed in Choral, as in the following snippet.
\begin{chorlisting}
class HelloRoles@(#A#,#B#) {
	public static void sayHello() {
		String@#A# a = "Hello from A"@#A#;
		String@#B# b = "Hello from B"@#B#;
		System@#A#.out.println(a);
		System@#B#.out.println(b);  
	}
}
\end{chorlisting}
The code above defines a class, \code{HelloRoles}, parameterised over two roles,
\code{#A#} and \code{#B#}. At line 3, we assign the string \code{"Hello from A"}
located at \code{#A#} (\code{"Hello from A"@#A#}) to variable \code{a} of type
`String at \code{#A#}' (\code{String@#A#}). At line 4, we do the same for a
string located at \code{#B#}. Then, at line 5, we print variable \code{a} by
using the \code{System} object at \code{#A#} (\code{System@#A#}); at line 6, we do the same for variable \code{b} at role \code{#B#}.

Roles are part of data types in Choral, adding a new dimension to typing.
For example, the statement \code{String@#A# a = "Hello from B"@#B#} would be
ill-typed, because the expression on the right returns data at a different role
from that expected by the left-hand side.

Formally, in Choral, \code{String} is a type of a higher kind (or type constructor): it takes a role in order to return a type of the same kind of Java types \citep{MPO08}. The code \code{String@#A#} represents the instantiation of \code{String} at role \code{#A#}. Any Java type is automatically liftable to a Choral type with a single role parameter by following the same reasoning, enabling interoperability.
Type constructors in Choral are not limited to a single role in general. We are going to see examples with multiple roles and more complex type parameters later in this section.

\paragraph{From Choral to Java}
Given class \code{HelloRoles}, the Choral compiler generates for each role parameter 
a Java class with the behaviour for that role, in compliance with
the source class. Here, the Java class for role \code{#A#} is
\code{HelloRoles_A} and the class for \code{#B#} is \code{HelloRoles_B}.
\trivlist\item\relax
\def\w{\dimexpr.5\textwidth-.7\parindent\relax}%
\begin{minipage}{\w}
\begin{commlisting}
class HelloRoles_A {
	public static void sayHello() {
		String a = "Hello from A";
		System.out.println(a);
	}
}
\end{commlisting}
\end{minipage}\hfill\begin{minipage}{\dimexpr \textwidth-\w-1.4\parindent\relax}
	\begin{commlisting}
class HelloRoles_B {
	public static void sayHello() {
		String b = "Hello from B";
		System.out.println(b);
	}
}
\end{commlisting}
\end{minipage}
\endtrivlist
Each generated class contains only the instructions that pertain that role. If Java developers want to implement the behaviour of method \code{sayHello} for a specific role of the \code{HelloRoles} choreography, say \code{#A#}, they just need to invoke the generated \code{sayHello} method in the respective generated class (\code{HelloRoles_A}). If all Java programs interested in participating to \code{HelloRoles} do that, then their resulting global behaviour complies by construction with the source choreography.

Notice that the code compiled for \code{#A#} and \code{#B#} will not interact and can therefore proceed fully concurrently, because the choreography does not prescribe so.
In general, choreographic programming languages are expected to generate code that interacts only to enact the communications that the programmer specified in the choreography \citep{M22}.
Choral follows this adequacy principle. We discuss how to program interactions in \cref{sec:interaction}.

\paragraph{Distributed state}
Fields of Choral classes carry state and can be distributed over different roles.
For example, a class \code{DPair} can define a distributed structure storing two values at different roles.
\begin{chorlisting}[breakable=true]
class DPair@(#A#,#B#)<L@#C#,R@#D#> {
	private L@#A# left; 
	private R@#B# right;
	public DPair(L@#A# left, R@#B# right) { this.left = left; this.right = right; }
	public L@#A# left() { return this.left; }
	public R@#B# right() { return this.right; }
}
\end{chorlisting}
Class \code{DPair} is distributed between roles \code{#A#} and \code{#B#} and has two fields, \code{left} and \code{right}. In general, every class or interface in Choral is always parameterised on at least one role, and is hence a type constructor.
The class is also parameterised on two data types, \code{L} and \code{R}, which exemplifies our support for generics \citep{NW06}. At line 1, \code{L@#C#} specifies that \code{L} is expected to be a data type parameterised over a single role, abstracted by \code{#C#}; similarly for \code{R@#D#}. Naming the role parameters of \code{L} and \code{R} does not add any information in this particular example (we only need to know that they have one parameter).
However, naming role parameters in generics is useful for expressing type bounds in \code{extends} clauses, as discussed later in this section.
Choral interprets role parameter binders as in Java generics: the first appearance of a parameter is a binder, while subsequent appearances of the same parameter are bound.
Observe that the scope of role parameters \code{#C#} and \code{#D#} is limited to the declaration of the type parameters \code{L} and \code{R}, respectively -- we use distinct names exclusively for readability.
At lines 2 and 3 we have the two fields, \code{left} and \code{right}, respectively located at \code{#A#} and \code{#B#} as stated by the types \code{L@#A#} and \code{R@#B#}, the constructor is at line 4, while accessors to the two fields are at lines 5--6.

Data structures like \code{DPair} are useful when defining choreographies where the data at some role needs to correlate with data at another role, as with distributed authentication tokens. We apply them to a use case in \cref{sec:use_cases_distAuth}.

\subsection{Interaction}
\label{sec:interaction}

Choral programs become interesting when they contain interaction between roles -- otherwise, they are simple interleavings of local independent behaviours by different roles, as in \code{HelloRoles}.

Choreography models typically come with some fixed primitives for interaction,
e.g., sending a value from a role to another over a channel \cite{CHY12}. 
Thanks to our data types parameterised over roles, Choral is more expressive: we can \emph{program} these basic building blocks and then construct more complex interactions compositionally. 
This allows us to be specific about the requirements of choreographies regarding communications, leading to more reusable code. For instance, if a choreography needs only a directed channel, then our type system can see by subtyping that a bidirectional channel is also fine.

\paragraph{Directed data channels}
We start our exploration of interaction in Choral from simple directed channels
for transporting data. In Choral, such a channel is just an object %
that takes data from one place to another. We
can specify this behaviour as an interface.
\begin{chorlisting}[][numbers=none]
interface DiDataChannel@(#A#,#B#)<T@#C#> { 
	<S@#D# extends T@#D#> S@#B# com(S@#A# m); 
}
\end{chorlisting}
A \code{DiDataChannel} is the interface of a directed channel between two roles, abstracted by \code{#A#} and \code{#B#}, for transmitting data of a given type, abstracted by the type parameter \code{T}, from \code{#A#} to \code{#B#} (hence the number of role parameters in \code{T}).
Data transmission is performed by invoking the only method of the interface:  \code{com} which takes any value of a subtype of \code{T} located at \code{#A#}, \code{S@#A#}, and returns a value of type \code{S@#B#}.
The type parameter \code{S} of method \code{com} has \code{T} as upper bound (we can read the expression \code{S@#D# extends T@#D#} as `for any role \code{#D#}, \code{S@#D# extends T@#D#}') and allows us to transmit data of types more specific than \code{T} without losing type information (as it would be if the signature of \code{com} was simply \code{T@#B# com(T@#A# m)}).

Parameterising data channels over the type of transferrable data (\code{T}) is important in practice for channel implementors, because they often need to deal with data marshalling. Choral comes with a standard library that offers implementations of our channel APIs for a few common types of channels, e.g., TCP/IP sockets supporting JSON objects and shared memory channels. Users can provide their own implementations.

Using a \code{DiDataChannel}, we can write a simple method that sends a string notification from a \code{#Client#} to a \code{#Server#} and logs the reception by printing on screen.
\begin{chorlisting}[][numbers=none]
notify(DiDataChannel@(#Client#,#Server#)<String> ch, String@#Client# msg) {
	String@#Server# m = ch.<String>com(msg);
	System@#Server#.out.println(m); 
}
\end{chorlisting}	
Note that \code{String} is a valid instantiation of \code{T} of \code{DiDataChannel} because we lift all Java types as Choral types parameterised over a single role.

\paragraph{Alien data types}
Compiling \code{DiDataChannel} to Java poses an important question: what should be the return type of method \code{com} in the code produced for role \code{#A#}?
Since the return type does not mention \code{#A#} (we say that it is \emph{alien} to \code{#A#}), a na\"ive answer to this question could be \code{void}, as follows.
\begin{tcblisting}{
	          enhanced, 
	          sharp corners=all,
	          boxrule=2pt,
	          boxsep=-.7em,
	          colframe=sbase1,
	          colback=sbase3,
	          detach title,
	          finish={\node[anchor=south east, font=\footnotesize\itshape,
	          text=sbase3,fill=sbase1] at (frame.south east) {Tentative Code};},
	          title={},
	          listing only,
	          listing options={language=Java,style=solarized-light,numbers=none},
	    }
interface DiDataChannel_A<T> { 
	<S extends T> void com(S m); 
}
\end{tcblisting}

It turns out that this solution does not work well with expressions that compose multiple method calls, including chaining like \code{m1(e1,e2).m2(e3)} and nesting like \code{m1(m2(e))}.
As a concrete example, consider a simple round-trip communication from \code{#A#} to \code{#B#} and back.
\begin{chorlisting}
static <T@#C#> T@#A# roundTrip(DiDataChannel@(#A#,#B#)<T> chAB,DiDataChannel@(#B#,#A#)<T> chBA,T@#A# msg) {
	return chBA.<T>com(chAB.<T>com(msg)); 
}
\end{chorlisting}

Method \code{roundTrip} takes two channels, \code{chAB} and \code{chBA}, which are directed channels respectively from \code{#A#} to \code{#B#} and from \code{#B#} to \code{#A#}.
The method sends the input \code{msg} from \code{#A#} to \code{#B#} and back by nested \code{com}s and returns the result at \code{#A#}.

A structure-preserving compilation of method \code{roundTrip} for role \code{#A#} would be as follows.
\begin{commlisting}
static <T> T roundTrip(DiDataChannel_A<T> chAB, DiDataChannel_B<T> chBA, T msg) {
	return chBA.<T>com(chAB.<T>com(msg)); 
}
\end{commlisting}
Observe how the inner method call, \code{chAB.com<T>(msg)}, should return something, such that it can trigger the execution of the outer method call to receive the response.
Therefore, the \code{com} method of \code{DiDataChannel_A} cannot have \code{void} as return type.

Programming language experts have probably guessed by now that the solution is to use unit values instead of \code{void}.
Indeed, Choral defines a singleton type \code{Unit} (a final class) that our compiler uses instead of \code{void} to obtain Java code whose structure resembles its Choral source code.

We now show the Java code produced by our compiler from \code{DiDataChannel} for both \code{#A#} and \code{#B#}.
\par\vspace{\abovedisplayshortskip}\noindent%
\def\w{\dimexpr.5\textwidth-.7\parindent\relax}%
\begin{minipage}{\w}%
\begin{commlisting}[][numbers=none]
interface DiDataChannel_A<T> { 
	<S extends T> Unit com(S m); 
}
\end{commlisting}
\end{minipage}\hfill\begin{minipage}{\dimexpr \textwidth-\w-1.4\parindent\relax}
\begin{commlisting}[][numbers=none]
interface DiDataChannel_B<T> { 
	<S extends T> S com(Unit m); 
}
\end{commlisting}
\end{minipage}\\[\belowdisplayshortskip]\noindent
Given these interfaces, the compilation of \code{roundTrip} for role \code{#A#} is well-typed and correct Java code.
An alternative to using \code{Unit} would have been to give up on preserving structure in the compiled code; we chose in favour of our solution because preserving structure makes it easier to read and debug the compiled code (especially when comparing it to the source choreography), and also makes our compiler simpler.

The users of our compiled libraries are not forced to passing \code{Unit} arguments to methods, as for method \code{com} of \code{DiDataChannel_B}: for methods like these, our compiler provides corresponding `courtesy methods' that take no parameters and inject \unit{}s automatically.

\paragraph{Bidirectional channels}
An immediate generalisation of directed data channels brings us to bidirectional data channels, specified by \code{BiDataChannel}.
\begin{chorlisting}[][numbers=none]
interface BiDataChannel@(#A#,#B#)<T@#C#,R@#D#> extends DiDataChannel@(#A#,#B#)<T>,DiDataChannel@(#B#,#A#)<R>{
  <S@#C# extends T@#C#> S@#B# com(S@#A# m); // inherited from DiDataChannel$\codeComment{@}$(A,B)<T>
  <S@#C# extends R@#C#> S@#A# com(S@#B# m); // inherited from DiDataChannel$\codeComment{@}$(B,A)<R>
}
\end{chorlisting}
A \code{BiDataChannel} is parameterised over two types: \code{T} is the type of data that can be transferred from \code{#A#} to \code{#B#} and, vice
versa, \code{R} is the type of data that can be transferred in the opposite direction. This is obtained by multiple type inheritance: \code{BiDataChannel} extends \code{DiDataChannel} in one and the other direction, which allows for using modularly a bidirectional data channel in code that has the weaker
requirement of a directed data channel in one of the two supported directions.
Distinguishing the two parameters \code{T} and \code{R} is useful for protocols that have different types for requests and responses, like HTTP.
Extending \code{DiDataChannel} twice does not result in any clashes since \code{#A#} and \code{#B#} play different roles in each supertype.
This `twin' inheritance results in the overload of method \code{com} one for each communication direction supported by the channel. 
This overload does not result in any clash in the generated code as illustrated by the code generated for \code{#A#} (code generated for \code{#B#} is symmetric).
\begin{commlisting}[][numbers=none]
interface BiDataChannel_A<T,R> extends DiDataChannel_A<T>, DiDataChannel_B<R> { 
  <S extends T> Unit com(S m); // inherited from DiDataChannel_A<T>
  <S extends R> S com(Unit m); // inherited from DiDataChannel_B<R>
}
\end{commlisting}
We discuss more types of channels (including symmetric channels) in \cref{sec:on-channels} and provide more details on inheritance and overloading in \cref{sec:implementation}.

\paragraph{Forward chaining}
We use bidirectional channels to define a choreography for remote procedure calls, called \code{RemoteFunction}, which leverages the standard Java interface \code{Function<T,R>}.

\begin{chorlisting}
class RemoteFunction@(#Client#,#Server#)<T@#A#,R@#B#> {
	private BiDataChannel@(#Client#,#Server#)<T,R> ch;
	private Function@#Server#<T,R> f;
	public RemoteFunction(BiDataChannel@(#Client#,#Server#)<T,R> ch, Function@#Server#<T,R> f) { 
		this.ch = ch; this.f = f;
	}
	public R@#Client# call(T@#Client# t) { 
		return ch.<R>com(f.apply(ch.<T>com(t))); 
	}
}
\end{chorlisting}

In the experience that we gained by programming larger Choral programs (as those in \cref{sec:usecases}), compositions of method invocations including data transfers
as in line 8 of \code{RemoteFunction} are rather typical. In these chains, we read data transfers from right to left (innermost to outermost invocation), but
most choreography models in the literature use a left-to-right notation (as in
`Alice sends 5 to Bob'). To make Choral closer to that familiar choreographic
notation, we borrow the forward chaining operator \code{>>} from
F\#~\cite{PS09}, so that \code{exp >> obj::method} is syntactic sugar for
\code{obj.method(exp)}. For example, we can rewrite method \code{call} of
\code{RemoteFunction} as follows, which is arguably more readable and recovers a
more familiar choreographic notation.
\begin{chorlisting}[breakable=true][numbers=none]
public R@#Client# call(T@#Client# t) {
	return t >> ch::<T>com >> f::apply >> ch::<R>com;
}
\end{chorlisting}

\paragraph{Using Choral libraries}
As mentioned for Channels, when we compile the \code{RemoteFunction} class
above, we obtain two Java classes: a \code{RemoteFunction_Client}, which sends
some data to the \code{#Server#} for processing and waits for its response, and
a \code{RemoteFunction_Server}, which, upon reception, feeds the data into a
\code{Function} and sends back to the \code{#Client#} its result.

The \code{RemoteFunction_Server} is an interesting example of how users interact
with Choral libraries. The code (snippet) generated from Choral is:
\begin{commlisting}
class RemoteFunction_Server<T,R> { 
	private BiDataChannel_B <T,R> ch; 
 private	Function <T,R> f; 
	public RemoteFunction_Server(BiDataChannel_B<T,R> ch, Function<T,R> f) { /*...*/	} 
	public Unit call() { /*...*/	} 
}
\end{commlisting}
A user of the \code{RemoteFunction_Server} can interact in the choreography by
providing the definition of the \code{Function} at the creation of the object.
In general, this is how we expect users to integrate Choral-generated code with
their `local code', i.e., code parametric to the choreography that users can
implement locally, without any coordination with the other participants (save
the APIs induced by Choral-generated code). For example, the snippet below is from
a Java class that uses \code{RemoteFunction_Server} to provide a remote
procedure for checking if an integer is even.
\begin{locallisting}
BiDataChannel_B<Integer,Boolean> channel = /*...*/;
new RemoteFunction_Server<Integer, Boolean>(channel, i -> i %
\end{locallisting}
Here, at line 2 (second argument of the constructor), we provide the definition
of the \code{Function} using Java Lambdas functional syntax.

\subsection{Knowledge of choice}
\label{sec:knowledge-of-choice}

Knowledge of choice is a hallmark challenge of choreographies: when a
choreography chooses between two alternative behaviours, roles should coordinate
to ensure that they agree on which behaviour should be implemented
\citep{CDP11}.

We exemplify the challenge with the following code, which implements the
consumption of a stream of items from a producer \code{#A#} to a consumer
\code{#B#}.
\begin{chorlisting}
// wrong implementation
consumeItems(DiDataChannel@(#A#,#B#)<Item@#C#> ch, Iterator@#A#<Item> it, Consumer@#B#<Item> consumer){
	if (it.hasNext()) { 
		it.next() >> ch::<Item>com >> consumer::accept;
		consumeItems(ch, it, consumer);
	}
}
\end{chorlisting}
Method \code{consumeItems} takes a channel from \code{#A#} to \code{#B#}, an
iterator over a collection of items at \code{#A#}, and a consumer function for items at \code{#B#}.
Role \code{#B#} works reactively, where its consumer function is invoked whenever the stream of \code{#A#} produces an element: if the iterator can provide an item (line 3), it is transmitted from
\code{#A#} to \code{#B#}, consumed at \code{#B#}, and the method recurs to consume the other items (line 4).

The reader familiar with choreographies should recognise that this method implementation is \emph{wrong}, due to (missing)
knowledge of choice: the information on whether the if-branch should be entered
or not is known only by \code{#A#} (since it evaluates the condition), so
\code{#B#} does not know whether it should run lines 4--5 (receive,
consume, and recur), or do nothing and terminate.

In choreographic programming, knowledge of choice is typically
addressed by equipping the
choreography language with a `selection' primitive to communicate constants drawn from a dedicated set of `labels' \citep{CM13,LNN16}.
This makes it possible for the compiler to build code that can react to choices
made by other roles, inspired by a theoretical operator known as merging
\citep{CHY12}.
In Choral, we adapt this practice to objects. Notably, Choral is expressive enough that we do not need to add a dedicated primitive, nor a dedicated set of labels.

We define a method-level annotation \annotation{@SelectionMethod}, which
developers can apply only to methods that can transmit instances of
enumerated types between roles (the compiler checks for this condition).
For example, we can specify a directed channel for sending such enumerated values with the following \code{DiSelectChannel} interface.
\begin{chorlisting}[][numbers=none]
interface DiSelectChannel@(#A#,#B#) { 
	$\annotation{@SelectionMethod}$
	<T@#C# extends Enum@#C#<T>¦¦> T@#B# select(T@#A# m);
}
\end{chorlisting}

Our compiler assumes that implementations of methods annotated with \annotation{@SelectionMethod} return at the receiver the same value given at the sender.
(This is part of the contract for channels, and it is a standard assumption in implementations of choreographies.)

Typically, channels used in choreographies are assumed to support both data communications and selections. We can specify this with \code{DiChannel} (directed channel), a subtype of both \code{DiDataChannel} and \code{DiSelectChannel} (we include inherited methods for convenience).
\begin{chorlisting}[][numbers=none]
interface DiChannel@(#A#,#B#)<T@#C#> extends DiDataChannel@(#A#,#B#)<T>, DiSelectChannel@(#A#,#B#) {
  <S@#C# extends T@#C#> S@#B# com(S@#A# m);          // inherited from DiDataChannel$\codeComment{@}$(A,B)
  <S@#C# extends Enum@#C#<T>¦¦> T@#B# select(T@#A# m); // inherited from DiSelectChannel$\codeComment{@}$(A,B)
}
\end{chorlisting}

Using \code{DiChannel}s, we can update \code{consumeItems} to respect knowledge of choice.

\begin{chorlisting}[][numbers=none]
enum Choice@#A# { GO, STOP }
\end{chorlisting}
\begin{chorlisting}
consumeItems(DiChannel@(#A#,#B#)<Item> ch, Iterator@#A#<Item> it, Consumer@#B#<Item> consumer) {
	if (it.hasNext()) {
		ch.<Choice>select(Choice@#A#.GO);
		it.next() >> ch::<Item>com >> consumer::accept;
		consumeItems(ch, it, consumer);
	} else { 
		ch.<Choice>select(Choice@#A#.STOP); 
	}
}
\end{chorlisting}

Differently from the previous broken implementation of \code{consumeItems}, now role \code{#A#} sends a selection of either \code{GO} or \code{STOP} to \code{#B#}. Role \code{#B#} can now inspect the received enumerated value to infer
whether it should execute the code for the if- or the else-branch of the conditional. This information
is exploited by our static analyser to check that \code{consumeItems} respects knowledge of choice, and also by our compiler to generate code for \code{#B#} that reacts correctly to the choice performed by \code{#A#}. (A more extensive example containing also the code compiled for the receiver is given in \cref{sec:use_cases_distAuth}.)

Our compiler supports three features to make knowledge of choice flexible. Firstly, our knowledge of choice check works with
arbitrarily-nested conditionals. Secondly, knowledge of choice can
be propagated transitively. Say that a role \code{#A#} makes a choice that determines that two other roles \code{#B#} and \code{#C#} should behave differently, and \code{#A#} informs \code{#B#} of the choice through a selection. Now either \code{#A#} or \code{#B#} can inform \code{#C#} with a selection, because our compiler sees that \code{#B#} now possesses knowledge of choice.
Thirdly, knowledge of choice is required only when necessary: if
\code{#A#} makes a choice and another role, say \code{#B#}, does not need to know because it performs the same actions (e.g., receiving an integer from \code{#A#}) in both branches, then no selection is necessary. We explain the technicalities behind this in \cref{sec:implementation}.

\subsection{The family of Choral channels}
\label{sec:on-channels}

\begin{figure}[t]
	\includegraphics[width=\textwidth]{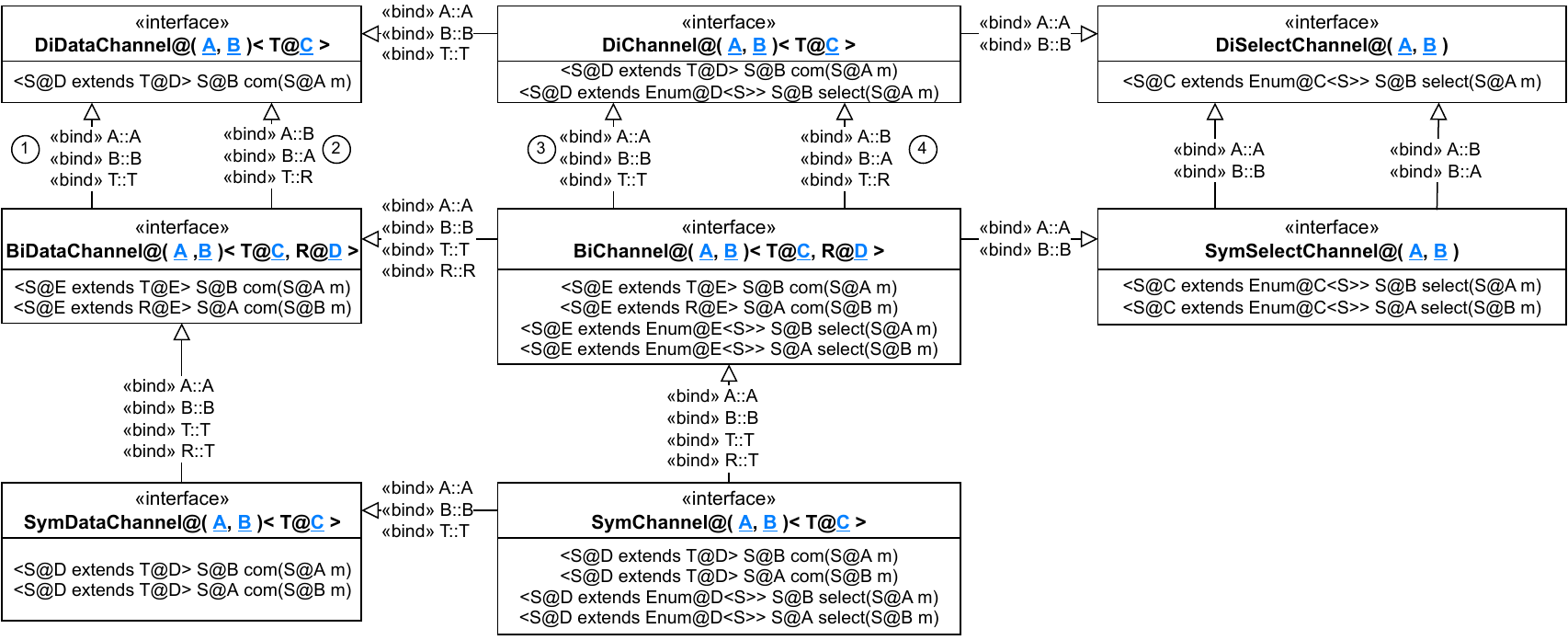}
	\caption{UML class diagram of the hierarchy of the \code{*Channel} interfaces.}
	\label{fig:channels_hierarchy}
\end{figure}

Choral types give us a new way to specify requirements on channels that prior
work implicitly assumed, leading to the definition of a family of channel interfaces
diagrammed in \cref{fig:channels_hierarchy}.

From the left-most column in \cref{fig:channels_hierarchy}, at the top, we find
\code{DiDataChannel}, representing a directed channel parameterised over
\code{T} (the type of the data that can be sent).
We obtain \code{BiDataChannel}, a bidirectional data channel, by extending
\code{DiDataChannel} once for each direction: \textcircled{\scalebox{.8}{\textsf{1}}}
it binds the role parameters of one extension in the same order given for the
role parameters of \code{BiDataChannel}, giving us a direction from \code{#A#}
to \code{#B#} and \textcircled{\scalebox{.8}{\textsf{2}}} it binds the role
parameters of the other extension in the opposite way, giving us a direction
from \code{#B#} to \code{#A#}. The result is that \code{BiDataChannel} defines
two \code{com} methods: one transmitting from \code{#A#} to \code{#B#}, the
other from \code{#B#} to \code{#A#}. The last lines in
\textcircled{\scalebox{.8}{\textsf{1}}} and \textcircled{\scalebox{.8}{\textsf{2}}} in
\cref{fig:channels_hierarchy} complete the picture: the first generic data type
\code{T} binds data from \code{#A#} to \code{#B#}, second generic data type
\code{R} binds data from \code{#B#} to \code{#A#}. 
The \code{SymDataChannel} in \cref{fig:channels_hierarchy}, by extending the
\code{BiDataChannel} interface and binding the two generic data types \code{T}
and \code{R} with its only generic data type \code{T}, defines a bidirectional
data channel that transmits one type of data, regardless its direction.

The right-most vertical hierarchy in \cref{fig:channels_hierarchy} represents
channels supporting selections and it follows a structure similar to that of
data channels. A \code{DiSelectChannel} is a directed selection channel and a
\code{SymSelectChannel} is the bidirectional version -- there is no \code{BiSelectChannel} since both directions exchange the same enumerated types.

The vertical hierarchy in the middle column of \cref{fig:channels_hierarchy} is the combination of the left-most and right-most columns. Interface \code{DiChannel} is a directed channel that supports both generic data communications and selections.
\code{BiChannel} is its bidirectional extension (\textcircled{\scalebox{.8}{\textsf{3}}} and \textcircled{\scalebox{.8}{\textsf{4}}} in \cref{fig:channels_hierarchy}), and \code{SymChannel} is the symmetric extension of \code{BiChannel}.
The snippet below contains the definition of the interface \code{BiChannel}.
\begin{chorlisting}[][numbers=none]
interface BiChannel@(#A#,#B#)<T@#C#, R@#D#> extends
    DiChannel@(#A#,#B#)<T>,         // A BiChannel is a pair of directed channels
    DiChannel@(#B#,#A#)<R>,         // in opposite directions
    BiDataChannel@(#A#,#B#)<T,R>,   // that supports data
    BiSelectChannel@(#A#,#B#)       // and selections
  { } // we do not define any new methods, since they are all inherited
\end{chorlisting}
This definition means that 
for any pair of distinct roles \code{#C#}, \code{#D#} and
for any pair of types \code{S}, \code{P} (with one role parameter),
\code{BiChannel@(#C#,#D#)<S,P>} is a subtype of
\code{DiChannel@(#C#,#D#)<S>},
\code{DiChannel@(#D#,#C#)<P>},
\code{BiDataChannel@(#C#,#D#)<S,P>}, and 
\code{BiSelectChannel@(#C#,#D#)}.

\pleaseCheck
\paragraph{Implementing Choral channels}
Our channel interfaces can be implemented directly in Choral or in Java.
We exemplify the latter case, which lets us also
show how one can carry out this task by fully leveraging the existing Java language
and ecosystem.

Let us say we want to implement a symmetric channel over TCP/IP that a client (say \code{#A#}) is going to use for transmitting strings to a server (say \code{#B#}).
We achieve this by writing a class \code{TCPClientChannel} that implements the interface generated for \code{#A#} from \code{SymChannel} (instantiating its generic with \code{String}).
The snippet below shows the structure of such an implementation.

\begin{locallisting}*[breakable=true]
public class TCPClientChannel implements SymChannel_A<String> {
	private final SocketChannel socketChannel;
	private TCPClientChannel(SocketChannel socketChannel) {
		this.socketChannel = socketChannel;
	}
	public void com(String m) { /*...*/ }
	public String com() { /*...*/ }
	/*...*/
	public static TCPClientChannel open(SocketAddress remote) throws IOException {
		return new TCPClientChannel(SocketChannel.open(remote));
	}
}
\end{locallisting}
The \code{TCPClientChannel} uses \code{SocketChannel} from the Java
standard library.
The class also offers a factory method for connecting to a remote address in the
standard way. This method creates a channel that the user can pass as argument
to the code generated from a choreography. In particular, the
\code{TCPClientChannel} is designed to provide a conventional socket set-up experience for
Java programmer, where the expected contract relies on opening a connection
based on a \code{SocketAddress}. Implementing the server-side counterpart of
\code{TCPClientChannel} is straightforward and follows the same approach.

\subsection{Handling exceptions}
\label{sec:handling-exceptions}
Typically, choreographic languages assume reliable communications~\citep{CHY08,CM13,DGGLM17}. The only exception is the language theory in \citep{MP17}, which shows that one can relax this assumption, by allowing the choreographic language to handle local exceptions. In Choral, we follow the same strategy, which we briefly illustrate here.

Choral can invoke Java code, which might raise an exception. Plain Java code is always located at one role, and therefore the same holds for exceptions (exceptions are `local').
We exemplify how we treat exceptions with the following choreography, where a role \code{#B#} uses the Java standard library to save on disk some text communicated from another role \code{#A#}.
\begin{chorlisting}[breakable]
public Result@#A#<String, String> save(
	SymDataChannel@(#A#,#B#)<String> chAB, String@#A# text, Path@#B# path
) {
	String@#B# textB = text >> chAB::<String>com;
	Result@#B#<String, String> result;
	try {
		Files@#B#.writeString(path, textB);
		result = Result@#B#.ok("Saved"@#B#);
	} catch(IOException@#B# ex) {
		result = Result@#B#.err(ex.getMessage());
	}
	return result >> chAB::<String>com;
}
\end{chorlisting}

Above, we start by communicating the text to be saved from \code{#A#} to
\code{#B#} (line 4). We then declare at \code{#B#} a \code{result} variable (line 5),
which will store either a success or error message -- that \code{#B#} later
communicates to \code{#A#} (line 12). At line 7, \code{#B#} attempts at saving
the received text.

This choreography might incur execution errors related to communication or file writing. Exceptions encapsulate these errors. For example, the invocation of method \code{writeString} might throw an \code{IOException} located at \code{#B#}, which we handle with the \code{try}-\code{catch} block at lines 6--11.

Also method \code{com} can throw exceptions, depending on the implementation of
channel \code{ch}. Channels for remote communication (e.g., based on TCP/IP
sockets) in the Choral library use the following strategy: the sender attempts
at sending a message until its network stack accepts the task (using exponential
backoff and bound to the number of attempts, to guarantee termination);
likewise, the receiver attempts at receiving until a timeout expires. If the
sender ultimately fails at relaying the message to its local network stack, the
channel throws a \code{SendException} at the sender (\code{#A#} in our example).
If the receiver timeouts before receiving a message, the channel raises a
\code{TimeoutException} at the receiver (\code{#B#} in our example).

As a design choice, we left the exceptions of method \code{com} unchecked. The
idea is that the implementation of channels should do their best to deliver
messages, and when this is not possible the local code that uses the code
generated from a choreography should deem the execution of the choreography
unsuccessful. The local code is free to catch these exceptions and attempt
recovery, for example by executing the choreography again (as in actor
frameworks~\citep{D13}).

However, our implementation of \code{com} in the Choral standard library is just
a default; Choral does not hardcode any communication semantics. The user is
free to implement alternative communication methods that expose an API which the
caller choreography might use to handle network errors, e.g., lost messages. For
instance, we might account for lossy communications between \code{#A#} and
\code{#B#} within the above choreography as follows.

\begin{chorlisting}[breakable]
public Result@#A#<String, String> save(
	LossySymDataChannel@(#A#,#B#)<String> chAB, String@#A# text, Path@#B# path
) {
	Optional@#B#<String> textB = text >> chAB::<String>lossyCom;
	Result@#B#<String, String> result;
	if(!textB.isEmpty()) {
		try {
			Files@#B#.writeString(path, textB.get());
			result = Result@#B#.ok("Saved"@#B#);
		} catch(IOException@#B# ex) {
			result = Result@#B#.err(ex.getMessage());
		}
	} else {
		result = Result@#B#.err("Network error");
	}
	return result >> chAB::<String>com;
}
\end{chorlisting}

Channel \code{chAB} is now a \code{LossySymDataChannel}, which in addition to method \code{com} offers also method \code{lossyCom}. The latter does not throw exceptions in case of communication failures, but rather returns an \code{Optional} value that contains the received value in case of success or is empty otherwise.

Choosing which errors a choreography should deal with and which errors
should be raised as unrecoverable exceptions to the local code is a design 
trade-off that derives from the usual tension between robustness versus
simplicity. This trade-off is typical of coordination protocols and exists
independently from Choral, which is why we decided to leave the programmer free
to navigate this spectrum. Gathering from our own experience with
Choral, we recognised the following design principles. 
Protocols whose design assumes a reliable network layer should not deal
with communication errors within the choreography (e.g., the Diffie-Hellman
protocol for key exchange, which we briefly describe below and that we implemented for our evaluation in
\cref{sec:evaluation}). Contrarily, choreographies implementing protocols designed
to deal with network errors should specify the handling of those errors (e.g., implementations of objects dedicated to data transfer, like our channels).
Choral is quite flexible regarding these aspects. A channel API can offer methods that both raise exceptions -- like method \code{com}, meaning that the communication is essential -- or wrap failures in data types (raising no exceptions) -- like method \code{lossyCom}, meaning that the communication is not essential and that the choreography can handle internally the failure.
The programmer can use the different methods within the same choreography to pinpoint which communications are deemed essential and which are not.

\subsection{What goes in a choreography?}

We have just looked at the design issue of deciding whether to deal with errors in the choreography or in the local code that uses the (communication code compiled from the) choreography. This is an instance of the more general issue of protocol design: what should be part of a choreography? This issue exists even when designing choreographies informally (without Choral) because one needs to choose what details are fixed in the protocol and what is instead left to the discretion of the local code.
There is no one-size-fits-all solution since these choices are influenced by the
concrete use case that the choreography deals with.

Consider, for example, the widely adopted Diffie-Hellman protocol for cryptographic key exchange~\cite{DH76}. Integral parts of the protocol specify both computation and communication.
In the protocol, two parties, e.g.,
\code{#Alice#} and \code{#Bob#}, use two pairs of keys (a private and a public
one) to generate a shared secret, which they can later use for symmetric
encryption. Formally, let \(p\) be a prime number and \(g\) be a primitive root modulo \(p\), {\color{smagenta}\(sA\)} be a secret key held by \code{#Alice#}, and {\color{smagenta}\(sB\)} be a secret key held by \code{#Bob#}.
First, \code{#Alice#} computes her public key \({\color{sgreen}pA} = (g^{\color{smagenta}sA} \bmod p)\) and, likewise, \code{#Bob#} computes his public key \({\color{sgreen}pB} = (g^{\color{smagenta}sB} \bmod p)\).
Then, \code{#Alice#} and \code{#Bob#} exchange their public keys, which they
can use to generate their shared secret \(s\):
\[
\overbrace{
{\underbrace{(g^{\color{smagenta}sB} \bmod p)}_{\text{\code{#Bob#}'s}\ \color{sgreen}{pB}}}^{\ \color{smagenta}sA}
\bmod p}^\text{\code{#Alice#}'s side}
\quad = \quad \underbrace{s}_{\text{shared secret}} 
\quad = \quad \overbrace{
{\underbrace{(g^{\color{smagenta}sA} \bmod p)}_{\text{\code{#Alice#}'s}\ {\color{sgreen}pA}}}^{\ \color{smagenta}sB}
\bmod p}^\text{\code{#Bob#}'s side}
.
\]

Since the computations performed by \code{#Alice#} and \code{#Bob#} are essential to the protocol, any faithful Choral implementation shall include those details too: doing otherwise would mean implementing a different protocol. The following is a snippet of the implementation that we have written for our evaluation in \cref{sec:evaluation} (with variables renamed to match our description above).
\begin{chorlisting}
BigInteger@#Bob# pA = g.modPow(sA, p) >> channel::<BigInteger>com;
BigInteger@#Bob# s = pA.modPow(sB, p);
\end{chorlisting}

Differently from these computational details, the Diffie-Hellman protocol does not fix the implementation of the channel used to communicate data. It is therefore reasonable that a Choral implementation of the protocol is parameterised on this implementation. In our case, this is reflected by the fact that \code{channel} is a parameter of the method that contains our code above.

An example of a choreography where the definition of computation is completely abstracted away is the \code{consumeItems} method in \cref{sec:knowledge-of-choice}. The choreography fixes the coordination between the participants, but not how they produce or consume the data to be exchanged. The latter is to be defined by either local code or another choreography that invokes \code{consumeItems}.

In general, how much computation should be defined in a choreography forms a spectrum. A `good' choreographic programming language should thus give freedom to define or abstract away computation at will. Choral provides this capability through the standard facilities of object-oriented programming (parameters, inheritance, etc.).

\section{Use cases}
\label{sec:usecases}

We dedicate this section to illustrating how the features of Choral contribute to writing realistic choreographies.
We start with a protocol for distributed authentication (\cref{sec:use_cases_distAuth}), which we then reuse modularly in another use case from the
healthcare sector that mixes cloud computing, edge computing, and Internet of
Things (IoT) (\cref{sec:use_cases_hospital}). Finally, we show a use case on
parallel computing, by showing a distributed implementation of merge sort (\cref{sec:use_cases_merge_sort}).

\subsection{Distributed authentication}
\label{sec:use_cases_distAuth}

We write a choreography for distributed authentication, inspired by the single sign-on authentication scheme: an \code{#IP#} (`Identity Provider', also known as central authentication service) authenticates a
\code{#Client#} that accesses a third-party \code{#Service#}.

We start by introducing an auxiliary class, \code{AuthResult}, that we will use
to store the result of authentication. The idea is that, after performing the
authentication protocol, both the \code{#Client#} and the \code{#Server#} should
have an authentication token if the authentication succeeded, or an `empty' value
if it failed. We model this by extending the \code{DPair} class
presented in \cref{sec:choral}.

\begin{chorlisting}[breakable=true]
public class AuthResult@(#A#,#B#) 
		extends DPair@(#A#,#B#)<Optional@#A#<AuthToken>,Optional@#B#<AuthToken>¦¦> {
	public AuthResult(AuthToken@#A# t1, AuthToken@#B# t2) {
		super(Optional@#A#.<AuthToken>of(t1), Optional@#B#.<AuthToken>of(t2)); 
	}
	public AuthResult() { 
		super(Optional@#A#.<AuthToken>empty(), Optional@#B#.<AuthToken>empty()); 
	}
}
\end{chorlisting}

The constructors of \code{AuthResult} guarantee that either both roles
(\code{#A#} and \code{#B#}) have an optional containing a value or both
optionals are empty (\code{Optional} is the standard Java type). Since
\code{AuthResult} extends \code{DPair}, these values are locally available
by invoking the \code{left} and \code{right} methods.

We now present the choreography for distributed authentication, as the
\code{DistAuth} class below.

\begin{chorlisting}[breakable=true]
enum AuthBranch { OK, KO }

public class DistAuth@(#Client#,#Service#,#IP#){
	private TLSChannel@(#Client#,#IP#)<Object> ch_Client_IP;
	private TLSChannel@(#Service#,#IP#)<Object> ch_Service_IP;
	
	public DistAuth(
		TLSChannel@(#Client#,#IP#)<Object> ch_Client_IP,
		TLSChannel@(#Service#,#IP#)<Object> ch_Service_IP ) { 
		this.ch_Client_IP = ch_Client_IP;
		this.ch_Service_IP = ch_Service_IP;
	}
	
	private String@#Client# calcHash(String@#Client# salt, String@#Client# pwd) { /*...*/ }

	public AuthResult@(#Client#,#Service#) authenticate(Credentials@#Client# credentials) {
		String@#Client# salt = credentials.username
		 >> ch_Client_IP::<String>com >> ClientRegistry@#IP#::getSalt >> ch_Client_IP::<String>com;
		Boolean@#IP# valid = calcHash(salt, credentials.password)
		 >> ch_Client_IP::<String>com >> ClientRegistry@#IP#::check;
		if (valid) {
			ch_Client_IP.<AuthBranch>select(AuthBranch@#IP#.OK);
			ch_Service_IP.<AuthBranch>select(AuthBranch@#IP#.OK);
			AuthToken@#IP# t = AuthToken@#IP#.create();
			return new AuthResult@(#Client#,#Service#)(
				ch_Client_IP.<AuthToken>com(t), ch_Service_IP.<AuthToken>com(t)
			);
		} else {
			ch_Client_IP.<AuthBranch>select(AuthBranch@#IP#.KO);
			ch_Service_IP.<AuthBranch>select(AuthBranch@#IP#.KO);
			return new AuthResult@(#Client#,#Service#)();
		}
	}
}
\end{chorlisting}

Class \code{DistAuth} is a \emph{multiparty} protocol parameterised over three roles: \code{#Client#}, \code{#Service#}, and \code{#IP#} (for Identity Provider).
It composes two channels as fields (lines 4--5), which respectively connect \code{#Client#} to \code{#IP#} and \code{#Service#} to \code{#IP#} -- hence, the interaction between \code{#Client#} and \code{#Service#} can only happen if coordinated by \code{#IP#}.
The channels are of type \code{TLSChannel}, a class for secure channels from the Choral standard library that uses TLS for security and the Kryo library~\citep{kryo} for marshalling and unmarshalling objects. Class \code{TLSChannel} implements interface \code{SymChannel}, from \cref{sec:choral}, so it can be used in both directions.
The private method \code{calcHash} (omitted) implements the local code that \code{#Client#} uses to hash its password.

Method \code{authenticate} (lines 16--33) is the key piece of \code{DistAuth}, which implements the authentication protocol. It consists of three phases.
In the first phase, lines 17--18, the \code{#Client#} communicates its \code{username} to \code{#IP#}, which \code{#IP#} uses to retrieve the corresponding salt in its local database \code{ClientRegistry}; the salt is then sent back to \code{#Client#}.
The second phase (lines 19--20) deals with the resolution of the authentication challenge. \code{#Client#} computes its hash with the received salt and its locally-stored password and sends the latter to \code{#IP#}. \code{#IP#} then checks whether the received hash is valid, storing this information in its local variable \code{valid}. The
result of the check is a \code{Boolean} stored in the \code{valid} variable located at \code{#IP#}.
The first two phases codify some best practices for distributed authentication and password storage \citep{GetAl17}: the identity provider \code{#IP#} never sees the password of the client, but only its attempts at solving the challenge (the \code{salt}), which
\code{#Client#} can produce with private information (here, its password).
In the third phase, (lines 21--32), \code{#IP#} decides whether the authentication was successful or not by checking \code{valid}. In both cases, \code{#IP#} informs the \code{Client} and the
\code{Service} of its decision, using selections to distinguish between success (represented by \code{OK}) or failure (represented by \code{KO}).
In case of success, \code{#IP#} creates a new authentication token (line 24) and communicates the token to both \code{#Client#} and \code{#Service#} (inner calls to \code{com} at line 26).
The protocol can now terminate and return a distributed pair (an \code{AuthResult}) that stores the same token at both \code{#Client#} and \code{#Service#}, which they can use later for further interactions (line 25).
In case of failure, the method returns an authentication result with empty \code{Optional}s (line 31).

New to choreographic programming, \code{DistAuth} is a higher-order choreography: the channels that it composes are choreographies for secure communication that carry state -- the result of the TLS handshake, which method \code{com} of \code{TLSChannel} uses internally. Taking this even further, we could overload method \code{authenticate} with a continuation-passing style alternative that, instead of returning a result, takes as parameters choreographic continuations (objects that involve \code{#Client#} and \code{#Service#}) to be called respectively in case of success (line 25) or failure (line 31).

\paragraph{Compilation}
We now discuss key parts of the compilation of \code{DistAuth} for role
\code{#Client#}, i.e., the Java library that clients can use to
\code{authenticate} to an identity provider and access a service.

\begin{commlisting}[breakable=true]
public class DistAuth_Client {
	private TLSChannel_A<Object> ch_Client_IP;
	
	public DistAuth_Client(TLSChannel_A <Object> ch_Client_IP) {
		this.ch_Client_IP = ch_Client_IP; 
	}
	
	private String calcHash( String salt, String pwd ) { /*...*/ }

	public AuthResult_A authenticate(Credentials credentials) {
		String salt = ch_Client_IP.<String>com(ch_Client_IP.<String>com(credentials.username));
		ch_Client_IP.<String>com(calcHash(salt, credentials.password));
		switch (ch_Client_IP.<AuthBranch>select(Unit.id)) { 
			case OK -> { return new AuthResult_A( ch_Client_IP.<AuthToken>com(Unit.id), Unit.id); }
			case KO -> { return new AuthResult_A(); }
			default -> { throw new RuntimeException( /*...*/ ); } 
		}
	}
}
\end{commlisting}

The field, constructor, and method at lines 2--8 are straightforward projections
of the source class for role \code{#Client#} -- fields and parameters pertaining only to other roles disappeared. The interesting code is at lines 10--17, which
defines the local behaviour of \code{#Client#} in the authentication protocol.
Note that forward-chaining sequences (\code{>>}) become plain nested calls in Java (lines
11 and 12). At line 11, the client sends its username to the identity provider
and receives back the \code{salt}. Recall from \cref{sec:choral} that the
innermost invocation of method \code{com} returns a \code{Unit} since the
client acts as the sender here. Once the username is sent, the innermost \code{com}
returns, and we run the outermost invocation of \code{com}, which received the
salt through the channel with the identity provider. At line 12, the Client
sends the computed hash to the identity provider.

At line 13, we see an example of how our compiler implements knowledge of choice for roles
that need to react to decisions made by other roles. The client receives an
enumerated value of type \code{AuthBranch}, which can be either \code{OK} or
\code{KO}, through the channel with the identity provider. Then, a \code{switch}
statement matches the received value to decide whether (case \code{OK}) we shall
receive an authentication token from the identity provider and store it as an
\code{AuthResult_A} or (case \code{KO}) the authentication procedure failed.

\subsection{A use case from healthcare: handling streams of sensitive vitals data}
\label{sec:use_cases_hospital}

In this use case, we exemplify how developers can locally compose the libraries generated by independent choreographies, using a healthcare use case inspired by previous works on edge computing and pseudonimisation \citep{SCGD19,GMSZ19}.

Suppose that a `healthcare service' in a hospital needs to gather sensitive data about vital signs (we call them vitals) from some IoT devices (e.g., smartwatches, heart monitors), and then upload them to the cloud for storage.
This is a typical scenario that requires the integration of libraries for participating in choreographies at the local level. We shall carry out the following two steps.
\begin{enumerate}
\item Define a new choreography class, called \code{VitalsStreaming}, that prescribes how data should be streamed from an IoT \code{#Device#} monitoring the vitals of a patient to a data \code{#Gatherer#}; this choreography shall enforce that the \code{#Gatherer#} processes only data that is (a) correctly cryptographically signed by the device and (b) pseudonymised.
\item Implement the healthcare service as a local Java class, called \code{HealthCareService}, that combines the Java library compiled from \code{VitalsStreaming} to gather data from the IoT devices with the Java library compiled from our previous \code{DistAuth} example, to authenticate at the cloud storage service through a third-party service (this could be, e.g., a national authentication system) and upload the data.
\end{enumerate}

\paragraph{Vitals choreography}
\code{VitalsStreaming} implements the choreography for streaming vitals.

\begin{chorlisting}[breakable=true]
public enum StreamState@#R# { ON, OFF }

public class VitalsStreaming@(#Device#,#Gatherer#) {
	private SymChannel@(#Device#,#Gatherer#)<Object> ch;
	private Sensor@#Device# sensor;
	
	public VitalsStreaming(SymChannel@(#Device#,#Gatherer#)<Object> ch, Sensor@#Device# sensor) {
		this.ch = ch;
		this.sensor = sensor; 
	}
	
	private static Vitals@#Gatherer# pseudonymise(Vitals@#Gatherer# vitals) { /*...*/ }
	private static Boolean@#Gatherer# checkSignature(Signature@#Gatherer# signature) {	/*...*/ }

	public void gather(Consumer@#Gatherer#<Vitals> consumer) {
		if (sensor.isOn()) {
			ch.<StreamState>select(StreamState@#Device#.ON);
			VitalsMsg@#Gatherer# msgOpt = sensor.next() >> ch::<VitalsMsg>com;
			if (checkSignature(msg.signature())) {
				msg.content() >> this::pseudonymise >> consumer::accept;
			}
			gather(consumer);
		} else {
			ch.<StreamState>select(StreamState@#Device#.OFF); 
		}
	}
}
\end{chorlisting}

At lines 3--5, the class \code{VitalsStreaming} composes a channel between the \code{#Device#} and the \code{#Gatherer#} and a \code{Sensor} object located at the \code{#Device#} (for obtaining the local vital readings).
At line 12, we define a method that pseudonymises personal data in \code{Vitals} at the \code{#Gatherer#}.
Likewise, at line 13 we have a method that the \code{#Gatherer#} uses to check that a message signature is valid. (We omit the bodies of these two static methods, which are standard local methods.)
The interesting part of this class is method \code{gather} (lines 15--26).
The \code{#Device#} checks whether its sensor is on (line 16) and informs the \code{#Gatherer#} of the result with appropriate selections for knowledge of choice (lines 17 and 24).
If the sensor is on, then \code{#Device#} sends its next available reading to \code{#Gatherer#} (line 18).
\code{#Gatherer#} now checks that the message is signed correctly (line 19); if so, it pseudonymises the content of the message and then hands it off to a local consumer function.
Notice that \code{#Gatherer#} does not need to inform \code{#Device#} of its local choice since it does not affect the code that \code{#Device#} needs to run.
We then recursively invoke \code{gather} to process the next reading.

\paragraph{Local code of the healthcare service}
The local implementation of the healthcare service acts as \code{#Gatherer#} in the \code{VitalsStreaming} choreography (to gather the data) and as the \code{#Client#} in the \code{DistAuth} choreography (to authenticate with the cloud storage). So we compose the compiled Java classes \code{VitalsStreaming_Gatherer} and \code{DistAuth_Client}, respectively.
\begin{locallisting}[breakable]
public class HealthCareService {
	public static void main() {
		TLSChannel_A toIP = HealthIdentityProvider.connect();
		MQTTClient toStorage = HealthDataStorage.connect();
		AuthResult_A authResult = new DistAuth_Client(toIP).authenticate(getCredentials());
		authResult.left().ifPresent(token -> {
			DeviceRegistry
				.parallelStream()
				.forEach(device ->
					Supervision.restart(() ->
						new VitalsStreaming_Gatherer(device.connect())
							.gather(data -> toStorage.com(new StorageMesg(token, data)))
					)
				);
		});
	}
	private static Credentials getCredentials() { /* ... */ }
}
\end{locallisting}

Above, the \code{main} method idiomatically combines Java standard libraries with those generated by our compiler.
At lines 3 and 4, we use auxiliary methods to connect to the identity provider (which implements \code{#IP#} in \code{DistAuth}) and the data storage service (which implements \code{#Service#} in \code{DistAuth}) -- these services are provided by third parties, e.g., the national health system and some cloud provider.
We choose a TLS channel to enact authentication. Instead, for communications with devices and storage, we use the MQTT protocol, which is typical for IoT applications~\cite{HTS08} -- \code{MQTTClient} implements (the projection of) interface \code{SymChannel}, dealing with possible connectivity issues.
At line 5, we run our distributed-authentication protocol as the \code{#Client#}.
At line 6, we check if we successfully received an authentication token by inspecting the optional result. If so, we obtain a parallel stream of \code{Device} objects from a local registry (lines 7--8). Each device (line 9) is handled by a worker that uses a restart supervision strategy (line 10), i.e., if the projection of the choreography \code{VitalsStreaming} encounters an unrecoverable error, it is restarted. At line 11, we create a new instance of \code{VitalsStreaming_Gatherer} (the code compiled for \code{#Gatherer#} from \code{VitalsStreaming}), which receives an MQTT channel for communicating with the device (obtained by \code{device.connect()}).
Finally, at line 12, we call the \code{gather} method to engage in the \code{VitalsStreaming} choreography with each device, passing a consumer function that sends the received data to the cloud storage service (including the authentication token).

Notice that we do not need to worry about pseudonymisation or signature checking
in the local code since the code compiled from \code{VitalsStreaming} manages
all these details.

\subsection{Merge sort}
\label{sec:use_cases_merge_sort}

The last use case that we present is a three-way concurrent implementation of merge sort \citep{K98}, which illustrates the design of parallel algorithms in Choral.
It also serves to showcase how role instantiation can help Choral programmers with writing load-distribution logic.

We define a class \code{MergeSort} parameterised on three roles: the \emph{merger} (\code{#M#}), which is the participant that holds the list of elements that have to be sorted, and two \emph{sorters} (\code{#S1#}, \code{#S2#}).
In the code of \code{MergeSort}, the merger splits its list into two halves and respectively communicates them to the two sorters.
The sorters are responsible for sorting their sublists and communicating the results back to the merger.
When the merger gets the ordered sublists, it merges them in the standard way to compute the complete ordered list.
The definition of \code{MergeSort} follows.

\begin{chorlisting}[breakable=true]
enum Choice@#R# { L, R }
public class Mergesort@(#M#,#S1#,#S2#){
	SymChannel@(#M#,#S1#)<Object> ch_MS1; 
	SymChannel@(#S1#,#S2#)<Object> ch_S1S2;
	SymChannel@(#S2#,#M#)<Object> ch_MS2;
	
	public MergeSort(
		SymChannel@(#M#,#S1#)<Object> ch_MS1,
		SymChannel@(#S1#,#S2#)<Object> ch_S1S2,
		SymChannel@(#S2#,#M#)<Object> ch_MS2 ) { 
		this.ch_MS1 = ch_MS1; this.ch_S1S2 = ch_S1S2; this.ch_MS2 = ch_MS2; 
	}

	public List@#M#<Integer> sort(List@#M#<Integer> a) {
		if (a.size() > 1@#M#) {
			ch_MS1.<Choice>select(Choice@#M#.L);
			ch_MS2.<Choice>select(Choice@#M#.L);
			Double@#M# pivot = a.size() / 2@#M# >> Math@#M#::floor >> Double@#M#::valueOf;
			MergeSort@(#S1#,#S2#,#M#) mb = new MergeSort@(#S1#,#S2#,#M#)(ch_S1S2, ch_MS2, ch_MS1);
			MergeSort@(#S2#,#M#,#S1#) mc = new MergeSort@(#S2#,#M#,#S1#)(ch_MS2, ch_MS1, ch_S1S2);
			List@#S1#<Integer> lhs = a.subList(0@#M#,pivot.intValue())
				>> ch_MS1::<List<Integer>¦¦>com >> mb::sort;
			List@#S2#<Integer> rhs = a.subList(pivot.intValue(), a.size())
				>> ch_MS2::<List<Integer>¦¦>com >> mc::sort;
			return merge(lhs >> ch_MS1::<List<Integer>¦¦>com, rhs >> ch_MS2::<List<Integer>¦¦>com);
		} else {
			ch_MS1.<Choice>select(Choice@#M#.R);
			ch_MS2.<Choice>select(Choice@#M#.R);
			return a;
		}
	}

	private List@#M#<Integer> merge(List@#M#<Integer> lhs, List@#M#<Integer> rhs) { /* ... */ }
}
\end{chorlisting}

In the code, the interaction logic that we have just described is implemented in method \code{sort}. Method \code{merge}, instead, merges two lists at \code{#M#}; we omit it, since it is entirely local and works as in the standard sequential algorithm.
Method \code{sort} consists of a conditional that checks whether the list should be split, which is determined by having more than one element (line 15).
If so, then the merger finds a pivot (line 18) and uses it to split the list and communicate the resulting sublists to the two sorters (lines 21--24).
The interesting part is that we recursively instantiate (lines 19--20) and use \code{MergeSort} (lines 22 and 24) to accomplish the two subtasks given to the sorters.
In these recursive invocations, roles are switched in order to put each sorter in charge of its own sublist: the sorter acts as the merger, and the other two nodes act as sorters.
This goes on until the list to be sorted cannot be split anymore.
The ordered sublists computed by each sorter are then merged at line 25.

The sequence diagram in \cref{fig:sd_mergesort} exemplifies the coordination pattern codified by our choreography (for brevity, we omit selections).
In the diagram we have three endpoint nodes -- Node\(_1\), Node\(_2\), and Node\(_3\) -- which engage in the choreography by playing the respective roles \code{#M#}, \code{#S1#}, and\\[2pt]
\begin{minipage}{.47\textwidth}
\code{#S2#}. We use numbered subscripts to denote the round (recursive step in the algorithm) that each interaction belongs to (sort$_1$, sort$_2$).
The initial input list is $[15,3,14]$; it is located at Node\(_1\), which in the beginning plays role \code{#M#}.
In the first round (first invocation of \code{sort}), Node\(_1\) asks Node\(_2\) and Node\(_3\) to sort the sublists obtained from the initial list -- respectively $[15,3]$ and $[14]$. This starts a recursive call (second round) where Node\(_2\) is the merger (\code{#M#}) and the other two nodes play the sorters. The lists communicated to the sorters in this round are $[15]$ and $[3]$. These lists contain only one element each, so they are dealt with by the else-branch of the conditional in method \code{sort} and we do not start other rounds.
Node\(_2\) now collects the lists from Node\(_1\) and Node\(_2\) and merges them locally, obtaining $[3,15]$.
Now that the recurs-{}\vspace{1pt}%
\end{minipage}\hfill\begin{minipage}{.51\textwidth}
  \vspace{-.5ex}
	\newcommand{\theline}{0}
	\newcommand{\setLine}[1]{\renewcommand\theline{#1}}
	\newcommand{\nodeA}{0}
	\newcommand{\nodeB}{3}
	\newcommand{\nodeC}{6}
	\begin{tikzpicture}
		\draw (\nodeA,.1) -- (\nodeA,-4.50) (\nodeB,.4) -- (\nodeB,-4.50) (\nodeC,.4) -- (\nodeC,-4.50);
		\setLine{.6}
		\node at (\nodeA,\theline) {\footnotesize{Node\(_1\)}}; 
		\node at (3,\theline) {\footnotesize{Node\(_2\)}}; \node at (6,\theline)
		{\footnotesize{Node\(_3\)}};
		\setLine{.3}
		\node at (0,\theline) {\footnotesize{[15,3,14]}};
		\setLine{-.2}
		\node at (0,\theline) [fill=white] {\footnotesize{plays \code{#M#}}}; \node
		at (3,\theline) [fill=white] {\footnotesize{plays \code{#S1#}}}; \node at
		(6,\theline) [fill=white] {\footnotesize{plays \code{#S2#}}};
		\setLine{-.7}
		\draw[-stealth] (\nodeA,\theline) -- node[above] {\footnotesize{sort\(_1\) [15,3]}} (\nodeB,\theline);
		\setLine{-.9}
		\draw[-stealth] (\nodeA,\theline) -- node[near end, above] {\footnotesize{sort\(_1\) [14]}} (\nodeC,\theline);
		\setLine{-1.4}
		\node at (0,\theline) [fill=white] {\footnotesize{plays \code{#S2#}}}; 
		\node at (3,\theline) [fill=white] {\footnotesize{plays \code{#M#}}}; 
		\node at (6,\theline) [fill=white] {\footnotesize{plays \code{#S1#}}};
		\setLine{-1.8}
		\draw[-stealth] (3,\theline) -- node[above] {\footnotesize{sort\(_2\) [15]}} (0,\theline);
		\setLine{-2}
		\draw[-stealth] (3,\theline) -- node[above] {\footnotesize{sort\(_2\) [3]}} (6,\theline);
		\setLine{-2.5}
		\draw[-stealth] (6,\theline) -- node[above] {\footnotesize{sort\(_2\) [3]}} (3,\theline);
		\draw[-stealth] (0,\theline) -- node[above] {\footnotesize{sort\(_2\) [15]}} (3,\theline);
		\setLine{-3}
		\node at (0,\theline) [fill=white] {\footnotesize{plays \code{#M#}}}; 
		\node at (3,\theline) [fill=white] {\footnotesize{plays \code{#S1#}}}; 
		\node at (6,\theline) [fill=white] {\footnotesize{plays \code{#S2#}}};
		\setLine{-3.75}
		\draw[-stealth] (6,\theline) -- node[near end, above]	{\footnotesize{sort\(_1\) [14]}} (0,\theline);
		\setLine{-4.25}
		\draw[-stealth] (3,\theline) -- node[above]	{\footnotesize{sort\(_1\) [3,15]}} (0,\theline);
		\setLine{-4.75}
		\node at (0,\theline) {\footnotesize{[3,14,15]}};
	\end{tikzpicture}
	\captionof{figure}{Sequence diagram of data exchanges in the three-way distributed	merge sort (selections are omitted).}\label{fig:sd_mergesort}
\end{minipage}
ive invocation for sorting $[15,3]$ is complete, we are back to completing the invocation of \code{sort} where Node\(_1\) plays \code{#M#}.
Node\(_1\) collects the lists $[3,15]$ and $[14]$ from Node\(_2\) and Node\(_3\), and then merges them to obtain the final result $[3,14,15]$.

Note that the code for parallel merge sort closely resembles the structure of a standard sequential merge sort, but thanks to role annotations we obtain a parallel implementation. We will come back to this aspect in \cref{sec:karatsuba}.
Another key benefit is that the compiled code is deadlock-free by construction, as usual for choreographic
programming \citep{CM13}.
A more detailed discussion of safety and liveness properties is given in \cref{sec:discussion}.

\section{Implementation}
\label{sec:implementation}\looseness=-1

We discuss the main elements of the implementation of Choral.
First, we show its syntax and comment on the main differences with Java's. 
Then, we present the Choral type checker, including examples of the main errors related to roles that it detects and related error messages.
Finally, we describe the key components of the Choral compiler.

\subsection{Language}

\begin{figure*}[t]
	\centering
  \begin{displaymath}
    \def\mlbr{\texttt{\{}}
    \def\mrbr{\texttt{\}}}
    \begin{array}{llrl}
      \\
      \text{Literals} & \cst & \Coloneqq & 
      \mcode{null}\diff{\mcode{@(}\many{\genW}\mcode{)}} \Div 
      \mcode{true}\diff{\mcode{@}{\genW}} \Div 
      \mcode{false}\diff{\mcode{@}{\genW}} \Div 
      \mcode{"a"}\diff{\mcode{@}{\genW}} \Div \dots  \Div
      \mcode{1}\diff{\mcode{@}{\genW}} \Div \dots
      \\
      \text{Program} & \program & \Coloneqq & 
        \program \s \interface 
        \Div \program \s \class 
        \Div \program \s \enum 
        \Div \program \s \grammarTerm{EOF}
      \\
      \text{Enum} & \enum & \Coloneqq & \opt{\ANN}\s\many{\MOD}\s \mcode{enum}\s\id\diff{\mcode{@}\genW }
      \mlbr \many{\id} \mrbr
      \\
      \text{Interface} & \interface & \Coloneqq & 
      \opt{\ANN}\s\many{\MOD}\s\mcode{interface}
      \s \id\diff{\at( \genWs )} 
        \opt{ \params{ \many{\FTP} } }\ 
        \opt{\mcode{extends}\s \TE\many{\mcode{,}\s\TE}}\mlbr \many{\MethodDef\mcode{;}} \mrbr
      \\
      \text{Annotation} & \ANN & \Coloneqq & \mcode{@}id\mcode{(}\many{id\ \texttt{=}\ \cst}\mcode{)}
      \\
      \text{Modifiers} & \MOD & \Coloneqq & \mcode{public} \Div \mcode{protected} \Div \mcode{private} \Div \mcode{abstract} \Div \mcode{final} \Div \mcode{static}
      \\
      \hspace{-5px}\begin{array}{l}
      \text{Formal}\\\text{Type Param.}
      \end{array}
      & \FTP & \Coloneqq & \id\diff{\mcode{@}( \genWs )}\s
      \opt{ \mcode{extends}\s \TE \s \many{\mcode{\&}\s\TE} }
      \\
      \text{Type Expr.} & \TE & \Coloneqq & 
      id\opt{\params{ \many{\TE} }} \Div id\diff{\mcode{@}(\genWs)}\opt{\params{ \many{\TE} }} \Div \mcode{void}
      \\
      \text{Method Def.} & \MethodDef & \Coloneqq & \opt{\ANN}\s\many{\MOD}\s 
      \opt{ \params{ \many{\FTP} } }\s \TE\s \id\s(\many{\VarDecl})
      \\
      \text{Class} & \class & \Coloneqq & \opt{\ANN}\s\many{\MOD}\s \mcode{class}\s 
      \id\diff{\mcode{@} ( \genWs )}\opt{\params{ \many{\FTP} }}\ 
      \opt{ \mcode{extends}\s \TE } \s
      \opt{ \mcode{implements}\s \TE\many{\mcode{,}\s \TE} }
      \\ &&&
      \mlbr \many{\ClassField}\ \many{\ClassConst}\ 
      \many{\MethodDef\mcode{;}}\ \many{\MethodDef\mlbr \Stm \mrbr } \mrbr
      \\
      \text{Class Field.} & \ClassField & \Coloneqq & 
      \opt{\ANN}\s\many{\MOD}\s \TE\s \many{id}\mcode{;}
      \\
      \text{Class Con.} & \ClassConst & \Coloneqq & 
      \opt{\ANN}\s\many{\MOD}\s \opt{ \params{ \many{\FTP} } } \s
      \id(\many{\VarDecl})\mlbr\Stm\mrbr
      \\
      \text{Statement} & \Stm & \Coloneqq & nil \Div \mcode{return}\s \opt{\Exp}\mcode{;} \Div \Exp\mcode{;} \Stm \Div \VarDecl\ \opt{\mcode{=}\   \Exp}\mcode{;} \Stm
      \\ && \Div & \Exp \s \AsgOp \s \Exp\mcode{;}\ \Stm \Div	\mcode{if} (\Exp)\mlbr\Stm\mrbr\opt{\mcode{else}\mlbr\Stm\mrbr}\ \Stm
      \\ && \Div &  \mlbr \Stm \mrbr\ \Stm \Div \mcode{try}\mlbr \Stm \mrbr 
      \many{\mcode{catch}(\VarDecl)\mlbr \Stm \mrbr}\ \Stm
      \\
      \text{Expression} & \Exp & \Coloneqq & \cst \Div \FA \Div \Exp \s \BinOp \s \Exp \Div \Exp\mcode{.}\Exp \Div \opt{\params{\many{\TE}}}\id( \many{ \Exp } )
      \\ && \Div & \new\s \opt{\params{\many{\TE}}}\id\diff{\at( \genWs )}\opt{\params{ \many{\TE} }}( \many{\Exp} ) \Div \id\diff{\at(\genWs)}.\opt{\params{\many{\TE}}}\id( \many{ \Exp } ) \Div \diff{\Exp \s \many{\mcode{>>} \s \ExpChain}}
      \\
      \text{Field Acc.} & \FA & \Coloneqq &
      \id \Div \id\diff{\at(\genWs)}.\id
      \\
      \text{Exp. Chain} & \ExpChain & \Coloneqq & 
      \FA  \opt{\many{\mcode{.}\id}} \mcode{::} \id \Div \id\diff{\mcode{@}(\genWs)}\opt{\params{\many{TE}}}\mcode{::}\new
      \\
      \text{Assign Op.} & \AsgOp & \in & \{ \mcode{=}, \mcode{+=}, \mcode{-=}, \mcode{*=}, \mcode{/=}, \mcode{\&\!=}, \mcode{|\!=}, \mcode{\%\!=} \}
      \\
      \text{Binary Op.} & \BinOp & \in & \{ \mcode{||}, \mcode{\&\&}, \mcode{|}, \mcode{\&}, \mcode{==}, \mcode{!=}, \mcode{<}, \mcode{>}, \mcode{<=}, \mcode{>=}, \mcode{+}, \mcode{-}, \mcode{*}, \mcode{/}, \mcode{\%} \}
    \end{array}
    \end{displaymath}    
	\caption{Syntax of the Choral language.}
	\label{fig:syntax}
\end{figure*}

\Cref{fig:syntax} displays the grammar of Choral; dashed underlines denote \hlOpt{optional terms} and solid overlines denote sequences of terms of the same sort. 
We omit syntax for packages and imports, which is as in Java.
Reserved identifiers like \mcode{super} and \mcode{this} are considered as identifiers in the grammar, but our compiler treats them like their Java counterparts.
The key syntactic novelties are \hlNew{underlined}; they consist of \emph{i}) syntax for declaring and instantiating role parameters and \emph{ii}) the forward chaining operator \mcode{>>} (cf.~\cref{sec:choral}).

Role parameters have a separate namespace, and always appear in expressions like
\code{@(#A$\textsubscript{1}$#,...,#A$\textsubscript{n}$#)} that follow the name
of a class, interface, enum, or type parameter e.g., \code{DiChannel@(#A#,#B#)}. 
Also, they are introduced only by the declaration of a type (e.g., \code{class Foo@(#A#,#B#)}) or a type parameter (e.g., \code{<T@(#A#,#B#) extends Foo@(#A#,#B#) & Bar@(#B#,#A#)>}) and their scope is limited to the defining type, similar to type parameters in Java.
The snippet below contains an example of shadowing of role parameters; for each use of role \code{#A#}, we show its binding site with an arrow.
\begin{chorlisting}*[top=13pt,bottom=13pt][numbers=none]
interface Foo@(¦\tikzmark{n0l}¦#A#¦\tikzmark{n0r}¦,#B#) extends Bar@(¦\tikzmark{n0-1l}¦#A#¦\tikzmark{n0-1r}¦,#B#) { <T@(¦\tikzmark{n1l}¦#A#¦\tikzmark{n1r}¦,#B#) extends Foo@(¦\tikzmark{n1-1l}¦#A#¦\tikzmark{n1-1r}¦,#B#) & Bar@(#B#,¦\tikzmark{n1-2l}¦#A#¦\tikzmark{n1-2r}¦)> T@(¦\tikzmark{n0-2l}¦#A#¦\tikzmark{n0-2r}¦,#B#) m();}
\end{chorlisting}
\begin{tikzpicture}[remember picture, overlay,ra/.style={<-,sblue,thick,rounded corners=2pt}]
\draw[ra] ([yshift=6pt]$(pic cs:n1l)!.5!(pic cs:n1r)$) -- ++ (90:5pt) -| ([yshift=6pt]$(pic cs:n1-1l)!.5!(pic cs:n1-1r)$);
\draw[ra] ([yshift=6pt]$(pic cs:n1l)!.5!(pic cs:n1r)$) -- ++ (90:5pt) -| ([yshift=6pt]$(pic cs:n1-2l)!.5!(pic cs:n1-2r)$);

\draw[ra] ([yshift=-4pt]$(pic cs:n0l)!.5!(pic cs:n0r)$) -- ++ (-90:5pt) -| ([yshift=-4pt]$(pic cs:n0-1l)!.5!(pic cs:n0-1r)$);
\draw[ra] ([yshift=-4pt]$(pic cs:n0l)!.5!(pic cs:n0r)$) -- ++ (-90:5pt) -| ([yshift=-4pt]$(pic cs:n0-2l)!.5!(pic cs:n0-2r)$);

\end{tikzpicture}

\ifsubmission
\newenvironment{comperror}[1][]{\noindent\tcblisting{
    enhanced, %
    before skip=\abovedisplayskip,
    after  skip=\belowdisplayskip,
    sharp corners=all,
    boxrule=2pt,
    boxsep=-.7em,
    colframe=red,
    colback=sbase3,
    detach title,
    finish={\node[anchor=north east, font=\footnotesize\itshape,
      text=sbase3,fill=red] at (frame.north east) {Compiler error};},
    title={},
    listing only,
    listing options={language=comperr,style=solarized-light,numbers=none},
    #1
}}{\endtcblisting\noindent}
\else
\newenvironment{comperror}[1][]{\noindent\tcblisting{
    enhanced, %
    before skip=\abovedisplayskip,
    after  skip=\belowdisplayskip,
    sharp corners=all,
    boxrule=2pt,
    boxsep=-.7em,
    colframe=red,
    colback=sbase3,
    title={},
    listing only,
    listing options={language=comperr,style=solarized-light,numbers=none},
    #1
}}{\endtcblisting\noindent}
\fi

The Choral type checker covers all common Java type errors (illegal type conversions, access to type members, etc.), as exemplified below.
Indeed, when checking a Choral program with exactly one role parameter, the Choral type checker acts exactly like its Java counterpart.
\begin{comperror}
Integer@#A# x = "foo"@#A#;
~------------^~
/~Incompatible types: expecting '~/Integer@#A#/~' found '~/String@#A#/~'.~/

return x.length();
~---------^~
/~Cannot resolve method '~/length/~' in '~/Integer@(#A#)/~'
\end{comperror}

\paragraph{Roles}
The novelties compared to Java compilers emerge when two or more roles are involved. In these settings, programmers can make new kinds of errors that are specifically about the misuse of role-parameterised types -- therefore, these errors are pertinent to Choral. In many of the examples discussed so far, we can think of role parameters as Java generics.
Although this is a reasonable approximation, some care is necessary when handling type instantiation due to some substantial differences between role and type parameters.

One type of these errors is that data types are instantiated using incompatible roles. Instances of the same type with different roles parameters represent values located at different roles, which restricts their usage.
\begin{comperror}
String@#A# x = "foo"@#B#; // error, same local type but at different roles
~-----------^~
/~Incompatible types: expecting '~/String@#A#/~' found '~/String@#B#/~'.~/
\end{comperror}
The order in which roles appear carries meaning, since role parameters are positional -- like type parameters in Java generics. 
\begin{comperror}
void m(SymChannel@(#A#,#B#)<T> x) {
  SymChannel@(#A#,#B#)<T> a = x; // matching roles
  SymChannel@(#B#,#A#)<T> b = x; // error, same roles but wrong positions
~------------------------^~
/~Incompatible types: expecting '~/SymChannel@(#B#,#A#)<T>/~' found '~/SymChannel@(#A#,#B#)<T>/~'.~/
\end{comperror}
Differently from Java generics, role parameters cannot appear multiple times in the same type since this corresponds to requiring that the same participant plays multiple roles in the same choreography. In the snippet below, \mcode{#A#} must play both the sender and receiver for the directed channel \mcode{c}.
\begin{comperror}
DiChannel@(#A#,#A#)<String> c;
~-------------^~
/~Illegal type instantiation: role '~/#A#/~' must play exactly one role in '~/DiChannel/~'.~/
\end{comperror}
Forbidding role aliasing is an established restriction in choreographic programming since aliasing introduces self-communication, which would potentially break deadlock-freedom and the capability to produce separate code for each role (unless roles are provably not-aliased).

\paragraph{Subtyping}
Choral types form a hierarchy defined following the same principles used by Java. This hierarchy is used to check if values are compatible type as expected. 
\begin{comperror}
void m(BiChannel@(#A#,#B#)<T,T> x) {
  DiChannel@(#A#,#B#)<T>  a=x; // BiChannel$\codeComment{@}$(A,B)<T,T> extends DiChannel$\codeComment{@}$(A,B)<T>
  DiChannel@(#B#,#A#)<T>  b=x; // BiChannel$\codeComment{@}$(A,B)<T,T> extends DiChannel$\codeComment{@}$(B,A)<T>  
  SymChannel@(#A#,#B#)<T> c=x; // error, BiChannel$\codeComment{@}$(A,B)<T,T> does not extend SymChannel$\codeComment{@}$(A,B)<T>
~-----------------------^~
/~Incompatible types: expecting '~/SymChannel@(#A#,#B#)<T>/~' found '~/BiChannel@(#A#,#B#)<T>/~'.~/
\end{comperror}

Classes and interfaces define their supertypes by extending and implementing other classes and their interfaces with the same set of roles.
This restriction provides a substitution principle that elicits all roles involved in a choreography.
\begin{comperror}
interface AuditedDiChannel@(#A#,#B#,#Auditor#)<T@#C#> extends DiChannel@(#A#,#B#)<T> {/*...*/}
~------------------------------------------------------^~
/~ Illegal inheritance: '~/AuditedDiChannel@(#A#,#B#,#Auditor#)/~' and '~/DiChannel@(#A#,#B#)<T>/~' must have the same roles.~/

interface ReplicatedList@(#A#,#Replica#)<T@#B#> extends List@#A#<T> {/*...*/}
~--------------------------------------------------^~
/~ Illegal inheritance: '~/ReplicatedList(#A#,#Replica#)/~' and '~/List@#A#<T>/~' must have the same roles.~/
\end{comperror}
In some cases, `hidden roles' in choreographies might be useful, e.g., to add external auditing or data replication as an extension of an existing choreography. 
Unfortunately, this introduces security concerns (channels may have hidden bystanders) or complex communication semantics (what is the meaning of sending a
\code{ReplicatedList@(#A#,#B#)} over a channel expecting a \code{List@#A#}?). These are general open problems for choreographies, left to future work.

Cyclic inheritance is not allowed and the type checker does not
discriminate over role parameters. As an example, consider the \code{SymChannel} interface; given its
symmetric nature, one might be tempted to force this equality by having 
\code{SymChannel@(#A#,#B#)} subtype \code{SymChannel@(#B#,#A#)}.
\begin{comperror}
interface SymChannel@(A,B)<T@#C#> extends SymChannel@(B,A)<T> { /* ... */ }
~----------------------------------------^~
/~Cyclic inheritance: '~/SymChannel/~' cannot extend '~/SymChannel/~'.~/
\end{comperror}
\noindent However, allowing declarations like the one above in Choral would result in cyclic inheritance errors in Java, as exemplified by the following `manual compilation' of the code above.
\begin{commlisting}*[][numbers=none]
interface SymChannel_A<T> extends SymChannel_B<T> { /* ... */ } // Projection for A
interface SymChannel_B<T> extends SymChannel_A<T> { /* ... */ } // Projection for B
\end{commlisting}
To have channels that are instances of both \code{SymChannel@(#A#,#B#)} and \code{SymChannel@(#B#,#A#)} one needs to define a subtype of both as in the snippet below.
\begin{chorlisting}[][numbers=none]
interface PeerChannel@(#A#,#B#)<T@#C#> extends SymChannel@(#A#,#B#)<T>, SymChannel@(#B#,#A#)<T> {
  <S@#C# extends T@#C#> S@#B# com(S@#A# m);          // inherited
  <S@#C# extends T@#C#> S@#A# com(S@#B# m);          // inherited
  <S@#C# extends Enum@#C#<S>¦¦> S@#B# select(T@#A# m); // inherited
  <S@#C# extends Enum@#C#<S>¦¦> S@#A# select(T@#B# m); // inherited
  
  PeerChannel@(#B#,#A#)<T> flip();               // roles A and B are interchangeable
}
\end{chorlisting}
By returning an instance of the same interface but with the roles flipped, the method \code{flip()} introduced by the interface \code{PeerChannel@(#A#,#B#)}, prescribes that the roles \code{#A#} and \code{#B#} are interchangeable peers.

Finally, primitive types (\code{int@#A#}, \code{bool@#A#}, etc.) follow the same rules of Java for subtyping, conversions, autoboxing, and autounboxing (when roles match, otherwise the compiler will return a role mismatch error).

\paragraph{Overloading}
The Choral type checker refines overload equivalence: it can discriminate overloaded methods by considering roles. For example, \code{m(Char@#B# x)} and \code{m(Char@#A# x)} can be distinguished because one parameter is located at \code{#A#} whereas the other at \code{#B#}.
However, we need to be careful with preventing potential clashes in the compiled Java code. Consider the following snippet and error message.
\begin{comperror}
class Foo@(#A#,#B#) {
  void m(Char@#B# x) { /* ... */ }  //        void m() at A and void m(Char x) at B
  void m(Char@#A# x) { /* ... */ }  //        void m(Char x) at A and void m() at B
  void m(Long@#A# x) { /* ... */ }  // error, void m(Long x) at A and void m() at B
~-------^~
/~Illegal overload: '~/m(Long@#A# x)/~' and '~/m(Char@#A# x)/~' have the same signature for role '~/#B#/~'.~/
\end{comperror}
The last two signatures are distinguishable in Choral, since each
method has different parameter types. However, this information is only
available to role \code{#A#}, while the projection of both signatures at role
\code{#B#} coincide (they would both be \code{void m()}).
This is an instance of knowledge of choice but, differently from
conditionals, it cannot be addressed locally (within the class/interface) because
extending classes may introduce new branches and new points of choice by overriding
and overloading, as in the example below.
\begin{chorlisting}[][numbers=none]
class Bar@(#A#,#B#) extends Foo@(#A#,#B#) {	void m(Integer@#A# x) { /* ... */ } }
\end{chorlisting}

\paragraph{Exceptions}
Like every other type lifted from Java, exceptions are located at one role.
This design choice allows us to preserve the expected type hierarchy in the generated code and have \code{java.lang.Exception} as the supertype of all exceptions.
The Choral compiler then enforces that a \code{try}-\code{catch} block is located at exactly one role.
\begin{comperror}
String@#A# fetch(DiChannel@(#A#,#B#)<T> ch, String@#B# file) {
  try { return RemoteReader@(#A#,#B#).read(ch,file); } catch (IOException@B e) { return null@A; }
~--^~
/~Non-local try-catch: try-catch must be at a single role, found '~/#A#/~' and '~/#B#/~'.~/
\end{comperror}
Allowing multiple roles in the mechanics of exceptions introduces a knowledge-of-choice situation where all roles need to obtain information about which handler to execute, if any, and when.
The general problem of exceptions and their choreographic handling has been investigated in some theories, but all models proposed so far assume reliable communications among all roles and either rely on specialised orchestration primitives in the target language (i.e., some form of middleware) \cite{CHY08,C09,FLMD19} or synthesise new communications for recovery \cite{NY17}.

\subsection{Compiler}
\label{sec:compiler}

The Choral compiler consists of several steps organised in a pipeline, which we illustrate in \cref{fig:comp_pipeline}.
From left to right, the first
step is (as expected) parsing the input Choral source code to obtain an Abstract Syntax Tree (AST) -- the chain operator \code{>>} is desugared in this step.
Next, we perform type checking as previously discussed.
This step also annotates the nodes in the AST with type information, which is used in the following projection step.
The next step -- projection -- transforms this annotated AST into a collection of Choral ASTs, each representing the implementation of a single role. At this stage, all types are located at exactly one role, representing the fact that all code is fully local.
Finally, in the last step, we output Java code by erasing all role annotations.

\newlength{\figBoxWidth}
\newcommand{\figBox}[2][.1]{\setlength{\figBoxWidth}{#1\textwidth}\begin{minipage}{\figBoxWidth}\begin{center}\footnotesize #2\end{center}\end{minipage}}
\newcommand{\figArrow}[2][.1]{\figBox[#1]{\(\xRightarrow{\parbox{1\textwidth}{#2}}\)}}
\begin{figure}[b]
\hspace{-1.8em}
\figBox{\includegraphics[width=.7\textwidth]{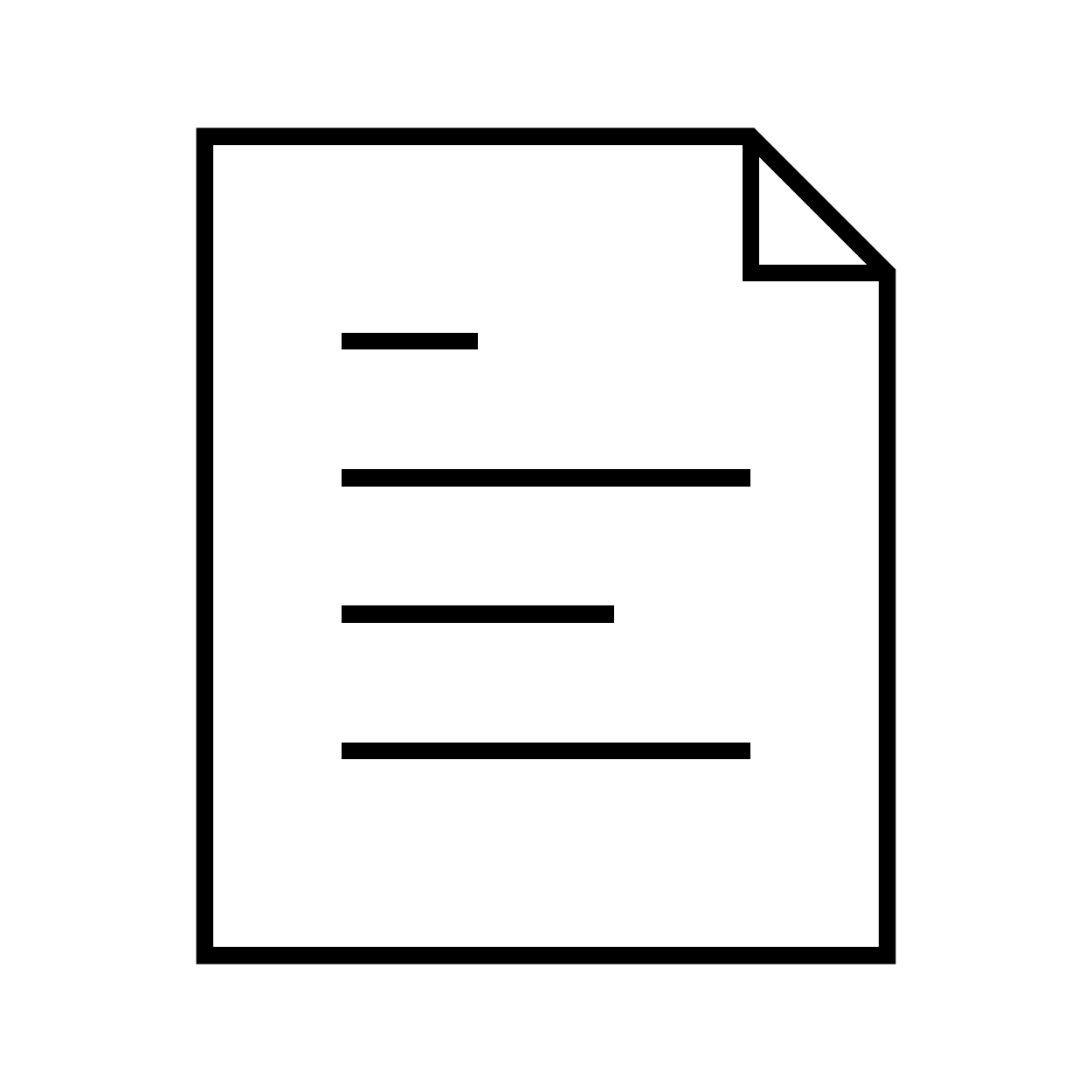} Choral\\ class}
\hspace{-1em}
\figArrow{Parsing \& \\ desugaring}
\figBox{\includegraphics[width=.7\textwidth]{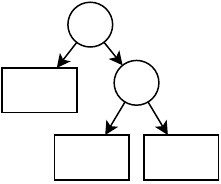} Choral\\ AST}
\hspace{-.8em}
\figArrow[.12]{Type checking \\ \& annotation}
\hspace{.1em}
\figBox{\includegraphics[width=.8\textwidth]{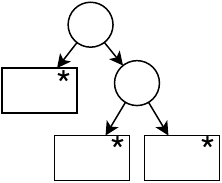} Annotated\\ Choral AST}
\figArrow[.09]{Projection}
\hspace{.1em}
\figBox[.14]{\includegraphics[width=\textwidth]{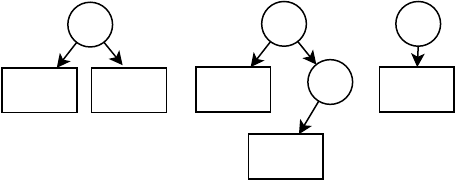} Single-role\\ Choral ASTs}
\hspace{.1em}
\figArrow{Role erasure \\ \& output}
\figBox{\includegraphics[width=.7\textwidth]{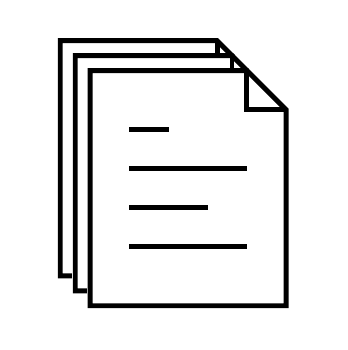} Java\\ classes}
\caption{\label{fig:comp_pipeline} Pipeline Schema of the Choral Compiler.}
\end{figure}

\newcommand{\epp}[2][\mcode{\#A\#}]{\sol{#2}^{#1}}
We discuss the most important aspects of projection. For clarity, we model and present it as a partial function from (well-typed) Choral terms to Java code: the projection of a Choral term $\mathit{\Term}$ on a role \code{#A#}, written $\epp{\mathit{\Term}}$, is a Java term that implements the behaviour of \code{#A#} in $\mathit{Term}$.
Intuitively, this models together the last two steps in \cref{fig:comp_pipeline}.
The full definition of projection is given in~\cref{sec:projection-full}.

The projection of a Choral \code{class}, \code{interface}, or \code{enum}
generates a corresponding Java term for each role parameter. If there are two or
more roles, each Java artefact name is suffixed with the role that it implements, e.g., the Java class compiled
from \code{class Foo(#A#,#B#)} for role \code{#A#} is called \code{Foo_A}.
If the Choral class has exactly one
role, then we use the same name, e.g., \mcode{class Integer@A} becomes \mcode{class Integer}.
(This erases friction for the integration of Java types within Choral.)

The projection $\epp{\TE}$ of a type expression $\TE$ at a role \code{#A#} is recursively defined below -- we use the auxiliary function $\mathrm{roleName}(id,i)$ to retrieve the name of the $i$-th
role parameter from the definition of $id$.
\[
\epp{id\mcode{@(}\many{\mcode{#B#}}\mcode{)<}\many{\TE}\mcode{>}} = 
\begin{cases}
	id\mcode{<}\many{\epp{\TE}}\mcode{>} & \text{if } \many{\mcode{#B#}} = \mcode{#A#}\\
	id\mcode{_}A'{<}\many{\epp{\TE}}\mcode{>} & \text{if } \mcode{#A#} \text{ is the $i$-th element of } \many{\mcode{#B#}} \text{ and } \mathrm{roleName}(id,i) = A' \\
	\mcode{Unit} & \text{otherwise}
\end{cases}
\]

\def\rolesOf{\mathrm{rolesOf}}\looseness=-1
The projection $\epp{\Exp}$ of an expression $\Exp$ at role \code{#A#} is defined following a similar intuition: it is a recursive stripping of role information as long as \code{#A#} occurs in the type of $\Exp$ or any of its subterms (written $\mcode{#A#} \in \rolesOf(\Exp)$), otherwise it is the only instance of the singleton \mcode{Unit} (stored in its static field \code{id}), as illustrated by the cases of static field access and constructor invocation below.
\begin{align*}
	\epp{id\mcode{@(}\many{\mcode{#B#}}\mcode{).}f} = {} &
	\begin{cases}
		\epp{\id\mcode{@(}\many{\mcode{#B#}}\mcode{)}}\mcode{.}f & 
		\text{if } \mcode{#A#} \in \rolesOf(f)\\
		\mcode{Unit.id} & \text{otherwise}
	\end{cases}
	\\
	\epp{\new\s \params{\many{\TE}}\id\mcode{@(} \many{\mcode{#B#}} \mcode{)}\params{ \many{\TE}}\mcode{(} \many{\Exp} \mcode{)}} = {} & \begin{cases}
	\new\s \params{\many{\epp{\TE}}}\epp{\id\mcode{@(} \many{\mcode{#B#}} \mcode{)}\params{ \many{\TE}}}\mcode{(} \many{\epp{\Exp}} \mcode{)} & \text{if \code{#A#}} \in \many{\mcode{#B#}} \\ 
	\mcode{Unit.id(} \many{\epp{Exp}} \mcode{)} & \text{otherwise}
	\end{cases}
\end{align*}

The projection $\epp{\Stm}$ of a statement $\Stm$ at \code{#A#} is defined following the above intuition, save for the cases of conditionals and selections, which require care to address knowledge of choice (cf.~\cref{sec:knowledge-of-choice}).
Specifically, the rule for projecting \code{if} statements:
for the role evaluating the guard (read from its type), it preserves the conditional;
for all other roles, the \code{if}
disappears and it is replaced by the projection of the guard (since it
might have side-effects) followed by the \emph{merging} \(\solMerge\) of the projections of the bodies of the two
branches and the projection of the continuation \(\Stm\).
\[
\scalemath{.93}{
\epp{\mcode{if}( \Exp )\{ \Stm_1 \} \mcode{else}\{ \Stm_2\} \Stm } =
\begin{cases}
  \mcode{if}( \epp{\Exp} )\{ \epp{\Stm_1} \} 
  \mcode{else}\{ \epp{\Stm_2}\} \epp{\Stm}
  & \mbox{if } \Exp\colon\mcode{boolean}\at\genW
  \\
  \epp{\Exp}; \left\{ \epp{Stm_1} \solMerge \epp{Stm_2} \right\} \epp{Stm} 
\  & \mbox{otherwise}
\end{cases}
}
\]
The merge operator \(\Stm \solMerge \Stm'\) is a partial operator that tries to combine branching code \citep{CHY12}, which we adapt to Java for the first time.
Essentially, given two Java terms, merging recursively requires them to be equivalent \emph{unless} they are switch statements.
\Cref{sec:projection-full} contains the full definition of merging.
Here we report its most interesting case: merging switch statements.
\[
{\def\arraystretch{1}
\begin{array}l
	\switch\s ( \Exp ) \{ 
	\\\quad\case\s \id_a \mcode{->} \{ \Stm_a \}
	\\\quad\cdots
	\\\quad\case\s \id_x \mcode{->} \{ \Stm_x \}
	\\\quad\many{\case\s \id_y \mcode{->} \{ \Stm_y \} }
	\\\quad\mcode{default} \mcode{->} \{ \Stm_{d1} \}
	\\\}\s \Stm 
	\end{array}
	\solMerge 
	\hspace{-5pt}\begin{array}l
		\switch\s ( \Exp ) \{ 
		\\\quad\case\s \id_a \mcode{->} \{ \Stm_a' \}
		\\\quad\cdots
		\\\quad\case\s \id_x \mcode{->} \{ \Stm_x' \}
		\\\quad\many{\case\s \id_z \mcode{->} \{ \Stm_z \} }
		\\\quad\mcode{default} \mcode{->} \{ \Stm_{d2} \}
		\\\}\s \Stm'
		\end{array}
	=
	\hspace{-5pt}\begin{array}l
		\switch\s ( \Exp \solMerge \Exp' ) \{
		\\\quad\case\s \id_a \mcode{->} \{ \Stm_a \solMerge \Stm_a' \}
		\\\quad\cdots
		\\\quad\case\s \id_x \mcode{->} \{ \Stm_x \solMerge \Stm_x' \}
		\\\quad\many{\case\s \id_y \mcode{->} \{ \Stm_y \} }
		\\\quad\many{\case\s \id_z \mcode{->} \{ \Stm_z \} } 
		\\\quad\mcode{default} \mcode{->} \{ \Stm_{d1} \solMerge \Stm_{d2} \}
		\\\}\s \Stm \solMerge \Stm'
	\end{array}
}
\]
Above, the merging of two switch statements is a switch whose guard is the merging of the original guards (\(\Exp \solMerge \Exp'\)).
Its cases consist of: for each case present in both the input switches (\(\id_a, \cdots, \id_x\)), we get a case in the result whose body merges the respective bodies of the original cases;
all cases that are not shared, which are simply put in the result as they are (the lists of cases $\many{\case\s \id_y\s \mcode{->} \s \Stm_y }$ from the first and $\many{\case\s \id_z: \Stm_z }$ from the second).

An example of the result of merging was presented for \code{DistAuth_Client} in \cref{sec:use_cases_distAuth}, where the cases for \code{OK} and \code{KO} are combined from the respective projections for \code{#Client#} of the two branches in the source choreographic conditional evaluated by \code{#IP#}.
These cases are produced by the rule for projecting selections, which applies to statements of the form
\(\Exp\mcode{;}\s\Stm\) when \(\Exp\) calls (possibly in a chain call) a method annotated with \annotation{@SelectionMethod}.
(Our type checker checks that these annotations are used only for methods that take enumerated types as parameters, cf.~\cref{sec:knowledge-of-choice}.)
For compactness, let $S = \many{\Exp.}{\params{\many{\TE}}}\id_1( \id_2@\mcode{#A#}'.\id_3 )$ where $\annotation{@SelectionMethod} \in \texttt{annotations}(\id_1)$.
\[
\epp{S\mcode{;} \Stm}
=
\begin{cases}
  \switch(\epp{S})
    \left\{ 
      \begin{array}l
        \mcode{case}\s id_3 \mcode{->} \{ \epp{\Stm} \}
        \\
        \mcode{default}\s \mcode{->} \{ \mcode{throw new ...} \}
      \end{array}
    \right\}
  & \mbox{if } S\colon \mcode{Enum<T>@#A#} \mbox{ for some } \mcode{T}
  \\
  \epp{S}; \epp{Stm} & \mbox{otherwise}
\end{cases}
\]
For the recipient of the selection (first case), the
statement becomes a switch on the projection of the \(\Exp\)ression that
will receive the selection, while the projection of the continuation \(\Stm\)
becomes the body of the corresponding case in the
argument. The projection for the other roles (second case) is standard,
projecting the \(\Exp\)ression followed by the projection of the
continuation \(\Stm\).

Our implementation of merging is smart enough to deal with some
`non-effectful' usages of \code{Unit}. For instance, consider the following
choreography.
\[
\mcode{if(true@#A#)\{System@#A#.out.println("true"@A);\}}
\]
If we project it at a role different from \code{#A#}, say \code{#B#}, we obtain
the code $\unit\uid( \unit\uid )$ for the then-branch, and $\blank$ for the
(missing) else-branch. These fragments are not mergeable, but our compiler uses
a \emph{unit-normalising operator}, given in \cref{sec:projection-full}, which
transforms also the first fragment into $\blank$ by removing the irrelevant
usages of $\unit$.

\section{Testing}
\label{sec:testing}

Testing implementations of choreographies is challenging since the distributed programs of all participants need to be integrated (integration testing).
We introduce ChoralUnit, a testing tool that enables the writing of integration tests as simple unit tests for choreographic classes.

\paragraph{Writing tests}
Following standard practice in object-oriented languages and inspired by JUnit, tests in ChoralUnit are defined as methods marked with a \annotation{@Test} annotation \citep{H04,GB06}.
For example, we can define the following unit test for the \code{VitalsStreaming} class from \cref{sec:use_cases_hospital}.
\begin{chorlisting}[breakable=true]
public class VitalsStreamingTest@(#Device#,#Gatherer#) {
	$\annotation{@Test}$
	public static void test1(){
		SymChannel@(#Device#,#Gatherer#)<Object> ch = 
			TestUtils@(#Device#,#Gatherer#).newLocalChannel("VST_channel1"@[#Device#,#Gatherer#]);
		new VitalsStreaming@(#Device#,#Gatherer#)(ch, new FakeSensor@#Device#())
			.gather(new PseudoChecker@#Gatherer#()); 
	}
}

class PseudoChecker@#R# implements Consumer@#R#<Vitals> {
	public void accept(Vitals@#R# vitals){ 
		Assert@#R#.assertTrue("bad pseudonymisation"@#R#, isPseudonymised(vitals)); 
	}
	private static Boolean@#R# isPseudonymised(Vitals@#R# vitals) { /* ... */ } 
}

class FakeSensor@#R# implements Sensor@#R# { /* ... */ }
\end{chorlisting}
The test method \code{test1} checks that data is pseudonymised correctly by \code{VitalsStreaming}.
Test methods must be annotated with \annotation{@Test}, be \code{static}, have no parameters, and
return no values.

At lines 4--5, we create a channel between the
\code{#Device#} and the \code{#Gatherer#} by invoking the
\code{TestUtils.newLocalChannel} method, which is provided by ChoralUnit as a library to simplify the creation of channels for testing purposes. This method returns an in-memory channel, which both \code{#Device#} and \code{#Gatherer#} will find by looking it up in a shared map under the key \code{"VST_channel1"}.
Thus, both roles must have the same key in their compiled code, which is guaranteed, here, by the fact that the expression \code{"VST_channel1"@[#Device#,#Gatherer#]} is syntactic sugar for \mbox{\code{"VST_channel1"@#Device#, "VST_channel1"@#Gatherer#}}. %

At lines 6--7, we create an instance of \code{VitalsStreaming} (the choreography we want to test). We use a \code{FakeSensor} object to simulate a sensor that sends some data containing sensitive information (omitted).
We then invoke the \code{gather} method, passing an implementation of a consumer that checks whether the data received by the \code{#Gatherer#} has been pseudonymised correctly.

Given a class like \code{VitalsStreamingTest}, ChoralUnit compiles it by invoking our compiler with a special flag (\code{-annotate}). This flag makes the compiler annotate each generated Java class with an \annotation{@Choreography} annotation that contains the name of its source Choral class and the role that the Java class implements.

When the compilation is finished, we can invoke ChoralUnit to run the tests from the class \code{VitalsStreamingTest}.
Once launched, the tool finds all the Java classes with an \annotation{@Choreography}
annotation whose \code{name} value corresponds to	\code{VitalsStreamingTest}.
By construction, each discovered class has a method with the same name, corresponding to the namesake
method from the source Choral test class (\code{test1}, in our example).
ChoralUnit exhaustively runs all the tests found in the test class. Namely, for
each \annotation{@Test}-annotated method and for each class generated from
the Choral source, ChoralUnit starts a thread that runs
the local implementation of that method implemented by that class.

In our example, \code{VitalsStreamingTest} is compiled to a class for \code{#Device#} and another for \code{#Gatherer#}, each with a \code{test1} method. Thus, ChoralUnit starts two threads, one running \code{test1} of the first generated Java class and the other running \code{test1} of the second generated Java class.

\paragraph{Multiparty assertions}
In the previous example, we have written an assertion (in \code{PseudoChecker}) that checks a condition at a single role (\code{#Gatherer#}).
Sometimes, it is useful to assert conditions that involve multiple roles.
A typical example is testing the correct implementation of protocols that aim at making two parties agree on a symmetric cryptographic key, like the Diffie-Hellman protocol \citep{DH76}.
In particular, after running the protocol, the two participants (say \code{#A#} and \code{#B#}) should have the same key.
We can express this assertion as follows.
\begin{chorlisting}[breakable=true]
Assert2@(#A#,#B#).assertEquals("key mismatch"@#B#, chAB, keyA@#A#, keyB@#B#);
\end{chorlisting}
Above, method \code{assertEquals} of class \code{Assert2} uses the channel \code{chAB} to communicate the key at \code{#A#} (\code{keyA}) from \code{#A#} to \code{#B#}, and then checks locally at \code{#B#} that it is equal to the key at \code{#B#} (\code{keyB}). If the check fails, an assertion error is raised at \code{#B#}.

Class \code{Assert2} can be user-defined, and likewise, developers can define classes that allow for assertions that involve more roles (e.g., \code{Assert3}, \code{Assert4}, etc.). In these implementations, the user can also freely code different protocols for communicating the data among the participants.

\section{Evaluation}
\label{sec:evaluation}

In \cref{sec:usecases}, we explored how Choral can be used to program
choreographies for a few realistic scenarios. In this section, we extend the
evaluation of our approach in three different directions:
\begin{enumerate}
\item In \cref{sec:from-java-to-choral}, we exemplify how one can use Choral to transition existing (Java) programs to choreographies.
In \cref{sec:karatsuba}, we show how Choral aids in
transitioning sequential algorithms into concurrent implementations. We consider a
Java algorithm and present the necessary steps to transform it into a Choral
program that distributes its computation over three nodes -- the number of nodes
follows naturally from the recursive structure of the algorithm. The steps are
straightforward, thanks also to the guidance offered by the Choral compiler.
In \cref{sec:retwisj}, we consider a complete, three-tier system: RetwisJ\footnote{\url{https://github.com/spring-attic/spring-data-keyvalue-examples/tree/master/retwisj}.}, a clone of Twitter implemented by the Spring team as an example of integration with the Redis data store. RetwisJ comes as a monolithic application that consists of mainly three components that respectively handle clients' invocations, the business logic, and the interaction with a data store. Following this design, in the transition to Choral (called ChoRetwis), each component appears as a role in the choreography. ChoRetwis is a drop-in replacement for RetwisJ (e.g., we can use RetwisJ and ChoRetwis with the same clients and data store). As an advantage of our choreographic refactoring, the architecture is more flexible wrt deployment: all components can be deployed on the same or different machines.
\item In \cref{sec:choral-vs-akka}, we compare Choral to a popular alternative
for concurrent and distributed programming: reactive actors. We use the Akka
framework for Java as a representative for reactive actors. Since Choral is
essentially an extension of Java, this choice helps in comparing our approach
against more standard approaches to concurrent programming at the net of
linguistic differences. In addition to the key qualitative advantage that Choral
provides choreography compliance, we find that Choral contributes to keeping the
codebase smaller. Furthermore, we find that the Java code generated by the
Choral compiler is not significantly different in size from manually-written
Java code. This opens the door to a quantitative evaluation, which we carry
out in \cref{sec:microbenchmarks}.
\item In \cref{sec:microbenchmarks}, we present a quantitative evaluation of how Choral impacts software development and execution performance.
In \Cref{sec:compilation-benchmarks}, we report relevant measurements on the performance of the Choral compiler. Our analysis shows that using Choral leads to smaller codebases and that the numbers of roles and conditionals in a choreography are significant determinants of how much our approach contributes to this aspect. We also observe that both our type checker and our compiler are fast for our example set -- they provide feedback and complete the compilation in a matter of milliseconds. Thus, our approach does not significantly reduce the performance of
development toolchains.
In \cref{sec:runtime-benchmarks}, we look at the runtime performance of the code that we generate. Specifically, we compare the execution times of Choral and Akka implementations of the Karatsuba algorithm presented, respectively, in \cref{sec:from-java-to-choral} and \cref{sec:choral-vs-akka}. We find the performance of the two models
comparable (and the Choral variant performs the best in the majority
of the tested cases).
\end{enumerate}

Overall, our results are encouraging. In addition to the advantage of choreography compliance, smaller codebases tend to host fewer bugs \citep{BBCCFH10}, and Choral appears rather approachable when we consider the context of existing practices.

The code used in this evaluation is available in two public repositories. The first contains the material for the comparison with Java and Akka (both quantitative and qualitative) and for benchmarking the performance of the compiler.\footnote{\url{https://github.com/choral-lang/evaluation}.} The second includes the original code and the Choral re-implementation of RetwisJ.\footnote{\url{https://github.com/choral-lang/choretwis}.}

\begin{figure}[t]
	\centering
  \begin{minipage}[b]{0.49\textwidth}
		\includegraphics[clip, trim=4.8cm 15.55cm 5.6cm 4.2cm, width=\textwidth]{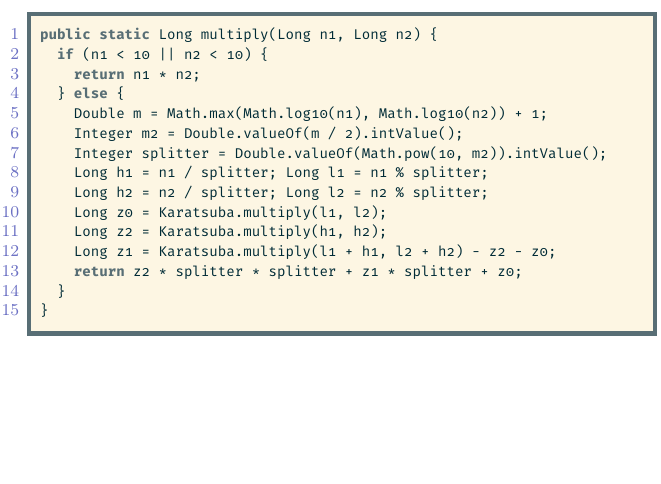}
  \end{minipage}
  \hfill
  \begin{minipage}[b]{0.5\textwidth}
		\includegraphics[clip, trim=5cm 15.5cm 4.7cm 4.2cm, width=\textwidth]{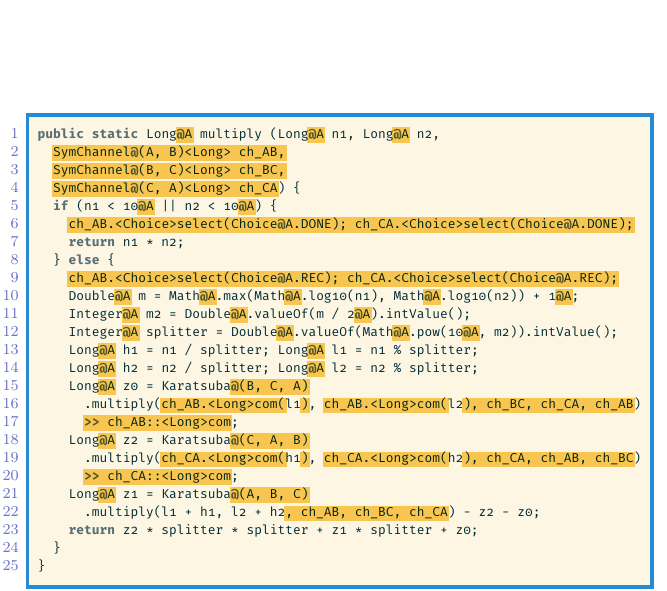}
  \end{minipage}
	\caption{Karatsuba algorithm. Left: Java (sequential). Right: Choral (choreographic).}
	\label{fig:karatsuba}
\end{figure}

\subsection{From Java to Choral}
\label{sec:from-java-to-choral}
Thanks to the fact that Choral is based on mainstream abstractions, we can use the implementation of a sequential algorithm in Java as a starting point to obtain a concurrent variant %
(in Choral).

\subsubsection{Transforming an algorithm: Karatsuba}
\label{sec:karatsuba}
Consider the algorithm for fast multiplication by \citet{KO62}. We report an implementation of that algorithm in Java on the left side of \cref{fig:karatsuba}.
Starting from the Java implementation, we can obtain a distributed implementation of the same algorithm in Choral by adding:
\begin{itemize}
\item Information on where the data is located.
\item Data transmissions for moving the data and implementing knowledge of choice.
\end{itemize}
We report the resulting Choral program on the right side of \cref{fig:karatsuba}, highlighting the additions from the original Java code in yellow.

The Choral program has three roles (\code{#A#}, \code{#B#}, and \code{#C#}), which distribute among themselves the three
sub-calculations of the algorithm.
In the parameters and return type, we added information on data locality (e.g., \code{Long@#A# n1}) and the necessary channels (e.g., \code{ch_AB}) for moving data in the implementation of the method. Given the original Java code, the type checker of the Choral compiler would assist the programmer by pointing out that data locality information must be added.
Likewise, in the implementation of the method, we added information on data locality for constant values and variables (e.g., \code{Double@#A# m}).
Additionally, we added the necessary data transmissions: selections to implement knowledge of choice for the conditional, and communications of values whenever they should move from a role to another. Again, the Choral compiler aids the programmer by asking for all this information.

\subsubsection{Transforming an entire system: RetwisJ (Redis-based Twitter clone)}
\label{sec:retwisj}
We now focus on the transformation of a complex, real-world system to provide a
more comprehensive view of the process. The transformation allows us to illustrate
how Choral can help developers transition from monolithic to distributed
implementations (\`a la microservices) while maintaining their options open wrt
code reuse, interoperability, and deployment configurations.

\begin{figure}
	\includegraphics[width=\textwidth]{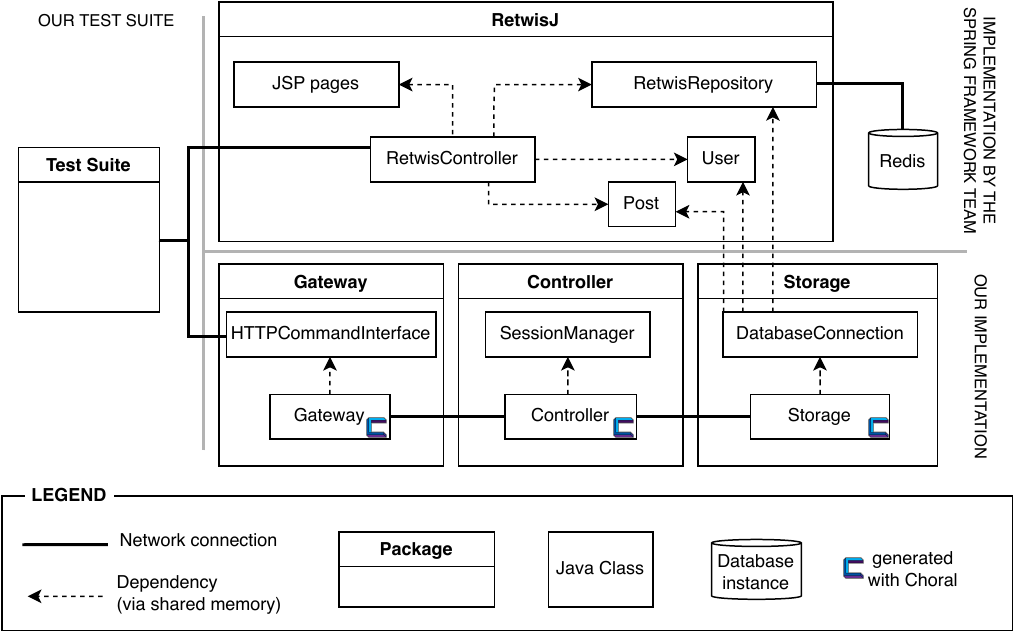}
	\caption{\label{fig:retwis}Diagram of the RetwisJ and ChorRetwis Systems (classes, packages, and deployment).}
\end{figure}

Concretely, we took RetwisJ\footnote{Source code available at
\url{https://github.com/spring-projects/spring-data-keyvalue-examples},
documentation available at
\url{https://docs.spring.io/spring-data/data-keyvalue/examples/retwisj/current/}.},
which is a Java, Spring-based port of Retwis\footnote{A Twitter clone originally
proposed by the Redis team to illustrate the capabilities of the data store,
described at \url{https://redis.io/topics/twitter-clone}.}, and re-implemented
its logic as a distributed application that consists of three separate modules.
We report, at the top of \cref{fig:retwis}, the simplified class diagram of
RetwisJ. The application is a monolith, where the central module
`RetwisController' works both as the gateway for serving webpages to the user
(the `JSP pages' module) and as the entry-point for user requests (e.g., to log in, to post tweets, etc.).
The classes `User' and `Post' model the main entities in the system, while the
`RetwisRepository' implements the logic for data persistence and retrieval.

Refactoring RetwisJ in Choral naturally follows well-known patterns from microservice architectures~\cite{DGLMMMS17,N21}: interaction with the client is handled by a `Gateway' component; business logic is managed by a `Controller' component; and data storage and access is managed by a `Storage' component. These components (depicted in the lower part of \cref{fig:retwis}) correspond to roles in our implementation, so we obtain a choreography that defines how these three roles collaborate to implement the application.

Each component is implemented by combining its respective code compiled from the choreography (for coordination) together with local code that implements the internal functionalities that are out of the scope of the choreography. For such internal functionalities -- e.g., the concrete data read/write operations on Redis -- we reuse existing code from RetwisJ. In \cref{fig:retwis}, we display which classes from RetwisJ have been reused as-is (e.g., `User' and `Post').
All three components are loosely coupled, in the sense that they interact purely via message passing (as instructed by the choreography).
Since the choreography uses our abstract channel interfaces, our implementation is more flexible than the original RetwisJ. Developers can choose to distribute the components by using, e.g., TCP/IP channels, or to deploy all of them as a single application using in-memory communication channels (which is the only option for RetwisJ).

The choreography is strategically parametric on a few notable aspects.
\begin{itemize}
\item The gateway receives API calls through a generic `CommandInterface', which allows us to expose the API over different media. In our concrete example, we implemented an `HTTPCommandInterface' for exposing a typical REST API (designed by us) and an alternative that acts as drop-in replacement for RetwisJ by implementing the API expected by the JSP pages provided in that project.
\item The controller delegates the storage and retrieval of session state to an abstract `SessionManager'. Our implementation stores state locally (in memory), but it can in principle be generalised to storing state on an external distributed store, to allow for replication.
\item The storage component relies on an abstract `DatabaseConnection' (a database abstraction layer), which determines how data is concretely represented, read, and written. Our implementation reuses the Redis-based code from RetwisJ. Thus, RetwisJ and our implementation can even be used in parallel. Storage is, however, not limited to using Redis and alternative implementations of `DatabaseConnection' can be provided.
\end{itemize}

To test our implementation and its consistency with the original RetwisJ, we developed a test suite that programmatically invokes the HTTP APIs of the two systems. The suite performs a series of tests that simulates usage, modifying state (e.g., creating users and posts) and then checking that the results are as expected.

\subsection{Programming paradigms: Choral and Akka}
\label{sec:choral-vs-akka}

We carry out a brief comparison between Choral and an established framework for concurrent programming: the Akka framework for the Java language.
Akka is a popular reactive framework based on actors for the `traditional' way of programming concurrent software; that is, software where each endpoint is programmed from a local viewpoint, in contrast with the global view on the expected interactions of choreographic programming. We use Akka version 2.6.18.

\begin{figure}
	\includegraphics[width=\textwidth]{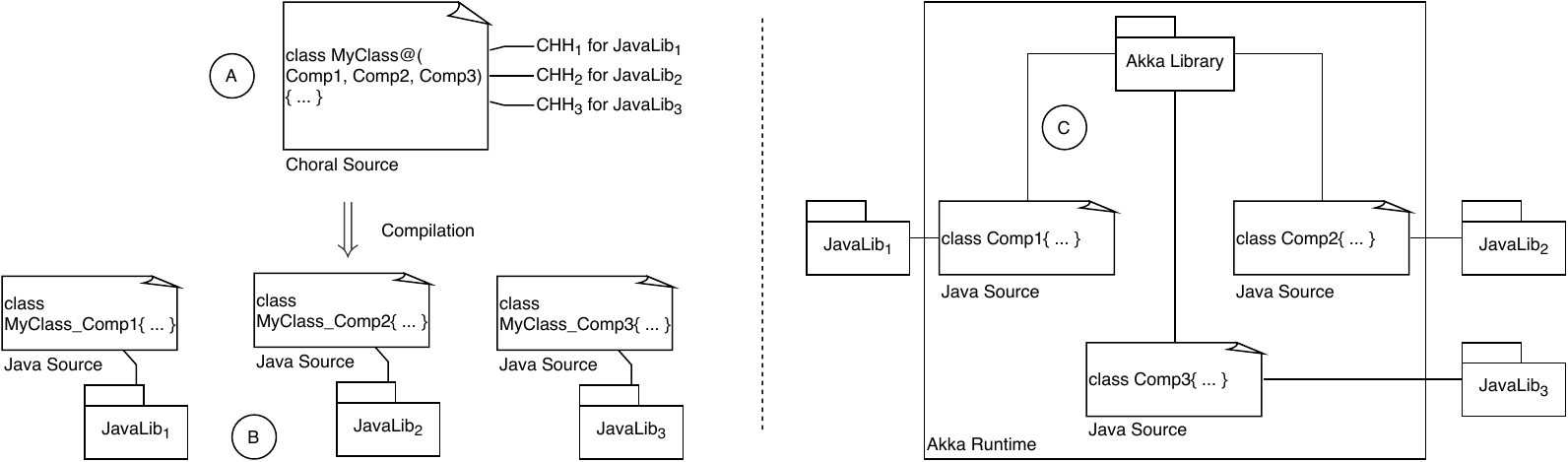}
	\caption{\label{fig:choral-vs-akka}Depiction of the programming approaches of Choral (left) and Akka (right).}
\end{figure}

We depict the development processes of Choral and Akka in \cref{fig:choral-vs-akka}, respectively on the left and on the right sides.
The processes are slightly different:
\begin{itemize}
\item In Choral, we implement a choreography using a single codebase (\textcircled{\scalebox{.8}{\textsf{A}}}). The codebase for each participant is then generated automatically by our compiler (\textcircled{\scalebox{.8}{\textsf{B}}}).
\item By contrast, in Akka, we implement the behaviour of each participant separately (\textcircled{\scalebox{.8}{\textsf{C}}}). There are no components to write choreographies.
\end{itemize}

Choral provides choreography compliance through its language and compiler. Akka provides no tools to express choreographies, nor to check for compliance -- these aspects must be handled manually by the programmer.

Both Choral and Akka require the programmer to adopt a few principles, respectively: for Choral, our notion of data types with multiple roles; for Akka, the design patterns and APIs expected by the Akka framework for user implementations, and the APIs of the Akka libraries.
Notably, Choral does not fix any APIs. The choice of APIs and implementations of channels or other methods are completely up to the user. Thus, Choral leaves more freedom to the developer in choosing what libraries to rely on. Furthermore, Choral requires no runtime library during execution, while Akka requires programmers to adopt the Akka runtime (represented in
\cref{fig:choral-vs-akka} by the \textsf{Akka Runtime} rectangle that surrounds the components).

Choral and Akka allow for reusing existing Java code and libraries, e.g., database drivers.
When Java code involves a single role, using it in Choral is straightforward -- we interpret any Java type as a type parametric at a single
role. In general, programmers can explicitly coerce Java types to arbitrary Choral types by using special header files called Choral Headers, shortened as CHH (\(\mathsf{CHH_1},\cdots,\mathsf{CHH_3}\) in \cref{fig:choral-vs-akka}), which isolate code that might be `unsafe' because manually written in Java.

To get more concrete observations and data, we manually implemented in Akka three choreographies
presented in this paper, namely DistAuth (see
\cref{sec:use_cases_distAuth}), MergeSort (see \cref{sec:use_cases_merge_sort}),
and Karatsuba (see the beginning in this section). In \cref{tab:choral-vs-akka}, we
compare the sizes of the three codebases for each example in terms of lines of code (LOC). Specifically, we report: the size of the Choral implementation,
the size of the Java code generated from the Choral implementation, and the size of the manually-written Akka implementation.

\begin{table}
	\begin{center} %
	\rowcolors{2}{gray!15}{white}
		\begin{tabular}{ l c c c }
		Program
		& Choral (LOC)
		& Java/Choral-generated (LOC)
		& Java/Akka (LOC)
		\\\midrule
		DistAuth           & 56 &  137 & 234 \\
		MergeSort          & 63 &  239 & 166 \\
		Karatsuba          & 31 &  92  & 118 \\
		\bottomrule
\end{tabular}
\end{center}
\caption{\label{tab:choral-vs-akka}Comparison of three codebases implemented in Choral and Akka.}
\end{table}

The comparison between the implementations of DistAuth is straightforward. The
main difference lies in the fact that Akka follows a reactive programming style
that dictates the usage of fields and messages of different types inside each
actor to track the (local) status of the protocol. The flow of interactions
becomes thus implicit, and needs to be reconstructed by the expected
asynchronous activations of methods at actors -- a similar observation has
already been made by \citet{WKS18}. Differently, in Choral the flow of
interactions is made explicit by our type system and sequencing of actions. This
difference makes the Choral codebase quite shorter, since the fields and message
types used to implement causal dependencies in Akka add boilerplate code. The
structures of the components between the generated Choral code and the Akka
implementation follow the same pattern, i.e., we have one class for each
participant (the Client, the Service, and the IP from
\cref{sec:use_cases_distAuth}). 

We then moved to implementing the MergeSort and Karatsuba examples, because their
recursive nature makes them a good fit for reactive programming as in Akka.
Indeed, differently from the DistAuth implementation that had a one-to-one
correspondence between the Choral-generated and Akka-based classes, the Akka
implementations of MergeSort and Karatsuba rely mainly on a single class that
defines the implementation of all roles. We find that the actor paradigm and the
programming style we used for Choral in this article promote different ways of
dealing with recursion. For the Akka implementation, we followed the idiomatic
approach of creating a new actor for each new recursive call of the distributed
algorithm.
By contrast, in the Choral implementations, it is natural to use the same participants in recursive calls by switching their roles as shown in \cref{sec:use_cases_merge_sort,sec:karatsuba} -- this lowers complexity and performance costs wrt coordination, since we avoid creating and managing new participants.
Adopting the `one class' style for Akka
helps keep the codebase small (the Java implementation generated from Choral is
bigger for MergeSort), but makes the implementation tightly coupled.

Interestingly, for the Karatsuba algorithm, the Choral-generated Java
implementation is smaller than the manually-written Akka implementation. This is
because Karatsuba requires more coordination and local tracking of the
distributed state, correspondent to boilerplate code that defines bookkeeping
message types and fields in Akka.

\paragraph{What went wrong}
There are valuable lessons that we learned from the exercise of writing implementations of the Karatsuba algorithm in Akka and Choral. In particular, writing the Akka implementation was trickier and more error-prone. We illustrate two representative issues, which arise from the lack of a choreographic view in Akka. These issues have been analysed by two of the authors working together to peer-review both the development process and the resulting code.

\begin{itemize}
\item For the first Akka implementation, we noticed that its performance was always the same for all tested inputs. It took us some time to notice that no communications were performed at all: all multiplications were resolved locally because there was a typo in the direction of the inequality check that decides whether to perform the \code{z0}-\code{z1}-\code{z2} decomposition of the product (cf.\ \cref{fig:karatsuba}) or perform it directly.
This was a very subtle bug, because the program was terminating successfully and returned the right results.

This issue was caused by the fact that the code for the Karatsuba algorithm needs to be distributed across different methods in Akka, which obfuscates the original structure of the algorithm and adds opportunities for banal bugs (e.g., swapping two arguments by mistake). We did not encounter this kind of issues with the Choral implementation, because we just needed to augment the original (correct) Java implementation with channel usage and roles (cf.\ \cref{fig:karatsuba}). The distribution of code in the final Java implementation has been carried out by the Choral compiler without mistakes.

\item Another bug that we encountered was due to a typo, too, which made the Akka implementation perform the calculation of \code{z2} instead of the one of \code{z1}.
However, more interestingly, this error manifested itself as a deadlock: an actor was waiting for the arrival of the three sub-calculations for \code{z0}, \code{z1}, \code{z2}, and instead received only those for \code{z0} and \code{z2} (the latter twice).
Fixing this kind of bugs is tricky in parallel programming, and indeed our first attempt at a fix introduced another bug: we fixed the deadlock, but we started obtaining wrong results because the fix caused a swap of the inputs for \code{z1} and \code{z2}.

The key reason behind these bugs was that, in Akka, we had to manually write code that reads and writes tags in messages to know if they contain the result of computing \code{z0}, \code{z1}, or \code{z2}. In this code, errors can be written in the code that creates messages at the sender and the code that processes them at the receiver. Implementations can therefore fall out of sync.
Choreographic programming (hence Choral) prevents this kind of bugs entirely because it is not possible to write mismatched communications at the choreographic level.
\end{itemize}

\subsection{Microbenchmarks}
\label{sec:microbenchmarks}
\begin{table}
	\begin{center}\footnotesize
		\rowcolors{2}{gray!15}{white}
		\resizebox{\textwidth}{!}{%
    \begin{tabular}{ l l l l l l l l l }
		\rotatebox{0}{Program}
		& \rotatebox{60}{Choral (LOC)}
		& \rotatebox{60}{\# Roles}
		& \rotatebox{60}{\# Conditionals}
		& \rotatebox{60}{Java (LOC)}
		& \rotatebox{60}{Size Increase (\%)}
		& \rotatebox{60}{Type Checking (ms)}
		& \rotatebox{60}{Proj. Checking (ms)}
		& \rotatebox{60}{Projection (ms)}
		\\\midrule
		HelloRoles         &  9 &  2 & 0 &  14 &  55\% & 5.915 & 0.334 & 0.187 \\
		ConsumeItems       & 16 &  2 & 1 &  49 & 206\% & 9.572 & 0.861 & 0.607 \\
		BuyerSellerShipper & 40 &  3 & 2 & 126 & 215\% & 8.204 & 1.274 & 1.015 \\
		DistAuth           & 56 &  3 & 1 & 137 & 144\% & 11.463 & 9.097 & 0.986 \\
		VitalsStreaming    & 47 &  2 & 1 &  78 &  65\% & 7.864 & 1.384 & 0.417 \\
		DiffieHellman      & 26 &  2 & 0 &  36 & 38\% & 5.911 & 0.232 & 0.152 \\
		MergeSort          & 63 &  3 & 4 & 239 & 279\% & 8.517 & 7.891 & 3.723 \\
		QuickSort          & 74 &  3 & 3 & 200 & 170\% & 7.213 & 6.204 & 2.806 \\
		Karatsuba          & 31 &  3 & 1 &  92 & 196\% & 6.491 & 2.566 & 1.078 \\
		DistAuth5          & 66 &  5 & 1 & 226 & 242\% & 10.581 & 5.573 & 1.036 \\
		DistAuth10         & 91 & 10 & 1 & 438 & 381\% & 10.576 & 5.643 & 3.011 \\
		\bottomrule
		\end{tabular}}
	\end{center}
	\caption{\label{tab:performance}Performance results for the Choral compiler.}
\end{table}

We now move to a more systematic and quantitative evaluation of how Choral impacts software development -- in addition to the key benefit of choreography compliance.
First, we evaluate the performance of the Choral compiler with microbenchmarks on 11 Choral programs. Then, since we implemented the same Karatsuba algorithm both in Java, Choral, and Akka, we provide some preliminary runtime benchmarks by contrasting their performance.

\subsubsection{Compilation Benchmarks}
\label{sec:compilation-benchmarks}
Regarding the performance of the Choral compiler, we report our results in
\cref{tab:performance}. There, for each program, we report (left to right): the
name of the Choral program, lines of code, number of roles, number of
conditionals (\code{if} and \code{switch} blocks), lines of code of the compiled
Java code (total for all roles), number of milliseconds to perform type
checking, the number of milliseconds to perform the check for projectability, and the number of milliseconds to perform the projection
(\cref{sec:compiler}). All code is well indented and the number of lines
just omits empty ones. We collected times on a machine equipped
with an Intel Core i5-3570K 3.4 GHz CPU and 12 GB of RAM, running macOS 10.15
and Java 17. The reported times are averages of 1000 runs each, after a warm-up
of 1000 prior runs.

\Cref{tab:performance} reports data for programs shown in this article, plus four other programs: BuyerSellerShipper is inspired by a recurring e-commerce example found in choreography articles \citep{CHY12,HYC16}; the Diffie–Hellman protocol for cryptographic key exchange \cite{DH76}; and DistAuth5 and DistAuth10, which are variants of the DistAuth class from \cref{sec:use_cases_distAuth},
where we respectively add 3 and 7 roles, 2 and 7 channels, and 4 and 14
selections for coordination.

Our preliminary data from \cref{sec:choral-vs-akka} points out that Choral programs are significantly smaller than the Java implementations compiled from them.
This is good in itself: recall that smaller codebases typically host fewer bugs \citep{BBCCFH10}.
In our microbenchmarks, compilation leads to an average increase of 181\% in codebase size
(going from the 38\% for \code{DiffieHellman} up to 381\% for \code{DistAuth10}): the difference between the sizes of the original Choral program and the generated Java code is a rough approximation of code that the programmer has been spared from writing manually.
Furthermore, our microbenchmarks suggest that there are two main parameters that affect this benefit:
\begin{itemize}
\item The number of roles involved in the source choreography. This is explained by the fact that each statement in Choral involving $n$ roles corresponds to $n$ statements in the generated Java code -- one for each role, implementing what the role has to do to follow the choreography. For example, a Choral statement invoking a method to communicate data from \code{#A#} and \code{#B#} would produce a statement in the code for \code{#A#} (for sending) and a statement in the code for \code{#B#} (for receiving).
There are examples of choreographies with many instructions that involve a single role, like \code{DiffieHellman}, which results in a smaller expansion.
\item The number of conditionals. Conditionals usually require performing selections to handle knowledge of choice (\cref{sec:knowledge-of-choice}). Then, the (code compiled for the) roles receiving selections has to inspect the type of the received message using a \code{switch} statement, which is automatically added by our compiler and is not present in the original Choral code. For example, MergeSort and QuickSort differ in that the former has 4 conditionals whereas the latter has 3 conditionals, and respectively reach an expansion of 279\% and 170\%.
\end{itemize}

As a final remark, we observe that type and projectability checking and projection do not add any
significant delay to the development experience: they respectively average ca.\
8.391ms, 3.732ms, and 1.365ms. This matches our own subjective programming experience with
Choral, where the compiler managed to feel quite responsive in providing quick
feedback while coding. Both operations are mostly influenced by the number of
conditionals and roles, matching our previous observations.

\subsubsection{Runtime Benchmarks}
\label{sec:runtime-benchmarks}
We conclude this section by comparing the execution times of the Choral
and Akka implementations of Karatsuba. We show the results in
\cref{fig:benchmark_runtime}. In the figure, we report in each of the six
plots the average execution time, in nanoseconds, of a sequence of 1000
multiplications. Each quadrant regards a specific `tier' of multiplication,
i.e., the \(10^9\) tier corresponds to the multiplication of two factors of the
shape, e.g., \(i*10^5\) and \(j*10^4\), which produce a result of that tier's
magnitude. All benchmarked implementations use the same inputs: we generated
and used six files, each containing 1000 pairs of random factors. Each file
corresponds to a tier. The considered tiers are: \(10^{9}\), \(10^{11}\),
\(10^{13}\), \(10^{15}\), \(10^{17}\), and \(10^{19}\) -- the latter nears the
maximal values managed by the Java \code{long} data type, but we make sure to
never produce overflows. We performed the benchmarks on the same machine used to
benchmark the Choral compiler above, with Java 17 and Akka 2.6.9. For each
implementation, we run the benchmark two times in sequence, discarding the data
of the first run to warm up the JVM and provide more stable results.

To benchmark both Akka and Choral implementations, we used in-memory
communications. We wrote an implementation of in-memory channels for Choral, while for Akka we used the default in-memory channels provided by the framework. In
\cref{fig:benchmark_runtime}, we report the execution times of the Choral and
Akka implementations both considering their setup (respectively, `ASetup' and
`CSetup') and without (respectively, `Akka' and `Choral'). The reason to
report setup times is to provide the reader with an indication of the wall-clock
times taken by the alternatives. The setup of Akka regards the creation of the
\code{ActorSystem} to execute the Karatsuba behaviour and its closure after
having obtained the result. The setup for Choral includes the creation of an
\code{Executor} pool of three threads, the creation of the three in-memory
channels, and the closure of the pool after having obtained the result.

The Choral implementation outperforms Akka's in all quadrants, both with and without the overhead from the setup -- indeed, although with a slight margin, the performance of the `Choral' implementation \emph{with} the setup times outperforms `Akka' without the
corresponding overhead. We also note that, contrarily to Choral, the wall-clock
execution times of Akka are largely dominated by its setup times. This
phenomenon is testified by the `ASetup' bar in \cref{fig:benchmark_runtime},
which has almost the same performance irrespective of the magnitude of the
computed result, while the alternatives show longer times for greater magnitudes.
We attribute this phenomenon to the different ways in which
the runtimes of the Choral and Akka in-memory implementations manage concurrency
and messaging. Indeed, skimming through the internal code of Akka, we found that
the framework puts in place threading models and advanced messaging systems
optimised for the execution of many actors that communicate in parallel, which
can take a high performance toll when implementing `lighter' computations,
like the one benchmarked here. In the future, we plan to investigate these aspects more in depth, identifying a set of benchmarks sensible to the peculiarities of
the chosen Choral and Akka runtimes and able to help to shed light on the
trade-offs of either approach.

\begin{figure}	
\resizebox{.34\textwidth}{!}{
\begin{tikzpicture}
\begin{axis}[
    ybar,
    bar width=1.2cm, %
    symbolic x coords={Choral, Akka, CSetup, ASetup},
    axis y line*=left,
    axis x line*=left,
    ylabel=time (ns),
    enlarge x limits=.2,
    xtick=data,
    title=Multiplications to \(10^{9}\)
 ]
    \addplot[
      draw=black,
      fill=sblue,
      error bars/.cd,
        y dir=both,
        y explicit,
        error bar style={line width=1pt,solid, black}
    ] 
    coordinates {
      (Choral,178181.981)
      (Akka,718318.628)
      (CSetup,390039.603)
      (ASetup,4673331.09)
    };
\end{axis} %
\end{tikzpicture}
}
\resizebox{.3\textwidth}{!}{
\begin{tikzpicture}
\begin{axis}[
    ybar,
    bar width=1.2cm, %
    symbolic x coords={Choral, Akka, CSetup, ASetup},
    axis y line*=left,
    axis x line*=left,
    enlarge x limits=.2,
    xtick=data,
    title=Multiplications to \(10^{11}\)
 ]
    \addplot[
      draw=black,
      fill=sblue,
      error bars/.cd,
        y dir=both,
        y explicit,
        error bar style={line width=1pt,solid, black}
    ] 
    coordinates {
      (Choral,216692.117)
      (Akka,712750.529)
      (CSetup,432934.4)
      (ASetup,4777349.722)
    };
\end{axis}
\end{tikzpicture}
}
\resizebox{.3\textwidth}{!}{
\begin{tikzpicture}
\begin{axis}[
    ybar,
    bar width=1.2cm, %
    symbolic x coords={Choral, Akka, CSetup, ASetup},
    axis y line*=left,
    axis x line*=left,
    enlarge x limits=.2,
    xtick=data,
    title=Multiplications to \(10^{13}\)
 ]
    \addplot[
      draw=black,
      fill=sblue,
      error bars/.cd,
        y dir=both,
        y explicit,
        error bar style={line width=1pt,solid, black}
    ] 
    coordinates {
      (Choral,284064.767)
      (Akka,715796.669)
      (CSetup,493363.831)
      (ASetup,4540766.345)
    };
\end{axis}
\end{tikzpicture}
}

\resizebox{.34\textwidth}{!}{
\begin{tikzpicture}
\begin{axis}[
    ybar,
    bar width=1.2cm, %
    symbolic x coords={Choral, Akka, CSetup, ASetup},
    axis y line*=left,
    axis x line*=left,
    ylabel=time (ns),
    enlarge x limits=.2,
    xtick=data,
    title=Multiplications to \(10^{15}\)
 ]
    \addplot[
      draw=black,
      fill=sblue,
      error bars/.cd,
        y dir=both,
        y explicit,
        error bar style={line width=1pt,solid, black}
    ] 
    coordinates {
      (Choral,344573.604)
      (Akka,840426.37)
      (CSetup,545932.884)
      (ASetup,4763393.688)
    };
\end{axis}
\end{tikzpicture}
}
\resizebox{.3\textwidth}{!}{
\begin{tikzpicture}
\begin{axis}[
    ybar,
    bar width=1.2cm, %
    symbolic x coords={Choral, Akka, CSetup, ASetup},
    axis y line*=left,
    axis x line*=left,
    enlarge x limits=.2,
    xtick=data,
    title=Multiplications to \(10^{17}\)
 ]
    \addplot[
      draw=black,
      fill=sblue,
      error bars/.cd,
        y dir=both,
        y explicit,
        error bar style={line width=1pt,solid, black}
    ] 
    coordinates {
      (Choral,431760.969)
      (Akka,741170.406)
      (CSetup,651810.152)
      (ASetup,4582646.977)
    };
\end{axis}
\end{tikzpicture}
}
\resizebox{.3\textwidth}{!}{
\begin{tikzpicture}
\begin{axis}[
    ybar,
    bar width=1.2cm, %
    symbolic x coords={Choral, Akka, CSetup, ASetup},
    axis y line*=left,
    axis x line*=left,
    enlarge x limits=.2,
    xtick=data,
    title=Multiplications to \(10^{19}\)
 ]
    \addplot[
      draw=black,
      fill=sblue,
      error bars/.cd,
        y dir=both,
        y explicit,
        error bar style={line width=1pt,solid, black}
    ] 
    coordinates {
      (Choral,517425.221)
      (Akka,774779.225)
      (CSetup,714182.889)
      (ASetup,4587585.262)
    };
\end{axis}
\end{tikzpicture}
}
\caption{\label{fig:benchmark_runtime}Benchmarks of Choral (Choral, CSetup), and Akka (Akka, ASetup) implementations of the Karatsuba algorithm.}
\end{figure}

\subsection{Threats to validity}
\label{sec:threats_to_validity}

Our evaluation covers a broad range of topics related to applicability and performance, which we believe indicate that Choral is a promising candidate to the programming of concurrent and distributed systems based on choreographies.
Here we discuss threats to the validity of our evaluation and directions for its future extension.

The experiments that we have presented include an algorithm (Karatsuba), a reference architecture (RetwisJ), one established programming framework (Akka), and several microbenchmarks.
These go on top of the other examples and use cases that we discussed in the previous sections, which included other concurrent and distributed scenarios.

Regarding Choral's applicability and usability, we could improve the validity of our evaluation by considering additional kinds of algorithms, architectures, and popular concurrent programming frameworks. Comparisons to other programming frameworks might also benefit from including a wider arrays of programs.
Another interesting direction would be to evaluate Choral's usability by conducting user studies that involve practitioners from academia and industry.
These user studies could provide precious input for the future growth and refinement of Choral.

Regarding the performance of Choral's compiler and generated code, a potential threat is that there might be performance bottlenecks that are not covered by our set of benchmarks.
An interesting future improvement could be to systematically extend this coverage, by developing a tool that synthesises Choral programs according to different patterns and constraints that are determined by input parameters -- such as those that we considered in~\cref{tab:performance}.
Another potential threat is that there might be parameters that significantly influence performance that we did not consider in~\cref{tab:performance}.
A tool for synthesising `random' choreographies could be useful in the discovery of such parameters.
\section{Related Work, Discussion, and Future Work}
\label{sec:discussion}

Choral is a \emph{choreographic programming language}, in that it makes the flow of interactions and their related computations manifest from a global viewpoint \citep{M13:phd}.
While Choral suffices already in tackling different kinds of use cases, as we have discussed in this article, the literature on choreographies is vast.
Choral includes and generalises many features found in previous work on choreographies. There are other features that we have not considered in this work, along with open problems that we pointed out when appropriate in the previous sections. We discuss related work and other potential future developments of Choral in the rest of this section.

\paragraph{Previous implementations of choreographic programming}
The idea of synthesising local participant specifications that comply with choreographies has been a hot research topic for more than 20 years, and work in this line of research is typically based on automata or process calculi abstractions \citep{AEY00,QZCY07,BBO12,HYC16,AIT18}.
Previous implementations of choreographic programming consist of Chor \citep{CM13} and AIOCJ \citep{DGGLM17}, which are based on process calculi and generate executable Jolie code. 
Compared to them, Choral solves the modularity problems mentioned in the Introduction, by revisiting choreographies under the light of mainstream abstractions.
Another advantage is that the types of channels needed by a choreography are made explicit and can be user-defined \citep{QZCY07,CHY12,CM13,HYC16}.

\paragraph{Other approaches to spatially-distributed programming}
The types that support our choreography-as-objects interpretation have been inspired by ideas found in modal logics for mobile ambients \cite{CG00} and, later, in the line of work on multitier programming \citep{MCHP04,NT05,CLWY06,SGL06,MCH07,LGVQWM09,WKS18}. In the words of \citet{MCHP04}, these works represent other approaches to `spatially-distributed computation'.
For example, in the most recent incarnation of multitier programming (ScalaLoci, by \citet{WKS18}), a distributed application is essentially defined as a single program that composes different functions, each localised at a single participant. A function can then invoke special primitives to request remote computation by another participant, whose implementation must always be ready for such requests \citep{WWS20}.
Differently from choreographies, this makes the flow of communications implicit and dependent on an underlying middleware -- indeed, multitier programming was not designed to address the problems of defining choreographies and addressing choreography compliance as an aim.
Choral generalises data types localised at a single participant to data types localised at many participants (roles), which enables our novel development process for choreography-compliant libraries.
\citet{CY20} explored the parallelisation of a simple multitier first-order functional language, for which they can infer abstract (computation is not included) choreographies of the communication flows that these programs can enact; Choral could be a candidate implementation language for this kind of models.
\citet{FLMD19} explored the idea of incorporating simple choreographic languages (without computation) to ensure that multitier code between two participants enacts specific protocols.

Choral's clear relation to the ideas found in logics for mobile ambients has already proven useful. In particular, \citet{GMPRSW21} use Choral to kickstart an investigation of the links and differences between choreographic and multitier programming, by taking Choral and ScalaLoci as representative languages.
After identifying the core abstractions that differentiate the two approaches, the authors provide algorithms for translating Choral code into ScalaLoci code and vice versa. Going from a multitier program to a choreographic one requires synthesising one of the many possible protocols (a choreography) that implements the necessary communications to execute the multitier program (which does not specify this aspect).
This connection paves the way for joint research and cross-fertilisation between the two communities \citep{GMPRSW21}.

\paragraph{Higher-order choreographies}
Interpreting choreographies as objects enables, for the first time, higher-order composition of choreographies that carry state (the fields of the objects): stateful choreographies (objects) can be passed as arguments.
Stateful choreographies have been investigated before without higher-order composition -- see, for example, the works by \citet{CHY12,CH12,CM20}.
\citet{DH12} studied how parameters that abstract choreographies can be expanded syntactically, but their choreographies cannot carry state and there must be a role that acts as an orchestrator to `enter' into a choreography (whereas in Choral, control is fully distributed).
In the setting of multitier programming, \citet{WS19} introduced a module system to write multitier programs as compositions of submodules. Differently from Choral and the work by \citet{DH12}, their approach requires fixing roles statically, whereas in our case roles can be freely instantiated at runtime -- for example, our merge sort example in \cref{sec:use_cases_merge_sort} exploits this feature when roles are exchanged in recursive calls. Our new data types might thus be interesting also in the setting of multitier programming.

\paragraph{Choral's principles in other settings}
The core idea in Choral's design is to have data types at multiple locations (the roles of the choreography): \code{T@(A_1, ... A_n)}. 
This information is the compass that guided us in lifting the various aspects of Java to the choreographic level and in designing our notion of projection.
We believe that this idea can be easily applied to other object-oriented languages -- C\#, Kotlin, Scala, etc. -- by extending their types in the same way and retracing our steps.

The applicability of this core idea goes even beyond object-oriented languages.
Choral's first technical report and release~\citep{GMP20,choral:website} have already inspired the investigation of theories and implementations of functional choreographic programming languages~\citep{CGLMP21,HG22,SKK23,GHM23}.
These studies confirm the generality of our approach: just like extending Java types with roles yields an object-oriented choreographic programming language (Choral), doing the same to the $\lambda$-calculus supports the development of a theory of functional choreographic programming~\citep{CGLMP21}.

\paragraph{Safety and liveness}
\label{sec:safety_and_liveness}
Theory of choreographic languages comes with strong safety and liveness properties, which stem from the high-level abstractions provided by choreographies and their projections to distributed code~\citep{M22}.
Under the same assumptions made (sometimes implicitly) in these proofs, Choral promises the traditional safety and liveness properties of choreographic programming languages.
We discuss these properties and how they relate to Choral's modularity and flexibility wrt communication middleware.
Furthermore, at the end, we report useful references and outline future research for the formalisation of these promises.

Choreographic programming languages guarantee \emph{communication safety}.
That is, processes never try to interact by performing incompatible communication actions -- if one action is `send' the other is a `receive', and the type of the sent message is always (a subtype of) the one expected by the receiver~\citep{M22,CHY12}.
Proofs of this result rely on the assumption that communications is reliable (messages are never lost, duplicated, or corrupted) or at least that message exchange primitives adopt suitable best-effort and timeout strategies~\citep{MP17,M22}.
Furthermore, communication primitives should respect a locality principle: messages should be dispatched to their intended recipients -- and vice versa, receiving from a role actually gives a message that was sent by that role~\citep{M13:phd,GMG18}.
In Choral, like in any previous implementation of choreographic languages, these assumptions form a contract for the implementation of channels.
This contract must be manually enforced, or communication methods might be
`wrong'.
A trivial example is a channel for communicating integers that always returns
the constant \code{1} at the receiver. In the future, it would be interesting to
explore the formalisation of contracts for channel middleware and libraries and
the development of verified implementations.

Another property of choreographic programming -- and probably the most known -- is \emph{deadlock-freedom by design}: the code compiled from a choreography is deadlock-free, because it is not possible to write mismatched communications in choreographies~\citep{CM13,DGGLM17,HG22,M22,JB22}.
The relevant assumptions, here, are that foreign code (in our case, Java code) used within a choreography always terminates, and that communication never blocks the sender or receiver indefinitely.
In the real world, these assumptions usually do not hold, so it is important to adopt timeout and supervision mechanisms to avoid divergence and deal with failures -- as exemplified in \cref{sec:handling-exceptions,sec:use_cases_hospital}.

Under the same assumptions for deadlock-freedom, a choreography that is tail-recursive gives the stronger liveness property of \emph{starvation-freedom} (or livelock-freedom): no role ever gets stuck, or `starves'~\citep{M22}. (In deadlock-freedom, it is sufficient that some part of the system keeps running~\citep{K00}.)
The same holds for Choral.

Lastly, choreographic programming languages typically guarantee
\emph{race-freedom}, provided that no role uses the same channel in parallel by
means of internal threads~\citep{HYC16,CHY12,M22,LGMZ08} -- unless the underlying
middleware can disambiguate messages, but this is not often the case. Choral
gives finer control over race freedom, thanks to its expressive (Channel) types
(cf. \cref{sec:on-channels}). For example, given a bidirectional channel, we can
pass it as a directed channel in one direction and as a directed channel in the
other direction to two separate threads. The API of directed channels would then
allow the two threads to use the channel, respectively, only for sending or for
receiving, which is safe to do in parallel. This control over channel usage was
found to be useful in the choreographic implementation of full-duplex
asynchronous communication (a pattern where roles can freely interleave different requests and responses in both directions), as found in the IRC client-server protocol. \citet{LM23} present a  detailed discussion
of this pattern and an interoperable implementation of IRC in Choral.

Since channel implementations play a key role in the properties discussed above, these implementations should ideally be simple, well-tested, or even verified.
Choral's features allow for encapsulating more sophisticated interaction behaviour as choreographies. \citet{LM23} started this activity, offering a reusable Choral library for programming asynchronous reactive protocols.

While, in principle, composing choreographies in Choral preserves safety and
liveness, one must pay attention to the interplay with local code. Most notably,
local code is free to compose and intertwine multiple instances of the same or
different choreographies, as we exemplified in \cref{sec:use_cases_hospital}.
This flexibility makes it possible to write two local programs that instantiate
two choreographies and try to communicate with each other in incompatible
orders, share the same channel across different instances of the same
choreography running in parallel, etc. Doing so can create communication errors
and make choreographies trivially timeout and fail. This problem can be tackled
in several ways, which we leave to future work. Briefly, one option
is to devise middleware for making local programs agree on which choreography
they want to engage in at any time. One such middleware can avoid some
incompatibilities, create dedicated channels, or, in general, throw a runtime
exception in case of disagreements. Other potential solutions include devising
static or runtime checks for how local code composes instances of
choreographies. Options for these checks encompass constraints such as keeping the graph of
connections among local programs acyclic; making (projections of) choreographies
that run in parallel have dedicated channels; ensuring that call to different
methods of different choreographic objects follow compatible orderings; and/or
synthesising an overall choreography that describes the collective behaviour of
how local programs combine their (sub)choreographies, thereby providing a
witness of safety and
liveness~\citep{MY13,CDYP16,VDG20,LTY15,CM13,HYC16,JBK22,CLM17}. Some of these
methods could be applied by means of structured concurrency libraries backed by
middleware.

`Bad' programming of code that interacts with projections of choreographies is
not a new issue, as it was already present in previous implementations of
choreographic languages~\citep{SDHY17,DGGLM17,CM13,M13:phd}. In fact, this is a
general issue when composing code from libraries that engage in communications,
even if these are not generated from choreographies.
At the very least, Choral already guarantees type safety: the APIs of
choreography projections are always respected.

Formalising Choral's ideas, implementation, and guarantees is an interesting area of research.
As already mentioned, the principles of compiling terms and data types equipped with roles as in Choral have been studied in the simple setting of the $\lambda$-calculus~\citep{CGLMP21,GHM23}.
The proofs in these models are not machine-checked.
A mechanisation could be achieved by building on the existing formalisations of the foundations of choreographic languages made with theorem provers~\citep{CMP21,CMP21b,CFGY21,PGSN22,HG22}.
These efforts are still focused on choreographic programming languages that are far less expressive than Choral, which by contrast is a large programming language that inherits all the complexities of both choreographic and object orientation.

\paragraph{Selection inference}
Choral requires the programmer to insert the necessary selections to achieve knowledge of choice (\cref{sec:knowledge-of-choice}).
Developing techniques for inferring these selections automatically is an ongoing research topic. Typically, these techniques either modify the source choreography to include extra selections or inject hidden communications in the generated endpoint code \cite{LMZ13,JSA15,BB16,DGGLM17,CM20}.
However, there is no silver bullet:
\begin{enumerate}
	\item
	In general, it is unfeasible to detect automatically what the optimal selection strategy is. This is a problem for both approaches (modifying the source choreography or injecting hidden communications in the generated code). Say that \code{#A#} needs to inform \code{#B#} and \code{#C#} of a choice by using point-to-point channels. Should \code{#A#} send the first selection to \code{#B#} or to \code{#C#}? That might depend on whether \code{#B#} has a longer task to perform in reaction to the selection compared to \code{#C#}, or vice versa. (Whichever has the longest task to start should get the selection first, to increase parallelism.) And what if multiple channels are available? For example, if \code{#A#} shares a fast channel with \code{#B#} but not with \code{#C#}, and \code{#B#} shares a fast channel with \code{#C#}, then it might be good that \code{#A#} informs \code{#B#} and subsequently \code{#B#} informs \code{#C#} (instead of \code{#A#} informing \code{#C#} directly). These issues become even more sophisticated when considering choreographies with more complex network topologies, scatter-gather channels, recursion, etc.
	\item
	If a compiler injects hidden communications in the generated code, then the source choreography program does not faithfully represent the communications enacted by the system any more. This makes the choreography less useful when reasoning about, for example, efficiency -- network communications like selections are especially a huge performance factor -- and security -- hidden extra communications might leak information in ways not intended by the designer of the original protocol.
\end{enumerate}
Since both issues are still the object of active investigation, we decided that a first version of Choral should be a base that future work can use to work on them.
A promising compromise could be a hybrid assisted way. That is, the programmer should be able to write a choreography including some selections deemed important, but also potentially missing some necessary other selections; then, a tool should detect the missing selections and propose a solution. The programmer could thus decide whether to accept the proposal or improve it manually to achieve their requirements.

In general, we believe that there is a lot of potential in future research on how to optimise communications in Choral. New algorithms might leverage annotations of channels, static analysis, and profiling data. Some algorithms might choose simpler approaches at the expense of parallelism, whereas others might take a more decentralised approach to spreading knowledge of choice to favour parallelism or energy saving (in the Internet of Things, spreading battery consumption evenly or in a focused way might be an advantage depending on the scenario).

\paragraph{Expressivity}
We discuss a few interesting directions for future work regarding the expressivity of Choral and of its type system. In general, we believe that our choreographic interpretation of OOP allows for importing established techniques from type theory to reason statically about roles in useful ways.

Choral can capture a variety of interaction patterns, like scatter-gather and producers-consumers with races. Nevertheless, there are cases where our types can be coarse.
For instance, the following could be an interface of a channel where two receivers, \code{#B#} and \code{#C#}, race to consume a message.
\begin{chorlisting}*[][numbers=none]
interface RaceDiChannel(#A#,#B#,#C#)<T@#X#> { 
	<S@#Y# extends T@#Y#> DPair@(#B#,#C#)<Optional<S>,Optional<S>¦¦> com(S@#A# m);
}
\end{chorlisting}
The return type of method \code{com} above guarantees that both receivers will have a value of type \code{Optional<S>} located at them.
However, depending on the behaviour that the programmer wishes to model, the interface above could be an over-approximation.
For example, implementations that simply discard the message sent from \code{#A#} or that deliver the message to both \code{#B#} and \code{#C#} will satisfy the return type.
Currently, we cannot express the type of a channel that forbids such implementations, e.g., a channel guaranteeing that exactly one between \code{#B#} and \code{#C#} will obtain the message. This means that, in such cases, we have to `pollute' the continuation of the choreography with local checks at both potential receivers.

A way to achieve a more specific type for races could be to extend Choral types with existential quantification over role parameters. For example, we could write \[\text{\code{S@#D# with #D# in [#B#,#C#]}}\] to express an instance of \code{S} located at some role \code{#D#} in the list \code{[#B#,#C#]} (i.e., \code{S@#D#} can be either \code{S@#B#} or \code{S@#C#}). With this type, we can write a more specific signature: method \code{com} returns a value of type \code{S@#B#} or of type \code{S@#C#}, but we cannot statically know which of the two types. 
\begin{chorlisting}*[][numbers=none]
interface RaceDiChannel(#A#,#B#,#C#)<T@#X#> { <S@#Y# extends T@#Y#> S@#D# with #D# in [#B#,#C#] com(S@#A# m); }
\end{chorlisting}
Although there is some work on the use of existential quantification in simple choreography languages \citep{JY20}, its application and integration with a general-purpose language like Choral poses some challenges and design choices. For instance, should roles that lose the race in method \code{com} be blocked? If so, is this specific of this method or the standard interpretation of every method with an existential return type? These and similar questions beg for a thorough investigation and go beyond Choral.
In fact, a satisfactory and general handling of races in choreographic programming languages is still missing.
In addition to the ideas just proposed, useful inspiration to address this aspect might come from nondeterministic choice in choreographies~\citep{LGMZ08,M22} and choreographic languages for the verification of message passing parallel programs~\citep{VMLY22}.

Another limitation of the current type system is that the number of role parameters of a choreography is fixed. This limitation is common to many choreographic languages. 
\cite{DY11} developed a theory for parameterising choreographies over
`collections of roles', whose sizes are determined at runtime. All roles in the
same collection must be treated uniformly (e.g., broadcast). We can import that
feature to Choral by allowing for role parameters to be collections. For example, we could prefix a role parameter declaration with \code{*}, as in \code{*#Ds#}, to specify that it is a collection of roles.
Then, we can write the type of a channel for broadcasting data from \code{#A#} to all roles in the collection \code{#Ds#} as follows.
\begin{chorlisting}*[][numbers=none]
interface BroadcastDiChannel(#A#, *#Ds#)<T@#X#> { /* ... */ }
\end{chorlisting}
The method \code{com} for this channel should take a value of type \code{S@#A#} and return a value of type \code{S@#D#} for every role \code{#D#} in the collection \code{#Ds#}. This would require investigating how Choral can be extended with types for distributed data collections, as well.

\paragraph{Error handling}
Choral supports exception handling at a single role, which can then propagate
errors to others via knowledge of choice. However, in our experience, it is more convenient
to represent failures in return types, like we did in \cref{sec:use_cases_distAuth} by using
\code{Optional}. The channel APIs that we showed in this paper are implemented by performing automatic
retries. These APIs also have equivalent versions that wrap results in
\code{Result} objects -- essentially sum types of the transmitted value type and
an error type, as in Go and Rust. Choosing among these
implementations is up to the choreography programmer, and programmers might also
devise channel implementations with their own strategies (e.g., exponential
backoff with bound on the number of retries).
Our compiler can, in principle, be extended to synthesise 
coordination for distributed exceptions, theorised by \citet{CHY08}.

\paragraph{Type and role inference}
Choral's intended audience consists of developers familiar with object-oriented programming (OOP) and, more specifically, Java.
For example, our syntax extends that of Java and our library of channels follows common OOP practices, like coding to interfaces~\citep{GHJV95}.
Our design choices rely heavily on generics to achieve reusability, similarly to what is done in the standard Java Collections framework~\citep{NW06}.
However, this comes at the cost of requiring that programmers have experience with generics and additional parameters in code.

Standard remedies to the verbosity of generics include shorthands like the diamond notation and type inference~\cite{SPT22}.
We plan to lift these features to Choral and expect that the standard solutions adopted for Java will be applicable in scenarios where roles are known, without major modifications.
For example, with such facilities, we could rewrite method \code{consumeItems} from \cref{sec:knowledge-of-choice} by removing all generic instantiations in its body, as follows.
\begin{chorlisting}
consumeItems(DiChannel@(#A#,#B#)<Item> ch, Iterator@#A#<Item> it, Consumer@#B#<Item> consumer) {
	if (it.hasNext()) {
		ch.select(Choice@#A#.GO);
		it.next() >> ch::com >> consumer::accept;
		consumeItems(ch, it, consumer);
	} else { 
		ch.select(Choice@#A#.STOP); 
	}
}
\end{chorlisting}

Similarly to generics, role parameterisation adds crucial flexibility at the cost of added verbosity.
It would therefore be interesting to explore inference mechanisms for role parameters, as well, in order the lighten the syntax of Choral even further.
At a basic level, programmers would be able to omit roles when these can be unambiguously determined from the context, e.g., in assignments and some method invocations.
The next snippets exemplify the potential gain in simplicity from the current situation (left) to one with this feature (right).
\trivlist\item\relax
\def\w{\dimexpr.5\textwidth-.7\parindent\relax}%
\begin{minipage}{\w}
\begin{chorlisting}*
// Without inference
List@#A#<String> l = 
  List@#A#.<String>of("Yes"@A, "No"@A);
String@#A# x = l.get(0@#A#);
\end{chorlisting}
\end{minipage}\hfill\begin{minipage}{\dimexpr \textwidth-\w-1.4\parindent\relax}
\begin{chorlisting}*
// With inference
List@#A#<String> l = 
  List.of("Yes", "No");
var x = l.get(0);
\end{chorlisting}
\end{minipage}
\endtrivlist
We conjecture that this feature can be achieved by a similar machinery to that already used for inferring generic parameters, because role information is available from type declarations and the typing context without ambiguity.

At a more advanced level, we could allow programmers to delegate to the Choral compiler decisions on the placement of data and computation.
Consider the snippet below.
\begin{chorlisting}*
int@C m(int@A x, int@B y) {
	int z = x + y;
	return z;
}
\end{chorlisting}
Since \code{x} and \code{y} reside at different roles, the location of \code{z} is ambiguous.
Even more, performing the addition at Line 2 requires communication, which the compiler would need to infer and inject by appropriately using channels available from the context.
This faces similar challenges to the ones previously discussed for selection inference.
We believe that exploring methods for the synthesis of communication strategies in choreographies is an interesting research line in general, with a scope that goes even beyond that of Choral's.

\paragraph{Other features from Java}
There are other features provided by Java (or other object-oriented languages) that Choral could benefit from. We do not discuss them in detail, because they we do not expect that lifting them to choreographies would pose significant challenges.
For example, we believe that adding wildcard types (\code{?}) would be a natural adaptation of Java's mechanism.

The patterns and libraries that support idiomatic Java programming, like streams (from the package \code{java.util.stream}), are immediately available in Choral thanks to our lifting of Java types to Choral types with a single role.
In some cases, it can be interesting to generalise these patterns to multiple roles.
For example, a possible interpretation of a `choreographic stream' at two roles \code{#A#} and \code{#B#} could be that of a stream of elements distributed at these roles (\code{Stream@(#A#, #B#)<T@(#C#, #D#)>}).
Methods for stream operations would then take choreographies at \code{#A#} and \code{#B#} as input.
Whether these choreographies would perform communications between the two roles or not would be left to the implementor and is irrelevant to the implementation of the stream.

\paragraph{Asynchronous programming}
The choreographies that we presented here use channel APIs as if they were
blocking. This does not mean that an endpoint must dedicate a thread for
participating in a choreography: future versions of Java will include fibers and
the asynchronous execution of blocking APIs (reactor pattern)
\citep{project_loom}. Choral is compatible with this direction. Should
programmers want to program a choreography explicitly for asynchronous execution
by using continuation-passing style, our channel
APIs can be extended to take choreographic continuations as parameters.

\paragraph{Fluid APIs from choreographic specifications}
As we mentioned and discussed in \cref{sec:introduction}, previous work explored the automatic generation of fluid APIs that enforces the code to follow a choreographic specification \citep{SDHY17}.
Such a choreographic language cannot include computation, so it cannot express any of our use cases, and its approach does not support modularity and API control, as we discussed more in detail in the Introduction.
Thus, Choral brings two improvements. First, our APIs are more reusable: they change only if the source API is changed, not if the communication behaviour of a method is simply updated.
Second, the APIs of our compiled Java code are more idiomatic: they are plain object APIs that look like the typical task-oriented APIs distributed by cloud vendors \citep{M08,W12}, which makes Choral a candidate drop-in replacement for current development practices.

\paragraph{Choreography-based verification and testing}
Choreographic languages that are less expressive than Choral (e.g., they cannot include computation) have been used also to verify that interactions among objects respect a protocol. This is obtained by statically checking method invocations, either by using typestates~\citep{KDPG18} or model checking~\citep{SYB19}.
As noted by \citet{HG22}, choreographic programming offers a simpler development method. Indeed, verification approaches require the programmer to design both a choreographic specification and then manually take care of writing a correct implementation of the projection of each role. On the contrary, choreographic programming (and hence Choral) generates the latter automatically.
Additionally, since our approach is based on compilation instead of verification, we can provide a more expressive choreographic language.

Choreographic languages without computation have been used also in a tool for testing abstract specifications of components given as labelled transition systems~\cite{CGT21}.
The purpose, there, is to test that the communication behaviour of a component (given as a labelled transition system) complies with a choreography.
By contrast, our testing tool is the first aimed at testing the functional correctness of a choreography and its generated implementation. Choral's capability of expressing internal computation is important to this end since it allows us to write arbitrary checks on the distributed state of the system.
\section{Conclusion}
\label{sec:conclusion}

With the increased adoption of cloud computing, edge computing, the Internet of
Things, and microservices, the need for libraries that implementors can use to participate correctly in choreographies is growing steadily \citep{M08,W12,AIM10,DGLMMMS17}.
Building on previous results on the theory of choreographies, choreographic programming came with the promise of aiding in the implementation of choreography-compliant concurrent and distributed software \citep{M15}. While the approach has been successfully applied in principle to different scenarios \citep{LN15,CM16,LNN16,LH17,DGGLM17}, the link between choreographic programming and mainstream programming has remained unexplored until now (all implementations rely on the Jolie programming language \citep{MGZ14}, which is based on the theory of CCS \citep{M80}). Among the most important consequences, no implementation of the paradigm so far properly supported modularity -- generating reusable libraries and controlling their APIs.

In this article, we have taken a fundamental step in the pursuit of the choreographic programming agenda. We have also shown that choreographies can be modelled by extending mainstream abstractions (in our case, objects) and that this leads to a choreographic programming language that supports modularity and can integrate with existing Java code. Choral is sufficiently expressive to capture use cases of different kinds, discussed the design of our compiler, and performed a first evaluation which points out that the approach is promising.
It is thus a step towards equipping programmers with a tool that safely ferries them from the design of choreographies to compliant implementations at the press of the proverbial button.

In the future, Choral could also be a useful vector for the application of research on choreographies based on other paradigms (automata, processes, etc.):
researchers could develop translations of their own choreography models to
Choral, and then leverage our compiler to obtain library implementations that
can be used in mainstream software (in Java).
Hopefully, this will lead to more implementations of choreography theories, allowing for their
evaluation \citep{Aetal16}.

\begin{acks}
This work was partially supported by Villum Fonden, grant no.\ 29518.
\end{acks}

\bibliography{biblio}

\ifappendix %
\appendix

\clearpage
\section{Projection to Java}
\label{sec:projection-full}
\newcommand{\bigepp}[2][\mcode{\#A\#}]{{\color{sblue}\left(\mspace{-2.5mu}\left|\normalcolor\mspace{-2mu}\array{l}#2\endarray\mspace{-2mu}\color{sblue}\right|\mspace{-2.5mu}\right)\normalcolor^{#1}}}
\allowdisplaybreaks
\def\rolesOf{\mathrm{rolesOf}}
\def\typeOf{\mathrm{typeOf}}
\def\head{\mathrm{head}}
\def\rest{\mathrm{rest}}
\subsection{Projection}
We omit modifiers ($\MOD$) and annotations ($\ANN$), they are rendered by the projection as they are.
\begin{alignat*}{4}
(\enum) & \quad && \sol{\mcode{enum}\s \id\mcode{@}\genW \s \{ \many{\id} \}}
		=
		\mcode{enum}\s \id \s \{ \many{\id} \}
\\
(\interface) &&& \sol{\mcode{interface}
		\s \id\at( \genWs )\s
			\params{ \many{\FTP} } 
			\s \mcode{extends}\s \TE \s \many{\mcode{\&}\s\TE}
			\s \{ \many{\MethodDef;} \}} = 
\\ &&& \left[\
			\mcode{interface}
			\s \mathrm{name}(\id,\genW,\genWs)\s
			\params{ \many{\sol{\FTP} } }
			\s\mcode{extends}\s \epp{\TE} \s \many{\mcode{\&}\s\epp{\TE}}\s
			 \{ \many{\epp{\MethodDef};} \}
			 \ \middle|\ \genW \in \genWs\ \right]
\\ 
(\class) &&& \solOpen\mcode{class}\s 
			\id\mcode{@} ( \genWs )\s \params{ \many{\FTP} } \s
			\mcode{extends}\s \TE \s
			\mcode{implements}\s \TE \many{\mcode{,}\TE}\ \{ 
			\many{\ClassField}\  
			\many{\ClassConst}\ 
			\many{\MethodDef\mcode{;} }\ 
\\    &&& 
			\many{\MethodDef\{ \Stm \} } \}\solClose =
			\left[\
			\mcode{class}\s \mathrm{name}(\id,\genW,\genWs)\s
			\mcode{extends}\s \epp{\TE} \s
			\mcode{implements}\s \epp{\TE} \s \many{\mcode{,}\s \epp{\TE}}
			\right.
\\ &&&\left.
			\{ \many{\epp{\ClassField}} \ 
			\many{\epp{\ClassConst}}\ 
			\many{\epp{\MethodDef}\mcode{;}}\ 
			\many{\epp{\MethodDef}\{ \unitSol{\epp{\Stm}} \}} \}
		  \ \middle|\ \genW \in \genWs\ \right]
\\
(\FTP) &&&
		\sol{\id\mcode{@}( \genWs )\s
			\mcode{extends}\s \TE \s \many{\mcode{,}\s\TE} }
		= \begin{cases}
		[\ \id\_{\genW} \s \mcode{extends} \s \epp{\TE} 
				\s \many{\mcode{&}\s \epp{\TE}}\ |\ \genW \in \genWs\ ] & \mbox{if } |{\genW}| \geq 1\\
		\id \s \mcode{extends} \s \epp{\TE}	& \mbox{otherwise}
		\end{cases}
\\
(\TE) &&&
		\epp{\mcode{void}} = \mcode{void}
\\ &&&
		\epp{id\mcode{@(}\many{\mcode{#B#}}\mcode{)<}\many{\TE}\mcode{>}} = 
		\begin{cases}
			id\mcode{<}\many{\epp{\TE}}\mcode{>} & \mbox{ } \many{\mcode{#B#}} = \mcode{#A#}\\
			id\mcode{_}A'{<}\many{\epp{\TE}}\mcode{>} & \mbox{ } \mcode{#A#} \text{ is the $i$-th element of } \many{\mcode{#B#}} \text{ and } \texttt{roleName}(id,i) = A' \\
			\mcode{Unit} & \text{otherwise}
		\end{cases}
\\
(\MethodDef) &&&
		\epp{\params{ \many{\FTP} } \s \TE\s \id\s(\many{\VarDecl})}
		=	\params{\sol{\many{\FTP}}}\s 
		\epp{\TE}\s \id\s(\many{\epp{\VarDecl}})
\\
(\ClassField) &&& \epp{\many{\VarDecl}\mcode{;}} = 
			\begin{cases}
				\epp{\TE} \id\mcode{;} & \mbox{if } \genW \in \rolesOf(\TE)\\
				\blank & \mbox{otherwise}
			\end{cases}
\\
(\ClassConst) &&& 
		\epp{\id\s (\many{\TE\s \id})\{\Stm\}}
		=
		\id\mcode{_}\genW (\many{\epp{\TE\s \id}})
		\{\unitSol{\epp{\Stm}}\}
\\(\Stm)&&& \epp{nil} = \epp{\blank} = \blank
\\&&& \epp{\mcode{return}\s\Exp\mcode{;}} = \mcode{return}\s\epp{\Exp}\mcode{;}
\\&&& \epp{\Exp\mcode{;}\Stm} = 
			\begin{cases}
				\!\!\array{l}
				\switch(\epp{\Exp}) \{ \\
			  \quad\mcode{case}\s id_3 \mcode{->} \{ \epp{\Stm} \}\\
			  \quad\mcode{default}\mcode{->} \{ \mcode{throw ...} \}\}
			  \endarray
			  &\!\!\!\!\array{l}
			  \mbox{if } \typeOf(\Exp) \mathrel{\texttt{<:}} \mcode{Enum@#A#}\mbox{,} \\
			  \quad\Exp = \many{\Exp'.}{\params{\many{\TE}}}\id_1( \id_2@\mcode{#A#}'.\id_3 ) \mbox{ and}\\
			  \quad\mcode{@SelectionMethod} \in \texttt{annotOf}(\id_1)
			  \endarray\\
				\epp{\Exp}\mcode{;}\epp{\Stm} & \mbox{if } \genW \in \rolesOf(\Exp)\\
				\epp{\Stm} & \mbox{otherwise}
			\end{cases}
\\&&& \epp{\TE\s\id \mcode{=} \Exp\mcode{;}\Stm} = 
			\begin{cases}
				\epp{\TE} \mcode{=} \epp{\Exp}\mcode{;}\epp{\Stm} & \mbox{if } \genW \in \rolesOf(\TE)\\
				\epp{\Exp}\mcode{;}\epp{\Stm} & \mbox{if } \genW \in \rolesOf(\Exp)\setminus\rolesOf(\TE)\\
				\epp{\Stm} & \mbox{otherwise}
			\end{cases}
\\&&& \epp{\Exp_1\s \AsgOp\s \Exp_2\mcode{;}\Stm} = 
			\begin{cases}
				\epp{\Exp_1} \AsgOp \epp{\Exp_2}\mcode{;}\epp{\Stm} & \mbox{if } \genW \in \rolesOf(\typeOf(\Exp))\\
				\epp{\Exp_1}\mcode{.id(}\epp{\Exp_2}\mcode{);}\epp{\Stm} & \mbox{if } \genW \in \rolesOf(\Exp_1,\Exp_2)\\
				\epp{\Stm} & \mbox{otherwise}
			\end{cases} 
\\&&& \epp{\mcode{if(}\Exp\mcode{)\{}\Stm_1\mcode{\}}\mcode{else\{}\Stm_2\mcode{\}}\Stm} = 
\\&&&
\begin{cases}
  \mcode{if}( \epp{\Exp} )\{ \epp{\Stm_1} \}
  \mcode{else}\{ \epp{\Stm_2}\} \epp{\Stm}
  & \mbox{if } \typeOf(\Exp) = \mcode{boolean}\at\genW
  \\
  \epp{\Exp}; \left\{ \unitSol{\epp{Stm_1}} \solMerge \unitSol{\epp{Stm_2}} \right\} \epp{Stm} 
  & \mbox{otherwise}
\end{cases}
\\&&& \epp{\mcode{\{}\Stm_1\mcode{\}}\Stm_2} = \mcode{\{}\epp{\Stm_1}\mcode{\}}\epp{\Stm_2}
\\&&& \epp{\mcode{try}\{ \Stm \} \many{\mcode{catch}(\VarDecl)\{ \Stm \}}\ \Stm} = \mcode{try}\{ \epp{\Stm} \} \many{\epp{\mcode{catch}(\VarDecl)\{ \Stm \}}}\ \epp{\Stm}
\\&&& \epp{\mcode{catch}(\VarDecl)\{ \Stm \}} = 
\begin{cases}
  \mcode{catch}(\epp{\TE}\s\id)\{ \epp{\Stm} \}
  & \mbox{if } \genW\in\rolesOf(\TE)
  \\
  \blank
  & \mbox{otherwise}
\end{cases}
\\(\Exp)&&& 
\epp{\cst} = \begin{cases}
l &\mbox{if } \cst = l\mcode{@(}\many{\mcode{#B#}}\mcode{)} \mbox{ and } \genW \in \many{\mcode{#B#}}\\
\mcode{Unit.id} & \mbox{otherwise}
\end{cases}
\\&&&\epp{\Exp\ \BinOp\ \Exp'} =
\begin{cases}
	\epp{Exp} \BinOp\ \epp{\Exp'} 
	& \mbox{if }
		\left( \begin{array}{l} 
			\BinOp \in \{\mcode{\&\&}, \mcode{||}\} 
		\\ \wedge\ \rolesOf(\Exp') = \{\mcode{#A}\}
		\end{array}
		\right) 
		\\ & \; \vee 
		\left( \begin{array}{l} 
			\BinOp \not \in \{\mcode{\&\&}, \mcode{||}\} 
			\\ \wedge\ \mcode{#A#} \in \rolesOf(\typeOf(\Exp))
		\end{array}
		\right)
	\\
	\mcode{Unit.id(}\epp{\Exp},\epp{\Exp'}\mcode{)}
	& \mbox{if }
			\left( \begin{array}{l} 
				\BinOp \in \{\mcode{\&\&}, \mcode{||}\} 
			\\ \wedge\ \rolesOf(\Exp') = \{\mcode{#A'#}\}
			\end{array}
		\right)
		\\ & \; \vee \BinOp \not \in \{\mcode{\&\&}, \mcode{||}\}
\end{cases}
\\&&&
	\mbox{let } \head(\Exp.\Exp') = \Exp_h,\ \rest(\Exp.\Exp') = \Exp_r
\\&&&
	\quad\epp{\Exp.\Exp'} = \quad\begin{cases}
		\epp{\Exp_h}\mcode{.id}(\epp{\Exp_r})
			& \mbox{if } \epp{\Exp_h} \in \{ \mcode{Unit.}\bullet\}
		\\
		\mcode{Unit.id(}\epp{\Exp_h}\mcode{)}\mcode{.}\many{\Exp_1}
			& \mbox{if } \epp{\Exp_r} \in \{ \mcode{Unit.}\many{\Exp_1} \}
		\\
		\epp{\Exp_h}.\epp{\Exp_r} & \mbox{otherwise}
	\end{cases}
\\&&&\epp{\params{\many{\TE}}id(\many{\Exp})} =
 \begin{cases}
  \params{\many{\epp{\TE}}}id(\many{\epp{\Exp}}) & \mbox{if \code{#A#}} \in \rolesOf(\typeOf(\params{\many{\TE}}id(\many{\Exp}))) \\ 
	\mcode{Unit.id(} \many{\epp{Exp}} \mcode{)} & \text{otherwise}
	\end{cases}
\\&&&\epp{id\mcode{@(} \many{\mcode{#B#}} \mcode{).}\params{\many{\TE}}id(\many{\Exp})} =
 \begin{cases}
	\epp{\id\mcode{@(} \many{\mcode{#B#}} \mcode{)}}\mcode{.}\params{ \many{\epp{\TE}}}\id\mcode{(} \many{\epp{\Exp}} \mcode{)} & \mbox{ \code{#A#}} \in \many{\mcode{#B#}} \\ 
	\mcode{Unit.id(} \many{\epp{Exp}} \mcode{)} & \text{otherwise}
	\end{cases}
\\&&&\epp{\new\s\params{\many{\TE}}id\params{\many{\TE}}(\many{\Exp})} =
 \begin{cases}
	\new\s \params{\many{\epp{\TE}}}\epp{\id\mcode{@(} \many{\mcode{#B#}} \mcode{)}\params{ \many{\TE}}}\mcode{(} \many{\epp{\Exp}} \mcode{)} & \mbox{ \code{#A#}} \in \many{\mcode{#B#}} \\ 
	\mcode{Unit.id(} \many{\epp{Exp}} \mcode{)} & \text{otherwise}
	\end{cases}
\\ (\FA) &&&\epp{id} = 
	\begin{cases}
		id & 
		\mbox{ } \mcode{#A#} \in \rolesOf(id)\\
		\mcode{Unit.id} & \text{otherwise}
	\end{cases}
\\&&&\epp{id\mcode{@(}\many{\mcode{#B#}}\mcode{).}f} =
	\begin{cases}
		\epp{\id\mcode{@(}\many{\mcode{#B#}}\mcode{)}}\mcode{.}f & 
		\mbox{ } \mcode{#A#} \in \rolesOf(f)\\
		\mcode{Unit.id} & \text{otherwise}
	\end{cases}
\end{alignat*}

\subsection{Merging}

\[\def\arraystretch{2}
\begin{array}{l}
	\bigMergeSign\ \many{\Stm} = 
	\bigMergeSign ( \Stm_1,\cdots,\Stm_n ) = 
	\unitSol{\Stm_1} \solMerge \cdots \solMerge \unitSol{Stm_n}
  \\
  \mbox{Statements}
  \\
	\mcode{return}\s \Exp \solMerge \mcode{return}\s \Exp' = 
	\mcode{return}\s \Exp \solMerge \Exp'
	\\
	\TE\s \id; \Stm \solMerge \TE\s \id; \Stm' 
	= \TE\s \id; ( \Stm \solMerge \Stm' )
	\\
	( \Exp_1 \s \AsgOp \s \Exp_2; \Stm ) \solMerge 
	( \Exp_1' \s \AsgOp \s \Exp_2'; \Stm' ) 
	\\ \qquad
	= ( \Exp_1 \solMerge \Exp_1' ) \s \AsgOp \s 
	( \Exp_2 \solMerge \Exp_2' ); ( \Stm \solMerge \Stm' )
	\\
 	( \Exp; \Stm ) \solMerge ( \Exp';\Stm' ) = ( \Exp \solMerge \Exp' ); ( \Stm \solMerge \Stm' )
	\\
	\{ \Stm_1 \}\s \Stm_2 \solMerge \{ \Stm_1' \}\s \Stm_2' = 
	\{ \Stm_1 \solMerge \Stm_1' \}\s ( \Stm_2 \solMerge \Stm_2' )
	\\
	\mcode{if} ( \Exp ) \{ \Stm_1 \} \mcode{else} \{ \Stm_2 \} \Stm
	\solMerge
	\mcode{if} ( \Exp' ) \{ \Stm_1' \} \mcode{else} \{ \Stm_2' \} \Stm'
	\\ \qquad =
	\mcode{if} ( \Exp \solMerge \Exp' ) \{ \Stm_1 \solMerge \Stm_1' \} 
	\mcode{else} \{ \Stm_2 \solMerge \Stm_2' \} ( \Stm \solMerge \Stm' )
	\\[1em]
	{
	\hspace{-5pt}
	\def\arraystretch{1}
	\begin{array}l
	\switch\s ( \Exp ) \{ 
	\\\quad\case\s \id_a \mcode{->} \{ \Stm_a \}
	\\\quad\cdots
	\\\quad\case\s \id_x \mcode{->} \{ \Stm_x \}
	\\\quad\many{\case\s \id_y \mcode{->} \{ \Stm_y \} }
	\\\quad\mcode{default} \mcode{->} \{ \Stm_{d1} \}
	\\\}\s \Stm 
	\end{array}
	\solMerge 
	\hspace{-5pt}\begin{array}l
		\switch\s ( \Exp ) \{ 
		\\\quad\case\s \id_a \mcode{->} \{ \Stm_a' \}
		\\\quad\cdots
		\\\quad\case\s \id_x \mcode{->} \{ \Stm_x' \}
		\\\quad\many{\case\s \id_z \mcode{->} \{ \Stm_z \} }
		\\\quad\mcode{default} \mcode{->} \{ \Stm_{d2} \}
		\\\}\s \Stm'
		\end{array}
	=
	\hspace{-5pt}\begin{array}l
		\switch\s ( \Exp \solMerge \Exp' ) \{
		\\\quad\case\s \id_a \mcode{->} \{ \Stm_a \solMerge \Stm_a' \}
		\\\quad\cdots
		\\\quad\case\s \id_x \mcode{->} \{ \Stm_x \solMerge \Stm_x' \}
		\\\quad\many{\case\s \id_y \mcode{->} \{ \Stm_y \} }
		\\\quad\many{\case\s \id_z \mcode{->} \{ \Stm_z \} } 
		\\\quad\mcode{default} \mcode{->} \{ \Stm_{d1} \solMerge \Stm_{d2} \}
		\\\}\s \Stm \solMerge \Stm'
	\end{array}}
	\\
	\mcode{try}\s\{ \Stm_1 \}\s 
	\many{\mcode{catch}\s(\VarDecl)\s\{ \Stm \}}\s \Stm_2
	\solMerge
	\mcode{try}\s\{ \Stm_3 \}\s 
	\many{\mcode{catch}\s(\VarDecl)\s\{ \Stm' \}}\s \Stm_4
	\\[-1em]\quad =
	\mcode{try}\s\{ \Stm_1 \solMerge \Stm_3 \}\s 
	\many{\mcode{catch}\s(\VarDecl)\s\{ \Stm \solMerge \Stm' \}}\s 
	\Stm_2 \solMerge \Stm_4
  \\
  \mbox{Expressions}
  \\
	\mbox{let } \bullet \in \{ nil, \blank, \mcode{null}, \mcode{this}, \mcode{super}, \id \}, \;
	\bullet \solMerge \bullet = \bullet
	\\
	\mbox{let } \bullet \in \{ \new\s\id \s{\params{ \many{\TE} }}, \id \s{\params{ \many{\TE} }} \s \}, \;
	
	\bullet\s( \many{\Exp} )
	\solMerge
	\bullet\s( \many{\Exp'} )
	=
	\bullet\s( \many{\Exp \solMerge \Exp'} )
	\\
	( \Exp_1 \s \BinOp \s \Exp_2 ) \solMerge 
	( \Exp_1' \s \BinOp \s \Exp_2'; ) =
	( \Exp_1 \solMerge \Exp_1' ) \s \BinOp \s ( \Exp_2 \solMerge \Exp_2' )
	
	\\
	\Exp_1.\Exp_2 \solMerge \Exp_3.\Exp_4 =
	( \Exp_1 \solMerge \Exp_3 )(.\Exp_2 \solMerge .\Exp_4 )
	\\
	.id \solMerge .id = .id 
	\qquad 
	.id{\params{ \many{\TE} }}(\many{\Exp}) 
	\solMerge .id{\params{ \many{\TE} }}( \many{\Exp'} ) =
		.id{\params{ \many{\TE} }}( \many{\Exp \solMerge \Exp'} )
	\\
	.\Exp_1.\Exp_2 \solMerge .\Exp_3.\Exp_4 =
		(.\Exp_1 \solMerge .\Exp_3)(.\Exp_2 \solMerge .\Exp_4) 
\end{array}
\]

\subsection{Normaliser}

\[
\def\arraystretch{1.5}
\begin{array}{l}
  \mbox{Statements}
  \\
  \unitSol{ nil } = nil
  \qquad
  \unitSol{ \mcode{return}\s \Exp; } = 
  \mcode{return}\s \unitSol{ \Exp };
  \qquad
  \unitSol{ \VarDecl; \Stm } = \VarDecl; \unitSol{ \Stm }
  \\
  \unitSol{ \Exp \s \AsgOp \s \Exp'; \Stm } = \unitSol{ \Exp } \s \AsgOp \s \unitSol{ \Exp' }; \unitSol{ \Stm }
  \\
  \unitSol{ \{ \Stm \}\s \Stm } = \{ \unitSol{ \Stm } \}\s \unitSol{ \Stm }
  \\
  \auxFn{noop}(\Exp) = \begin{cases}
    \blank & \mbox{if } \Exp \in \{\unit\uid,\ id\ \many{.id}, \mcode{this}, \mcode{null} \}
    \\
    \Exp & \mbox{otherwise}
  \end{cases}
  \\
  \unitSol{ \Exp; \Stm } = 
  \begin{cases}
    \unitSol{ \Stm } & \mbox{if } \auxFn{noop}(\unitSol{\Exp}) = \blank
    \\
    \unitSol{ \Exp }; \unitSol{ \Stm } &\mbox{otherwise}
  \end{cases}
  \\
  \unitSol{ \mcode{if}( \Exp )\{ \Stm_1 \} \mcode{else} \{ \Stm_2 \} \Stm }
  = \mcode{if}( \unitSol{ \Exp } )\{ \unitSol{ \Stm_1 } \} \mcode{else} \{ \unitSol{ \Stm_2 } \} \unitSol{ \Stm }
  \\
  \unitSol{ \mcode{switch}( \Exp )\{ \many{\case\s \id \mcode{->} \{ \Stm \}} \s {\mcode{default} \mcode{->} \Stm' } \}\s \Stm }
  \\ \qquad =
  \mcode{switch}( \unitSol{ \Exp } )\{ \many{\case\s \id \mcode{->} \{ \unitSol{ \Stm } \} } \s {\mcode{default} \mcode{->} \unitSol{ \Stm' } } \}\s \unitSol{ \Stm }
  \\
  \unitSol{\mcode{try}\s\{ \Stm \}\s \many{ \mcode{catch}\s( \VarDecl )\s\{ \Stm \} } \s \Stm }
  =
  \mcode{try}\s\{ \unitSol{ \Stm } \}\s \many{ \mcode{catch}\s( \VarDecl )\s\{ \unitSol{ \Stm } \} } \s \unitSol{ \Stm }
  \\
  \def\arraystretch{1.5}
  \mbox{Expressions}
  \\
  \unitSol{ \mcode{null} } = \mcode{null}
  \quad
  \unitSolChange{ \mcode{null} } = \langle \mbox{false}, \mcode{null} \rangle
  \quad
  \unitSol{ \mcode{this} } = \mcode{this}
  \quad
  \unitSolChange{ \mcode{this} } = \langle \mbox{false}, \mcode{this} \rangle
  \\
  \unitSol{ id } = id
  \qquad
  \unitSolChange{ id } = \langle \mbox{false}, id \rangle
  \qquad
  \mbox{let } \unitSolChange{ id{\params{\many{TE}}}(\many{\Exp}) } = \langle \bullet, \diamond \rangle,\
  \unitSol{ id{\params{\many{TE}}}(\many{\Exp}) } = \diamond
  \\
  \mbox{let } \many{\unitSolChange{Exp}} = 
  \many{\langle \bullet, \diamond \rangle},\
  \unitSolChange{ id{\params{\many{TE}}}(\many{\Exp}) } = 
  \langle 
    \bigvee \many{\bullet}, 
    id{\params{\many{TE}}}(\many{\diamond})
  \rangle
  \\
  \mbox{let } \unitSolChange{ \mcode{new}\s id{\params{\many{TE}}}(\many{\Exp}) } = \langle \bullet, \diamond \rangle,\
  \unitSol{ \mcode{new}\s id{\params{\many{TE}}}(\many{\Exp}) } = \diamond
  \\
  \mbox{let } \many{\unitSolChange{Exp}} = 
  \many{\langle \bullet, \diamond \rangle},\
  \unitSolChange{ \mcode{new}\s id{\params{\many{TE}}}(\many{\Exp}) } =
  \langle 
    \bigvee \many{\bullet}, 
    \mcode{new}\s id{\params{\many{TE}}}(\many{\diamond})
  \rangle
  \\
  \unitSol{ \Exp \s \BinOp \s \Exp' } = \unitSol{ \Exp } \s \BinOp \s \unitSol{ \Exp' }
  \\
  \mbox{let } \unitSolChange{\Exp.\Exp'} = \langle \bullet, \diamond \rangle,\
  \unitSol{ \Exp.\Exp' } = 
  \begin{cases}
    \unitSol{\diamond} & \mbox{if } \bullet = \mbox{true}
    \\
    \diamond & \mbox{otherwise}
  \end{cases}
  \\
  \unitSolChange{ \Exp.\Exp' } = 
  \begin{cases}
    \langle \mbox{true}, \unit\uid(\many{\Exp}) \rangle
    & \mbox{if } \Exp.\Exp' = \unit\uid\uid(\many{\Exp})
    \\
    \langle \mbox{true}, \Exp \rangle
    & \mbox{if } \Exp.\Exp' = \unit\uid( \Exp )
    \\
    \langle \mbox{false}, \unit\uid \rangle
    & \mbox{if } \Exp = \unit \mbox{ and }  
    \unitSolChange{.\Exp'} = \langle \bullet, \blank \rangle
    \\
    \langle \bullet \vee \bullet', \diamond\ \diamond' \rangle
    & \mbox{otherwise, let } \unitSolChange{\Exp} = \langle \bullet, \diamond \rangle
    \\ &\mbox{ and } \unitSolChange{.\Exp'} = \langle \bullet', \diamond' \rangle
  \end{cases}
  \\
  \unitSolChange{.id} = \langle \mbox{false}, .id \rangle
  \qquad
  \mbox{let } \unitSolChange{.\Exp} = \langle \bullet, \diamond \rangle,\
  \unitSolChange{.id.\Exp} = \langle \bullet, .id\ \diamond\rangle
  \\
  \mbox{let }
  \many{\unitSolChange{\Exp}} = \many{\langle \bullet, \diamond \rangle},\
  \newcommand{\blackdiamond}{
    \rotatebox[origin=c]{45}{\scalebox{.5}{\(\blacksquare\)}}
  }
  \unitSolChange{.id(\many{\Exp})} =
  \begin{cases}
    \langle \bigvee \many{\bullet}, .id(\many{\diamond}) \rangle
    & \mbox{if } .id \neq \uid
    \\
    \langle \mbox{true}, \blank \rangle
    & \mbox{if } \many{\auxFn{noop}(\diamond)} = \many{\blank}
    \\
    \langle \bigvee \many{\bullet} 
    \vee |\many{\diamond}| \neq |\many{\star}|,
    \uid(\star) \rangle
    & \mbox{otherwise}, 
    \mbox{let } \many{\auxFn{noop}(\diamond)} = \many{\star}
  \end{cases}
  \\
  \mbox{let } 
  \unitSolChange{.id{\params{\many{TE}}}(\many{\Exp})} =
  \langle \bullet, \diamond \rangle \mbox{ and }
  \unitSolChange{.\Exp} = \langle \bullet', \diamond' \rangle,\\
  \qquad
  \unitSolChange{.id{\params{\many{TE}}}(\many{\Exp}).\Exp} =
  \langle \bullet \vee \bullet', \diamond \ \diamond' \rangle
\end{array}
\]
\fi

\end{document}